\documentclass[aps,floatfix,reprint,superscriptaddress]{revtex4-2}
\usepackage{amsmath,amssymb,graphicx,mathtools}
\usepackage{float}

\usepackage[T1]{fontenc}

\usepackage{color}
\usepackage{colortbl}
\usepackage{nicefrac}
\usepackage[colorlinks]{hyperref}
\usepackage{tabularx}

\usepackage{calligra}

\usepackage{tikz-cd}

\usepackage{multirow}

\usepackage{diagbox}

\usepackage{scalerel}
\usepackage{stackengine,wasysym}

\newcommand\reallywidetilde[1]{\ThisStyle{%
  \setbox0=\hbox{$\SavedStyle#1$}%
  \stackengine{-.1\LMpt}{$\SavedStyle#1$}{%
    \stretchto{\scaleto{\SavedStyle\mkern.2mu\AC}{.5150\wd0}}{.6\ht0}%
  }{O}{c}{F}{T}{S}%
}}

\makeatletter
\DeclareRobustCommand{\cev}[1]{%
  {\mathpalette\do@cev{#1}}%
}
\newcommand{\do@cev}[2]{%
  \vbox{\offinterlineskip
    \sbox\z@{$\m@th#1 x$}%
    \ialign{##\cr
      \hidewidth\reflectbox{$\m@th#1\vec{}\mkern4mu$}\hidewidth\cr
      \noalign{\kern-\ht\z@}
      $\m@th#1#2$\cr
    }%
  }%
}
\makeatother

\newcounter{theoremcounter}
\stepcounter{theoremcounter}

\newtheorem{definition}{Definition}

\newcommand{\bs}{\boldsymbol}
\newcommand{\bra}[1]{\langle #1|}
\newcommand{\ket}[1]{|#1\rangle}

\newcommand{\ketbra}[2]{\ket{#1}\bra{#2}}
\newcommand{\partder}[2]{\frac{\partial #1}{\partial #2}}
\newcommand{\der}[2]{\frac{\text d #1}{\text d #2}}
\DeclareMathOperator{\Tr}{Tr}
\DeclareMathOperator{\pf}{\text{pf}}
\DeclareMathOperator{\Ai}{\text{{Ai}}\,}

\let\Re\relax

\DeclareMathOperator{\Re}{\text{Re}}

\definecolor{blue(ryb)}{rgb}{0.01, 0.28, 1.0}
\definecolor{scarlet}{rgb}{1.0, 0.13, 0.0}

\makeatletter
\newcommand{\vast}{\bBigg@{4}}
\newcommand{\Vast}{\bBigg@{5}}
\makeatother

\begin{document}

\title{Octonions and Quantum Gravity through the Central Charge Anomaly in the Clifford Algebra}%
\author{Lucas Kocia Kovalsky}
\affiliation{Quantum Algorithms and Applications Collaboratory, Sandia National Laboratories, Livermore, California 94550, U.S.A.}
\date{\today }
\begin{abstract}
  We derive a theory of quantum gravity containing an AdS$_3$ isometry/qubit duality. The theory is based on a superalgebra generalization of the enveloping algebra of the homogeneous AdS$_3$ spacetime isometry group and is isomorphic to the complexified octonion algebra through canonical quantization. %
  Its first three quaternion generators correspond to an \(\hbar\)-quantized AdS$_3$ embedded spacetime and its remaining four non-quaternion generators to a \(G\)-quantized embedding \(2+2\) Minkowski spacetime. %
  The quaternion algebra's expression after a monomorphism into the complexified Clifford algebra produces a two-dimensional conformal operator product expansion with a central charge anomaly, which results in an area-law \(\hbar G\) scaling satisfying the holographic principle and defines an ``arrow of time''. This relationship allows us to extend the theory through supersymmetry- and conformal-breaking \(\mathcal O(G)\) transformations of the embedding to produce perturbed AdS$_3$ spacetimes and derive a resolution to the black hole information paradox with an explicit mechanism. %
\end{abstract}
\maketitle

The relationship between general relativity and quantum theory has benefited recently from the study of the correspondence between anti-de Sitter (AdS) spacetime and conformal field theories (CFTs)~\cite{Maldacena99,Aharony00,Hayden07} and particularly promising results have been found regarding the thermodynamic and entropic properties of black holes~\cite{Almheiri13,Almheiri15,Brown16}. This correspondence elegantly encapsulates the holographic principle, wherein the properties of a larger dimensional bulk geometry are encoded in a lower-dimensional field theory that resides on its boundary~\cite{Ryu06,Ryu06_2,Headrick07,Brown16,Brown16_2,Susskind16}.

\begin{figure}[h]
\includegraphics[]{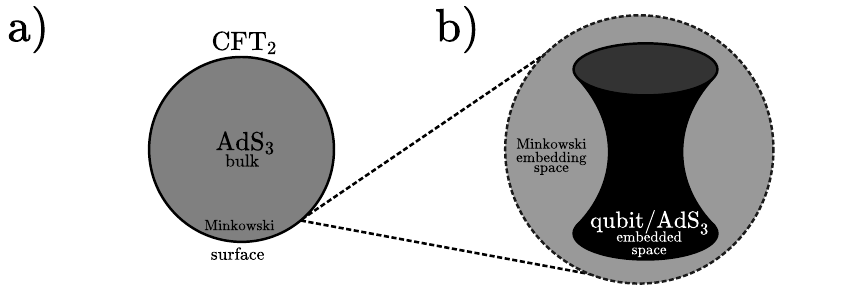}
\caption{a) The Poincar\'e surface of section for the AdS$_3$/CFT$_2$ correspondence. b) The embedding and embedded space for the AdS$_3$ isometry/qubit correspondence. In the conjectured AdS$_3$/CFT$_2$ correspondence, the AdS$_3$ bulk spacetime asymptotes to a Minkowski spacetime. Therein, the Minkowski spacetime inherits the AdS$_3$ \(\text{SO}(2,2)\) isometry group of the bulk and corresponds to a conformal field theory symmetric under \(\text{SO}(2,2)\). The correspondence is exact at this asymptotic boundary, where it is equivalent to embedding a hyperboloid onto a larger-dimensional Minkowski embedding spacetime that inherits its isometries and defines an AdS$_3$ embedded spacetime. The latter is the setting for the algebraic AdS$_3$ isometry group/qubit correspondence presented here.} %
\label{fig:AdS/qubit}
\end{figure}

AdS/CFT dualities are based on approximate postulates that form a sort of ``dictionary''~\cite{Witten98,Gubser98,Banks98,Susskind98,Polchinski99,Giddings99}. 
Here we present a quantum gravity theory with an AdS$_3$ isometry group/qubit duality that produces a very similar dictionary but instead depends on relations between finitely-many Lie superalgebra generators corresponding to spacetime and spin quantized degrees of freedom (see Figure~\ref{fig:AdS/qubit}). This is a supersymmetric (SUSY) extended model produced by considering the embedding of a spacetime in one of higher dimension~\cite{Mckeon04,Mckeon05,Mckeon13}. It is also a superalgebra generalization of a canonical (anti-) commutation relation (CAR/CCR) algebra using the complexified Clifford algebra \(\text{Cl}_{3,0}(\mathbb C)\), which provides a basis-free formulation of the commutation relations of the indefinite special orthogonal Lie groups \(\text{SO}(n,m)\) and the canonical qubit anticommutation relations. This proves to be a natural way to link the geometric covariant formulation of general relativity to a similar basis-independent formulation of quantum mechanics.

Our results can be summarized into the finding that the quaternion subalgebra of the octonion algebra emergently defines an AdS$_3$ spacetime on an embedding space and that a perturbed AdS$_3$ spacetime can be derived as an \(\mathcal O(G)\) transformation of this embedding space. %

We offer a short technical summary highlighting our particular contributions in Section~\ref{sec:summary} for the benefit of readers well-versed with Clifford algebras %
and would like to quickly evaluate how our results compare to prior work.

Otherwise, this paper can be divided into three distinct parts. The first concentrates on the qubit subtheory of the AdS$_3$/qubit correspondence, the second part focuses on the AdS$_3$ dual subtheory, and the third part develops further transformations on the AdS$_3$ dual subtheory to produce a perturbed AdS$_3$. More specifically, the first part introduces \(\text{SL}(2,\mathbb R)\) symmetry in complexified quaternions %
and a Majorana-to-qubit monomorphism from \(\text{Cl}_{3,0}(\mathbb C)\) to the complexified quaternion algebra isomorphic to \(\text{Cl}_{3,0}(\mathbb R)\) (Sections~\ref{sec:SLinoctonions}-\ref{sec:conformal}). The second part introduces octonions, whose non-quaternion elements define dual AdS$_3$ spacetime isomorphic to \(\text{Cl}^{[0]}_{2,0}(\mathbb C) \cong \text{Cl}_{3,0}(\mathbb R)\) and expressed as a subalgebra in \(\text{Cl}^{[0]}_{4,0}(\mathbb C)\) (Sections~\ref{sec:octonions}-\ref{sec:dictionary}). The third part covers extensions to a perturbed AdS$_3$ spacetime and the information paradox (Sections~\ref{sec:perturbedAdS3}-\ref{sec:conc}). See Figure~\ref{fig:CliffordAlgebramorphisms} in Section~\ref{sec:summary} for an overview of the relationships between these Clifford algebras. Note that there is really only one unique Clifford algebra in most of this construction since all are either isomorphic to the Clifford algebra \(\text{Cl}_{3,0}(\mathbb R)\) or its complexification. %

More specifically, %
in Section~\ref{sec:SLinoctonions} we introduce the \(\text{SL}(2, \mathbb R)\) symmetry group that covers the symmetry group of non-unitary qubit quantum mechanics and the AdS$_3$ isometry group, \(\text{SO}^+(2,2)/\text{SO}^+(1,2)\), in terms of the basis elements of the complexified quaternion algebra \(\mathbb H \otimes \mathbb C\). %
In the Sections~\ref{sec:Grassmannalgebra} we establish an monomorphism from the Clifford algebra \(\text{Cl}_{3,0}(\mathbb C)\) to \(\mathbb H \otimes \mathbb C \cong \text{Cl}_{3,0}(\mathbb R)\) that is dependent on a dimensionless constant \(\hbar\) %
We derive how to relate the elements of \(\text{Cl}_{3,0}(\mathbb C)\) under this monomorphism by orders of \(\hbar\) and define the associated Lagrangians and expectation values in this Clifford algebra that are equivalent to their quantum mechanical analogs. In Section~\ref{sec:hbarCliffordtheory} we derive an operator product expansion (OPE) in \(\text{Cl}_{3,0}(\mathbb C)\) that is isomorphic to multiplication in the quaternion algebra. We show how expressing this in \(\text{Cl}_{3,0}(\mathbb C)\) after the \(\hbar\)-dependent monomorphism makes this OPE equivalent to the commutation relations of the Virasoro algebra, i.e.~the OPE is conformal and exhibits the expected central charge anomaly dependent on \(\hbar\). %
In Section~\ref{sec:conformal}, we formally establish how this Clifford algebra respects conformal symmetry in its complexified Clifford algebra. %
In Section~\ref{sec:octonions} we introduce the octonion algebra as a supersymmetric \(G\)-extension of the \(\hbar\)-quantized quaternion algebra. %
We define the monomorphism from \(\text{Cl}_{2,0}(\mathbb C) \cong \text{Cl}_{3,0}^{[0]}(\mathbb R)\) subalgebra of \(\text{Cl}^{[0]}_{4,0}(\mathbb C)\) to \(\text{Cl}^{[0]}_{4,0}(\mathbb C)\). This allows us to define a dual operator product expansion in this larger \(\text{Cl}_{4,0}(\mathbb C)\) algebra. 
In Section~\ref{sec:holographicprinciple} we show how the holographic principle is present in the octonion theory. %
In Section~\ref{sec:dictionary} we summarize the ``dictionary'' of the AdS$_3$/qubit correspondence. %

\begin{figure}[t]
\includegraphics[]{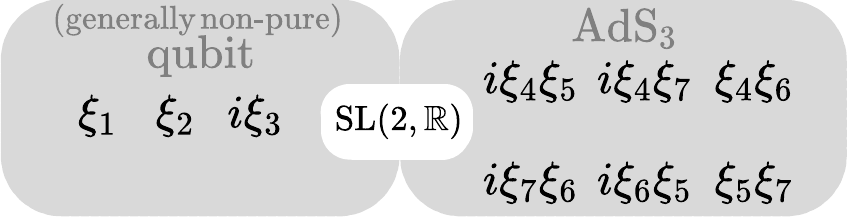}
\caption{AdS$_3$/qubit duality is governed by the algebraic identities that relate the first three non-identity generators of the octonion algebra (or any cyclic permutation w.r.t.~to imaginary coefficient \(i\)), which span the complexified quaternion algebra and exhibit \(\text{SL}(2,\mathbb R)\) symmetry, and the remaining four generators in paired groups.}
\label{fig:intro}
\end{figure}

This isomorphic dual super-subalgebra allows us to find that the central charge anomaly necessarily introduces eternal large black holes in AdS$_3$ through Killing horizons and singularities. %
Upon Wick rotation into imaginary time to produce a dual AdS Euclidean spacetime, the thermal AdS state becomes a Euclidean BTZ black hole.
We also show that the octonion theory satisfies the property that large thermodynamically stable black holes in AdS$_3$ are eternal and relate it to the renormalized subalgebra's complete description at \(\mathcal O(G^0)\).

In Section~\ref{sec:perturbedAdS3} we derive a perturbed AdS$_3$ spacetime through a \(\mathcal O(G)\) perturbative transformation of the octonion algebra. %
In Section~\ref{sec:RT} we extend this to derive the Ryu-Takayanagi formula for AdS$_3$ and the perturbed AdS$_3$ spacetime. %
In Section~\ref{sec:irreversibility} we discuss how evolution in the dual formalism is irreversible when the RG flow is taken into account, effectively defining an ``arrow of time''. %
We discuss more general properties of the full theory in Section~\ref{sec:discussion}.

In this part we link radiative black holes with \(\mathcal O(G)\) perturbations of homogeneous curved spacetime. Prior constructions have required a bath-coupling at the CFT boundary~\cite{Almheiri19,Penington20} to render eternal AdS black holes evaporative. This added construct in prior efforts can lead to uncertainty in its interpretation~\cite{Karlsson20,Karlsson21,Yoshinori21,Geng22}. The \(\mathcal O(G)\) perturbation of the embedding spaced requires no such coupling and allows for an explicit derivation of a mechanism for the turnaround of the Page curve, thereby providing a mechanistic resolution of the black hole information paradox. In particular, we show that Hawking pair production at the event horizon must be accompanied by wormhole creation along perpendicular isometries.

This approach also allows us to formally justify the maximin principle used to define quantum extremal surfaces when calculating RT surfaces. We show that as evaporation proceeds along the Page time the quantum extremal surfaces transition to higher-dimensional surfaces that loop progressively more times around singularities. We accomplish this by relating the Page time to the magnitude of the evaporating holographic area. This allows us to uniquely define which order in a semiclassical \(G \hbar\) expansion duality is dominant, equivalent to critical points in the associated RG flow. Since RG flow is irreversible in two-dimensions, this also specifies an ``arrow of time''.

We conclude in Section~\ref{sec:conc} and discuss future directions and implications.

The entire paper focuses only on the single-qubit Hilbert space. %

\subsection{Technical Summary}
\label{sec:summary}

We present here a technical high-level summary of the results presented in this paper, highlighting the novel contributions that we make, which we roughly organize into seven results.

\begin{figure}[h]
\begin{equation*}
  \begin{tikzcd}[remember picture]%
    {\color{black}\begin{tabular}{c}$\overbrace{\text{Cl}^{[0]}_{3,0}(\mathbb C)}^{\mathbb H \otimes \mathbb C} \cong \text{Cl}_{3,0}(\mathbb R)$\\\scriptsize{(qubit)}\\$\mathcal O(\hbar^1)$\end{tabular}} \arrow[r,hook,black, "{\color{black}\hbar > 0}"] \arrow[dd, leftrightarrow, black, start anchor={[yshift=6ex]}, start anchor={[xshift=-6ex]}, end anchor={[xshift=-6.4ex]}] & \substack{\text{\small{Cl$_{3,0}(\mathbb C)$}}\\\text{\tiny{(conformally}}\\\text{\tiny{symmetric}}\\\text{\tiny{cover of SL($2, \mathbb R$))}}\\{\color{black}\mathcal O(\hbar^0)}} \\
    \begin{tabular}{c}$\text{Cl}_{2,0}(\mathbb C)$\\\scriptsize{(AdS$_3$)}\\$\mathcal O(G^{1})$\end{tabular} \arrow[r, hook, "G>0"] \arrow[u, leftrightarrow, "\cong"] \arrow[u, leftrightarrow, "{\color{black}\hbar} \leftrightarrow G", swap] & \substack{\text{\small{Cl$_{4,0}^{[0]}(\mathbb C)$}}\\\text{\tiny{(non-}}\\\text{\tiny{conformally}}\\\text{\tiny{symmetric}}\\\text{\tiny{cover of SL($2, \mathbb R$)}}\\\text{\small{$\mathcal O(G^0)$}}} %
    \arrow[u, leftrightarrow, "\sim"] \arrow[u,leftrightarrow,"\substack{\text{non-}\\\text{canonical}}",swap] \\
    \underbrace{{\color{black}\text{Cl}^{[0]}_{3,0}(\mathbb C)} \oplus \text{Cl}_{2,0}(\mathbb C)}_{\mathbb O \otimes \mathbb C} \arrow[u, leftrightarrow, start anchor={[xshift=5ex]}, start anchor={[yshift=-0.3ex]}, end anchor={[yshift=0ex]}, end anchor={[xshift=0.5ex]}] & %
\end{tikzcd}
\begin{tikzpicture}[overlay,remember picture]
  \draw[black, thick, rounded corners] (-3.4,1.4) -- (-3.4,-0.95) -- (-6.25,-0.95) -- (-6.25,3.15) -- (0,3.15) -- (0,1.4) -- cycle;
  \draw[->, thick] (-1,3.4) -- (-4,3.4);
  \node[text width=4cm] at (-1.85,3.6) {irreversible RG flow};  
\end{tikzpicture}
\end{equation*}
\caption{The qubit quaternion theory can be expressed in \(\text{Cl}_{3,0}(\mathbb C)\) with three Grassmann generators through an \(\hbar\)-dependent monomorphism to the Pauli algebra \(\text{Cl}_{3,0}(\mathbb R)\) isomorphic to the complexified quaternions. This is a conformal theory when expressed in \(\text{Cl}_{3,0}(\mathbb C)\) since it contains the Witt algebra and the monomorphism centrally extends it to the Virasoro algebra. This monomorphism is defined by \(\hbar\)-quantization that produces a conformal operator product expansion (OPE), which when expressed in \(\text{Cl}_{3,0}(\mathbb C)\) produces an \(\mathcal O(\hbar^0)\) and conformal anomaly \(\mathcal O(\hbar^1)\) term, where \(\hbar\) is proportional to the central charge. This also defines an irreversible two-dimensional renormalization group (RG) flow, where \(\hbar\) plays the role of running coupling. This means that OPEs are dominated by their \(\mathcal O(\hbar^0)\) term in early evolution and \(\mathcal O(\hbar^1)\) term in later evolution, defining an evolution ``direction''. The Pauli algebra is isomorphic to \(\text{Cl}_{2,0}(\mathbb C)\), which when expressed as a subalgebra of \(\text{Cl}^{[0]}_{4,0}(\mathbb C)\), produces a dual algebra that is a universal cover of the AdS$_3$ isometry group. This is defined through a \(G\)-dependent monomorphism from \(\text{Cl}^{[0]}_{4,0}(\mathbb C)\) to \(\text{Cl}_{2,0}(\mathbb C)\) that allows the conformal OPE to be equivalently expressed in terms of \(\mathcal O(G^{-1})\) and \(\mathcal O(G^0)\) terms. The direct sum of the \(\text{Cl}^{[0]}_{3,0}(\mathbb C)\) and dual \(\text{Cl}_{2,0}(\mathbb C)\) algebras contain the complexified octonion algebra. An \(\mathcal O(G)\) transformation can extend \(\text{Cl}_{2,0}(\mathbb C)\) to the full \(\text{Cl}^{[0]}_{4,0}(\mathbb C)\) algebra.} %
\label{fig:CliffordAlgebramorphisms}
\end{figure}
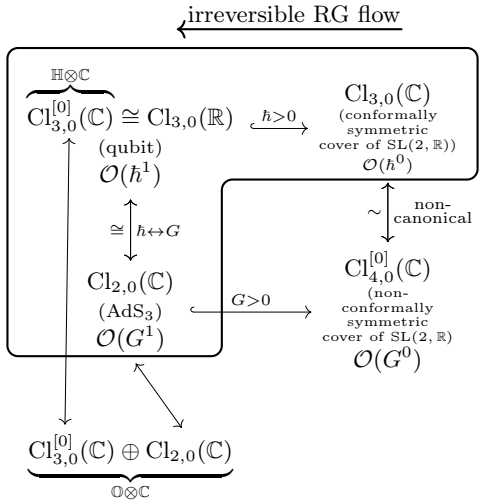

We begin by generating the universal cover \(\text{SL}(2,\mathbb R)\) of \(\text{SO}^+(2,2)\) in the usual manner, with a mix of even and odd Grassmann elements of the Clifford algebra (the Pauli algebra) \(\text{Cl}_{3,0}(\mathbb R)\), \(\xi_1\), \(\xi_2\), \(\xi_1 \xi_2\), expressed in the larger complexified Clifford algebra \(\text{Cl}_{3,0}(\mathbb C)\). We consider the complexified algebra because it also contains the Witt algebra generators that are closely related to the conformal symmetry group. Therefore, we consider a subalgebra of \(\text{Cl}_{3,0}(\mathbb C)\) that is monomorphic to \(\text{Cl}_{3,0}(\mathbb R)\), which is isomorphic to the complexified quaternions \(\mathbb H \otimes \mathbb C\) when all three permutations of its possible bases are included. These basis elements of \(\mathbb H \otimes \mathbb C\) are the quantized Grassmann generators we denote \(\{\hat \xi_i\}_{i=1}^3\) with non-zero anticommutator we define to be the dimensionless constant \(\hbar\).
Since \(\mathbb H \otimes \mathbb C\) has a non-zero anticommutator proportional to \(\hbar\), 
the isomorphism %
relates the even and odd terms of \(\text{Cl}_{3,0}(\mathbb R)\) (and thus \(\text{Cl}_{3,0}(\mathbb C)\) after complexification) by powers of \(\hbar\) %
and allows us to show that a conformal operator product expansion (OPE) of this algebra can be written in \(\text{Cl}_{3,0}(\mathbb C)\) equivalent to the product of the \(\mathbb H \otimes \mathbb C\) algebra. This OPE produces a finite series with an \(\mathcal O(\hbar^0)\) and \(\mathcal O(\hbar^1)\) term (see Eq.~\ref{eq:WeylSymbolGroenewoldRuleLeadingTerms_combined}). We identify \(\hbar/2\) as the central charge associated with the algebra's Casimir operator and show that the monomorphism to \(\text{Cl}_{3,0}(\mathbb R)\) produces an extension of the Witt algebra to the Virasoro algebra in \(\text{Cl}_{3,0}(\mathbb C)\), where the \(\mathcal O(\hbar^1)\) term is the central charge anomaly. This identifies \(\hbar\) as the running constant in an irreversible renormalization group flow of the two-dimensional conformal symmetry group \(\text{SO}^+(2,2)\). See Figure~\ref{fig:CliffordAlgebramorphisms} for a sketch of these morphisms. We believe this relationship with \(\hbar\) marks the first novel result of this paper, though similar results have been found in other contexts~\cite{Kocia17_3}.

\(\text{SL}(2, \mathbb R)\) is also the universal cover of the homogeneous AdS$_3$ isometry group. This can be shown by noting that \(\text{Cl}_{3,0}(\mathbb R)\) is also isomorphic to \(\text{Cl}_{2,0} (\mathbb C)\), which can be expressed as a subalgebra of \(\text{Cl}_{4,0}(\mathbb C)\) to define a non-associative extension of the three complexified basis elements of \(\mathbb H \otimes \mathbb C\) to four new elements \(\hat \xi_{\mu=4,\ldots,7}\), \(\hat \xi_1 = \alpha \hat \xi_4 \hat \xi_5 = \alpha \hat \xi_7 \hat \xi_6\), \(\hat \xi_2 = \alpha \hat \xi_4 \hat \xi_6 = \alpha \hat \xi_5 \hat \xi_7\), and \(\hat \xi_3 = \alpha \hat \xi_4 \hat \xi_7 = \alpha \hat \xi_6 \hat \xi_5\). These elements have an anticommutator set to a new non-zero dimensionless constant, we define to be \(G\). Taking into account the first commutator relation for the \(\mathbb H \otimes \mathbb C\) algebra, this sets \(\alpha = i \sqrt{2 \hbar}/G\). We show that these \(\hat \xi_\mu \hat \xi_\nu\) correspond to the isometries of AdS$_3$ spacetime when written as a homogeneous symmetric space, which is symmetric under the quotient group \(\text{SO}^+(2,2)/\text{SO}^+(1,2)\) and isomorphic to \(\text{SL}(2, \mathbb R)\). Considering the remaining algebraic multiplication rules, \(i \hat \xi_\mu = \hat \xi_i \hat \xi_\nu\), this identifies the \(\hat \xi_\mu\) as the translations in the embedding space, as has been found before in supersymmetric extensions~\cite{Mckeon04,Mckeon05,Mckeon13}. %

However, by restricting to the homogeneous symmetric spacetime%
, we are further able to find that together the elements \(\{\hat \xi_i\}_{i=1}^3\) and \(\{\hat \xi_\mu\}_{\mu=4}^7\) define the complexified octonion algebra, \(\mathbb O \otimes \mathbb C\). As far as we can find in prior literature, this is the first time that this association has been noticed and so we believe this marks the second novel result of this paper. This insight proves key to transforming this extension to spacetimes with non-negative curvature while preserving the property of a norm. [Moreover, as we shall show, the quotient structure of the symmetric spacetimes reduces the number of asymptotically timelike Killing vectors to one, which allows for a well-defined time-conjugate energy and thus stress-energy tensor.]

By showing that \(\hbar\) can be interpreted as a dimensionless Planck's constant due to a relationship between contextuality and classical integrable probability theories, we are then able to show that this larger octonion Lie superalgebra's nested anti-commutation relations satisfy the two-dimensional holographic principle if \(G\) is interpreted as a dimensionless gravitational constant. This marks the third novel result of this paper.

We can then define an expansion in the \(\text{Cl}_{2,0}(\mathbb C)\) subalgebra of \(\text{Cl}^{[0]}_{4,0}(\mathbb C)\) in terms of only the dual constant \(G\) that is isomorphic to the \(\hbar\)-dependent monomorphism from \(\text{Cl}_{3,0}(\mathbb C)\) that we defined before (see Eq.~\ref{eq:AdSWeylSymbolGroenewoldRule_combined}). This transforms the \(\mathcal O(\hbar^0)\) and \(\mathcal O(\hbar^1)\) terms in \(\text{Cl}_{3,0}(\mathbb C)\)'s operator product expansion to \(\mathcal O(G^{0})\) and \(\mathcal O(G^1)\), respectively. This is the fourth novel result of this paper.

Using this dual to the operator product expansion, we can find the dual of entanglement entropies in terms of the subalgebra \(\text{Cl}^{[0]}_{4,0}(\mathbb C)\) (see Eq.~\ref{eq:Weylentanglemententropydual}). We obtain agreement with the Ryu-Takayanagi generalized formula and Page curve for AdS$_3$ black holes with an explicit mechanism for the turn-around based on wormhole formation. We show that states in the dual subalgebra can be associated with symplectic areas defined by isometry group elements, which allows us to show that new quantum extremal surfaces are associated with each term of the dual operator product expansion. This is the fifth novel result of this paper.

We then proceed to show that, when considering the full octonion algebra, \(\mathcal O(G)\) transformations on the \(\hat \xi_\mu\) produce isometries that define a perturbed AdS$_3$ spacetime. %
This extends the dual algebra from \(\text{Cl}^{[0]}_{4,0}(\mathbb C)\) to \(\text{Cl}_{4,0}(\mathbb C)\) and adds a third \(\mathcal O(G^1)\) term to the dual operator product expansion. This is the sixth novel result of this paper and ultimately allows us to derive our final novel result: a mechanistic resolution to the blackhole information paradox.

\subsection{\(\text{SL}(2,\mathbb R)\) Symmetry in the Octonions}
\label{sec:SLinoctonions}

We begin by briefly reviewing unitary qubit quantum mechanics, which is generally more familiar compared to the non-unitary quantum theory. Unitary qubit quantum mechanics is defined within \(\text{SU}(2)\). An important rotation group is doubly covered by this unitary qubit group:
\begin{align}
  \text{SU}(2) \approx \text{SO}(3),
\end{align}
where \(\approx\) denotes a surjective homomorphism (i.e.~a double cover).

This can be shown by identifying \emph{real}-valued linear combinations of the Wick rotated quaternions of unit norm (i.e.~multiplied by \(i\)) as a parameterization of \(\text{SU}(2)\):
\begin{align}
  \hat g &= X_{0} \hat \xi_0 + i X_1 \hat \xi_1 + i X_2 \hat \xi_2 + i X_3 \hat \xi_3\\
  &= \left( \begin{array}{cc} X_{0} + i X_2 & i X_3 + iX_1\\ -i X_3 + i X_1& X_{0} - iX_2 \end{array} \right),
\end{align}
where
\begin{equation}
  \label{eq:SO3embeddingspace}
  X^2_{0} + X_1^2 + X_2^2 + X_3^2 = 1,
\end{equation}
\(X_i \in \mathbb R\), and \((\hat \xi_1, \hat \xi_2, \hat \xi_3) \equiv (\hat \sigma_x, \hat \sigma_z, \hat \sigma_y)\) are the Pauli operators.

Eq.~\ref{eq:SO3embeddingspace} is a restriction on a four-dimensional embedding space with metric signature \(({+}{+}{+}{+})\) to a three-dimensional embedded manifold (a sphere).

We can now identify \(\mathbb R^3\) with the span of \(\{\hat \xi_i\}_{i=1}^3\), and show that \(v \in \mathbb R^3\) is rotated around \(\bs X \equiv (X_1,X_2,X_3)\) with angle \(2 \theta\), where \(\cos \theta = X_0\) and \(|\sin \theta| = X^T X\), by
\begin{equation}
  \hat g v \hat g^{-1} = (-\hat g) v (-\hat g)^{-1} \in \mathbb R^3.
\end{equation}

Complexification of the corresponding algebra \(\mathfrak{su}(2)\) produces \(\mathfrak{sl}(2,\mathbb C)\) and reveals that the universal cover of this space is spanned by the basis
\begin{align}
  \hat T_3 &= -\frac{i}{2} \hat \xi_2,\\
  \hat T^{\pm} &= \frac{1}{2}(\hat \xi_1 \pm i \hat \xi_3).
\end{align}
This produces the Casimir operator
\begin{equation}
  \hat C_2 = \hat T^+ \hat T^- + \hat T_3^2 - T_3.
\end{equation}

Invariance under a Casimir operator (whose highest-weight eigenvalue is called the central charge) is the main property of conformal field theories that exhibit spacetime duals. However, such a spacetime duality requires the Casimir operator to correspond to a symmetry group that is non-Euclidean (i.e.~its metric signature cannot have the same signs like \((+++)\) for \(\text{SO}(3)\)).

We can obtain such a non-Euclidean space with spin-\(1/2\) qubits by noting that a non-unitary subspace can be defined on their Hilbert space by Wick rotating only two of their basis elements.

In particular, replacing \(i \hat \sigma_x\) and \(i \hat \sigma_z\) with \(\hat \sigma_x\) and \(\hat \sigma_z\), respectively, produces
\begin{align}
  \label{eq:SUisoSL}
  \text{SL}(2,\mathbb R) &\cong \frac{\text{SL}(2,\mathbb R) \times \text{SL}(2,\mathbb R)}{\text{SL}(2,\mathbb R)}\\
                                    &\approx \frac{\text{SO}^+(2,2)}{\text{SO}^+(1,2)} \\
                                    &\cong \frac{\text{SO}^+(1,2) \times \text{SO}^+(1,2)}{\text{SO}^+(1,2)}
\end{align}
where
\begin{equation}
  \text{SL}(2,\mathbb R) \equiv \{ A \in GL(2) : \det A = 1 \}.
\end{equation}

This relationship can be proven by verifying the following correspondence between group elements:
\begin{align}
  \label{eq:SLpm2Riso1}
  \hat \xi_1 &= \left( \begin{array}{cc}0& 1\\ 1& 0 \end{array} \right)  = \left( \begin{array}{cccc} 0& -1& 0& 0\\ -1& 0& 0& 0\\ 0& 0& 0& 1\\ 0& 0& 1& 0 \end{array} \right) = K^{01} \oplus K^{32},\\
  \hat \xi_2 &= \left( \begin{array}{cc}1& 0\\ 0& -1 \end{array} \right) = \left( \begin{array}{cccc} 0& 0& 0& -1\\ 0& 0& -1& 0\\ 0& -1& 0& 0\\ -1& 0& 0& 0 \end{array} \right) =  K^{21} \oplus K^{03},\\
  \label{eq:SLpm2Riso3}
  -i\hat \xi_3 &= \left( \begin{array}{cc}0& -1\\ 1& 0 \end{array} \right) = \left( \begin{array}{cccc} 0& 0& 1& 0\\ 0& 0& 0& 1\\ -1& 0& 0& 0\\ 0& -1& 0& 0 \end{array} \right) = J^{02} \oplus -J^{13},
\end{align}
where
\begin{equation}
  (K^{ij})_{kl} = \delta_{ik} \delta_{jl} + \delta_{il} \delta_{jk}
\end{equation}
are boosts on a spacetime with metric signature \(({-}{+}{-}{+})\) and
\begin{equation}
  (J^{ij})_{kl} = \epsilon_{ij} (\delta_{ik} \delta_{jl} + \delta_{il} \delta_{jk})
\end{equation}
are rotations on the same spacetime~\cite{Mckeon04}.

The Lorentz group \(\text{SO}^+(2,2)/\text{SO}^+(1,2)\) is the homogeneous symmetry group of AdS$_3$ spacetime without parity or time reversal~\cite{Ballesteros17}. The same correspondence can be found by cyclically permuting the \(\hat \xi_i\) w.r.t.~the coefficient \(i\). Therefore, Eqs.~\ref{eq:SLpm2Riso1}-\ref{eq:SLpm2Riso3} mean that the algebra defined by a non-normalized (spin-\(1/2\)) qubit on a Hilbert space contains three overlapping universal covers of AdS$_3$ spacetime.

\(\text{Spin}(3)\) is defined to be the three even elements \(\text{Cl}^{[0]}_{3,0}(\mathbb R)\) of the Clifford algebra \(\text{Cl}_{3,0}(\mathbb R)\) (defined in Section~\ref{sec:Grassmannalgebra}). These are the three generators of the four-dimensional quaternion algebra, \(\mathbb H\): \(\{i \hat \xi_1, i \hat \xi_2, i \hat \xi_3\}\). The odd elements of \(\text{Cl}_{3,0}(\mathbb R)\) do not form a subalgebra. The generators of the one-dimensional real algebra \(\mathbb R\) and the two-dimensional complex algebra \(\mathbb C\) can be similarly shown to be the even elements of \(\text{Cl}_{1,0}(\mathbb R)\) and \(\text{Cl}_{2,0}(\mathbb R)\), respectively. Considering larger Clifford algebras \(\text{Cl}_{n,0}(\mathbb R)\) reveals that the only remaining normed division algebra must have eight dimensions~\cite{Hurwitz98,Baez02}. %

Thus, these three Clifford subalgebras, \(\text{Cl}^{[0]}_{1,0}(\mathbb R)\), \(\text{Cl}^{[0]}_{2,0}(\mathbb R)\) and \(\text{Cl}^{[0]}_{3,0}(\mathbb R)\), can be associated with the normed division algebras \(\mathbb R\), \(\mathbb C\) and \(\mathbb H\). %
The same association does not hold for the largest eight dimensional normed algebra, but we shall see its corresponding lower dimensional Clifford algebra, \(\text{Cl}_{3,0}(\mathbb R)\), can still be used to build it in a different manner after complexification.

Identifying \(i\hat \xi_3 \equiv \hat \xi_1 \hat \xi_2\), it is clear from Eqs.~\ref{eq:SLpm2Riso1}-\ref{eq:SLpm2Riso3} that a mix of even and odd elements of \(\text{Cl}_{3,0}(\mathbb C)\) form the group %
\(\text{SL}(2, \mathbb R)\): \(\{\hat \xi_0\equiv \hat I_2, \hat \xi_1, \hat \xi_2, \hat \xi_1 \hat \xi_2\}\). Since these possess an irreducible direct sum representation given by Eqs~\ref{eq:SLpm2Riso1}-\ref{eq:SLpm2Riso3}, it is possible to construct a larger normed algebra by introducing four new elements \(\hat \xi_4\), \(\hat \xi_5\), \(\hat \xi_6\), and \(\hat \xi_7\), where \(\hat \xi_\mu^2 = 1\), and non-associatively equate them (with proportionality constants \(\pm 1\) and \(-i\)) to these irreducible elements of \(\text{SL}(2,\mathbb R)\):
\begin{align}
  \label{eq:AdSirreducibleelements1}
  -iK^{01} &= \hat \xi_4 \hat \xi_5\\
  \label{eq:AdSirreducibleelements2}
  -iK^{32} &= \hat \xi_7 \hat \xi_6\\
  \label{eq:AdSirreducibleelements3}
  -iK^{21} &= \hat \xi_6 \hat \xi_5\\
  \label{eq:AdSirreducibleelements4}
  -iK^{03} &= \hat \xi_4 \hat \xi_7\\
  \label{eq:AdSirreducibleelements5}
  J^{02} &= \hat \xi_4 \hat \xi_6\\
  \label{eq:AdSirreducibleelements6}
  -J^{13} &= \hat \xi_5 \hat \xi_7.
\end{align}

This is equivalent to assigning the multiplication rules
\begin{align}
  \label{eq:Octonionmultrules1}
  -i \hat \xi_1 \,&\substack{=\\12}\, \hat \xi_4 \hat \xi_5 \, \substack{=\\34} \, \hat \xi_7 \hat \xi_6\\
  \label{eq:Octonionmultrules2}
  -i \hat \xi_2 \, &\substack{=\\14}\, \hat \xi_4 \hat \xi_7 \, \substack{=\\23} \, \hat \xi_6 \hat \xi_5\\
  \label{eq:Octonionmultrules3}
  -i \hat \xi_3 = \hat \xi_1 \hat \xi_2 \, &\substack{=\\13} \, \hat \xi_4 \hat \xi_6 \, \substack{=\\24} \, \hat \xi_5 \hat \xi_7,
\end{align}
where \(\substack{=\\ij}\) denotes equality restricted to the subspace spanned by the \(i\)th and \(j\)th elements. We will drop this notation and have the subspace implied by the bare equality henceforth.

By construction, these inherit a norm and (thus inverses) from the norm of the elements on the left-hand-side (which as we stated before, are elements of the normed algebra \(\text{Cl}^{[0]}_{3,0}(\mathbb R) \cong \mathbb H\)). For example,
\begin{align}
  \label{eq:Onorminherited1}
  \hat \xi_4 &= -i \hat \xi_1 \hat \xi_5^{-1}\\
  \label{eq:Onorminherited2}
                  &= -i \hat \xi_1 (-i \hat \xi_6^{-1} \hat \xi_2)^{-1}\\
  \label{eq:Onorminherited3}
                  &= -(\hat \xi_1 \hat \xi_2) \hat \xi_6\\
  \label{eq:Onorminherited4}
                  &= -i \hat \xi_3 \hat \xi_6\\
  \label{eq:Onorminherited5}
                  &= -i \hat \xi_3 (i \hat \xi_3 \hat \xi_4^{-1})\\
  \label{eq:Onorminherited6}
                  &= (\hat \xi_3 \hat \xi_3 ) \hat \xi_4^{-1}\\
  \label{eq:Onorminherited7}
                  &= \hat \xi_4^{-1},
\end{align}
where in line~\ref{eq:Onorminherited1} we used Eq.~\ref{eq:Octonionmultrules1}, in line~\ref{eq:Onorminherited2} we used Eq.~\ref{eq:Octonionmultrules2}, and in line~\ref{eq:Onorminherited5} we used Eq.~\ref{eq:Octonionmultrules3} (in between we also used the \emph{alternate} property of these non-associative elements: \((\hat \xi_i \hat \xi_j) \hat \xi_\rho = -\hat \xi_i (\hat \xi_j \hat \xi_\rho)\)). In this manner the norm is dependent on the \emph{full} irreducible direct sum structure of the multiplication rules assigned in Eq.~\ref{eq:Octonionmultrules1}-\ref{eq:Octonionmultrules3}. All remaining multiplication rules in this normed algebra can be found from these:
\begin{align}
  i \hat \xi_4 =& \hat \xi_1 \hat \xi_5 = \hat \xi_2 \hat \xi_6 = \hat \xi_3 \hat \xi_7,\\
  i \hat \xi_5 =& \hat \xi_4 \hat \xi_1 = \hat \xi_2 \hat \xi_7 = \hat \xi_6 \hat \xi_3,\\
  i \hat \xi_6 =& \hat \xi_7 \hat \xi_1 = \hat \xi_4 \hat \xi_2 = \hat \xi_3 \hat \xi_5,\\
  i \hat \xi_7 =& \hat \xi_1 \hat \xi_6 = \hat \xi_5 \hat \xi_2 = \hat \xi_4 \hat \xi_3.
\end{align}
Taking into account the imaginary constants in Eqs.~\ref{eq:AdSirreducibleelements1}-\ref{eq:AdSirreducibleelements4} that relate these to the spacetime, it follows that \(-(i\hat \xi_4)^2 = -(i \hat \xi_6)^2 = \hat \xi_5^2 = \hat \xi_7^2 = 1\) and we will see that this defines the metric signature \(({-}{+}{-}{+})\) of the embedding space.

The resulting algebra can be identified as the complexified octonion algebra \(\mathbb O\).

It is perhaps not surprising that such a construction cannot be iterated once more since these elements can be found to contain no new irreducible subgroups; the octonion algebra \(\mathbb O\) is the largest normed algebra possible.

We have thus seen that the corresponding lower dimensional Clifford algebra \(\text{Cl}_{3,0}(\mathbb R)\) of the octonion algebra \(\mathbb O\) can still be used to construct it, but this requires an association to both its even and odd complexified elements. Specifically, Eqs.~\ref{eq:Octonionmultrules1}-\ref{eq:Octonionmultrules3} define \(\text{Cl}_{2,0}^{[0]}(\mathbb C)\) (as can be seen by noting that \(\xi_1\), \(\xi_2\) and \(\xi_1 \xi_2\) are sufficient to define the equalities), which is isomorphic to \(\text{Cl}_{3,0}(\mathbb R)\).

As we noted before, representations of the algebra \(\mathfrak{sl}(2,\mathbb C)\) produces operators which possess weights and are invariant under a Casimir operator. \(\mathfrak{sl}(2,\mathbb C)\) is also the universal enveloping algebra of \(\text{SL}(2,\mathbb R)\), and so can be used to produce its representation too.

A parameterization of a group element \(g\) of \(\text{SL}(2,\mathbb R)\) can be given by
\begin{align}
  \hat g &= X_{0} \hat \xi_0 + X_1 \hat \xi_1 + X_3 \hat \xi_2 + X_2 i \hat \xi_3\\
  &= \left( \begin{array}{cc} X_{0} + X_3 & X_2 + X_1\\ -X_2 + X_1& X_{0} - X_3 \end{array} \right),
\end{align}
where
\begin{equation}
  \label{eq:AdS3embeddingspace}
  X^2_{0} - X_1^2 + X_2^2 - X_3^2 = 1.
\end{equation}

Eq.~\ref{eq:AdS3embeddingspace} is similar to Eq.~\ref{eq:SO3embeddingspace}, except that it is a restriction on a four-dimensional embedding space with metric signature \(({+}{-}{+}{-})\) to a three-dimensional embedded manifold (a hyperboloid) (see Figure~\ref{fig:AdS/qubit}b).

Complexification of its algebra produces \(\mathfrak{sl}(2,\mathbb C)\) and reveals that the universal cover of this three-dimensional hyperboloid is spanned by the basis~\cite{Maldacena01}
\begin{align}
  \hat T_3 &= -\frac{i}{2} \hat \xi_3,\\
  \hat T^{\pm} &= \frac{1}{2}(\hat \xi_2 \pm i \hat \xi_1).
\end{align}
This produces the Casimir operator
\begin{equation}
  \hat C_2 = \frac{1}{2}(\hat T^+ \hat T^- + \hat T^- \hat T^+) - \hat T_3^2.
\end{equation}

In this sense, there is a correspondence between every two-dimensional qubit's associated conformal symmetry group and three-dimensional AdS spacetime's symmetry group. Applying these relationships to continuous field theories with actions that are invariant under \(\text{SL}(2, \mathbb R)\) and with metrics \(\text d s^2 = - \text d X_{-1}^2 - \text d X_0^2 + \text d X_1^2 + \text d X_2^2\), produces the AdS$_3$/CFT$_2$ correspondence. One such example is the bosonic string theory on AdS$_3 \times \mathcal M$ and the \(\text{SL}(2,\mathbb R)\) WZW model~\cite{Maldacena01}, where \(\mathcal M\) is compact.

This exposition reveals that the symmetry group \(\text{SL}(2,\mathbb R)\) underlying the AdS$_3$/CFT$_2$ correspondence is inherently part of the octonion algebra \(\mathbb O\); \(SL(2,\mathbb R)\) defines the multiplication rules of the octonion algebra's quaternion algebra generators and also of the remaining non-quaternion generators. The former define (generally non-pure) qubit quantum mechanics and the latter define the isometries of an AdS$_3$ spacetime. In this sense, we call this an AdS$_3$/qubit correspondence.

We will use this observation in the following to develop a representation of qubit quantum mechanics through a Majorana-to-qubit mapping that produces a monomorphism from the complexified Clifford algebra, \(\text{Cl}_{3,0}(\mathbb C)\) (the complexified Pauli algebra), to the complexified quaternion algebra, \(\text{Cl}^{[0]}_{3,0}(\mathbb C) \cong \mathbb H\). Working with the larger complexified Clifford algebra instead of directly with the quaternions will allow us to define operator product expansions similar to those in conformal field theory. This will allow us to derive an isomorphism to a dual algebra after we extend this algebra by four generators to produce the octonion algebra. The operator product expansion in the dual algebra proves key in deriving many of the gravitational results here, as well as the further extension to a perturbed anti-de Sitter spacetime. The differing contextuality or conformal anomaly of the two-dimensional field theory compared to the full octonion theory will also guide us on how quantization related subsets of elements in this algebra.

\subsection{The Clifford Algebra \(\text{Cl}_{3,0}(\mathbb C)\)}
\label{sec:Grassmannalgebra}

This section reviews the basics of interest to us of superalgebra generalizations of CCR/CAR algebras.

We begin with a definition of the Clifford algebra \(\text{Cl}_{n,0}(K)\) over the field \(K\): %
\begin{definition}[Clifford algebra \(\text{Cl}_{n,0}(K)\)]
  We define the Clifford algebra \(\text{Cl}_{n,0}(K)\) to be an associative algebra over the field \(K\) with \(n\) Grassmann generators \(\{\xi_i\}_{i=1}^n\) that have an antisymmetric multiplication operation,
  \begin{equation}
    \label{eq:anticommutation}
    \{\xi_i, \xi_j\} \equiv \xi_i \xi_j - \xi_j \xi_i = 0,
  \end{equation}
  called their anticommutator where \(i \in \{1, \ldots, n\}\).
\end{definition}
This subset of Clifford algebras are also called exterior or Grassmann algebras. They are associative algebras, which  means that multiplication simply produces higher order monomials. However, Eq.~\ref{eq:anticommutation} implies that \(\xi_i^2 = 0\) and so any monomial cannot have higher than quadratic order in any particular generator. For \(n\) generators, this means that the maximal monomial power is \(n\).

In the following part of this section we will begin by considering the Clifford algebra \(\text{Cl}_{3,0}(\mathbb R)\), which is generated by three Grassmann generators.

Any element \(g \in \text{Cl}_{3,0}(\mathbb R)\) can be represented as a finite sum of homogeneous monomials consisting of these three generators:
  \begin{equation}
    \label{eq:gelement}
    g(\bs \xi) = g_0 \xi_0 + \sum_{i=1}^3 g_i \xi_i + \sum_{i,j=1}^3 g_{ij} \xi_i \xi_j + \sum_{i,j,k=1}^3 g_{ijk} \xi_i \xi_j \xi_k,
  \end{equation}
  where \(g_0,g_i,g_{ij},g_{ijk} \in \mathbb R\).

Grassmann-valued fields are commonly used in fermion quantum field theories where the number of their independent modes is taken to be infinite and they are also often paired into independent sets associated with a creation and annihilation operators~\cite{Peskin18,Schwartz13}. Instead, we will only use a single Grassmann algebra with three generators (per qubit). %

See Appendix~\ref{sec:Paulialgebra} for an introduction into \(\text{Cl}_{3,0}(\mathbb R)\).

  \subsubsection{The Monomorphism from \(\text{Cl}_{3,0}(\mathbb C)\) to the Complexified Quaternion Algebra \(\text{Cl}^{[0]}_{3,0}(\mathbb C)\)}%
  \label{sec:hbarquantization}

  We can define an isomorphism from the Clifford algebra \(\text{Cl}_{3,0}(\mathbb R)\) to the complexified quaternion algebra \(\mathbb H \otimes \mathbb C \cong \text{Cl}^{[0]}_{3,0}(\mathbb C)\) by quantizing the Clifford algebra \(\text{Cl}_{3,0}(\mathbb R)\) in the usual manner by replacing its ``Poisson bracket'' by an anti-commutator and setting it equal to a non-zero (dimensionless) constant that we set to be \(\hbar\):
  \begin{equation}
    i \sum_j \left( \xi_k \frac{\cev{\partial}}{\partial \xi_j} \right) \left( \frac{\vec{\partial}}{\partial \xi_j} \xi_l \right) \equiv \{\xi_k, \xi_l\}_{\text{P.B.}} \rightarrow \{\hat \xi_k, \hat \xi_l\} = \hbar \delta_{kl}.
  \end{equation}
  This produces the complexified quaternion algebra \(\mathbb H \otimes \mathbb C\).

  At the moment, we do not mean to imply anything physical by assigning \(\hbar\) to be this dimensionless constant. We will justify its interpretation as a dimensionless analog to Planck's constant in Appendix~\ref{sec:hbar0qutritCliffordtheory}.

  This anticommutation group structure means that the \(\hat \xi_i\) are proportional to the Pauli group elements (operators),
  \begin{equation}
    \hat \xi_k = \sqrt{\frac{\hbar}{2}} \hat \sigma_k,
  \end{equation}
  which obey
  \begin{equation}
    \{\hat \sigma_k, \hat \sigma_l\} = 2 \delta_{kl}.
  \end{equation}
  Thus we can make the identification: \(\hat \xi_1 \equiv X\), \(\hat \xi_2 \equiv Z\), and \(\hat \xi_3 \equiv Y\). %

  All these relations can be summarized with the usual (rescaled) Pauli operator identity:
\begin{equation}
  \hat \xi_j \hat \xi_k = \begin{cases} \hat \xi_j & \text{if $k = 0$}\\ \hat \xi_k & \text{if $j=0$}\\ \delta_{jk} \hat \xi_0 - i \epsilon_{jkl} \hat \xi_l & \text{otherwise}. \end{cases}
\end{equation}

  Therefore, this quantization simply transforms any element \(g(\bs \xi) \in \text{Cl}_{3,0}(\mathbb R)\) given by Eq.~\ref{eq:gelement} to
  \begin{equation}
    \label{eq:quantizedWeylsymbol}
    \hat g = g_0 \hat \xi_0 + \sum_{i=1}^3 g_i \hat \xi_i + \sum_{i,j=1}^3 g_{ij} \hat \xi_i \hat \xi_j + \sum_{i,j,k=1}^3 g_{ijk} \hat \xi_i \hat \xi_j \hat \xi_k.
  \end{equation}

  This allows us to find a relationship between the even and odd elements in the Clifford algebra \(\text{Cl}_{3,0}(\mathbb R)\) and even and odd powers in \(\mathbb H\) that we can use to define this isomorphism between \(\text{Cl}_{3,0}(\mathbb R)\) and \(\mathbb H\) without appealing to quantization directly: %
\begin{align}
  \hat \xi_0 =& i \frac{2}{\hbar} \hat \xi_i \hat \xi_j \hat \xi_k,\\
  \hat \xi_i \hat \xi_j =& i \sqrt{\frac{\hbar}{2}} \hat \xi_k.
\end{align}

  In \(\text{Cl}_{3,0}(\mathbb R)\), this can be equivalently expressed using the Berezin calculus by defining the Grassmann Fourier transform to be
\begin{equation}
    \label{eq:GrassmannFouriertransform}
  \tilde g(\bs \rho) = \mathcal F(g(\bs \xi)) = \sqrt{\frac{\hbar}{2}} \int e^{i\sqrt{2/\hbar}\bs \xi \cdot \bs \rho}g(\bs \xi) \mbox d^3 \bs \xi,
\end{equation}
and its inverse to be
\begin{equation}
    \label{eq:GrassmanninverseFouriertransform}
  g(\bs \xi) = \mathcal F^{-1}(\tilde g(\bs \rho)) = \sqrt{\frac{2}{\hbar}} \int e^{-i\sqrt{\hbar/2}\bs \xi \cdot \bs \rho}\tilde g(\bs \rho) \mbox d^3 \bs \rho.
\end{equation}

  \(\tilde g(\bs \rho) \in \text{Cl}_{3,0}(\mathbb R)\) has the same form as \(g(\bs \xi)\) as we have defined it so far, namely
  \begin{equation}
    \tilde g(\bs \rho) = \tilde g_0 \rho_0 + \sum_{i=1}^3 \tilde g_i \rho_i + \sum_{i,j=1}^3 \tilde g_{ij} \rho_i \rho_j + \sum_{i,j,k=1}^3 \tilde g_{ijk} \rho_i \rho_j \rho_k.
  \end{equation}
The Grassmann Fourier transform relates the odd coefficients of \(\tilde g(\bs \rho)\) to the even coefficients of \(g(\bs \xi)\) in the following manner:
\begin{align}
  \label{eq:Fouriercoeffrelationship1}
    i\sqrt{\hbar/2} \tilde g_1 &\substack{=\\\mathcal F^{-1}} (g_{32} - g_{23}),\\
    i\sqrt{\hbar/2} \tilde g_2 &\substack{=\\\mathcal F^{-1}} (g_{13} - g_{31}),\\
    i\sqrt{\hbar/2} \tilde g_3 &\substack{=\\\mathcal F^{-1}} (g_{21} - g_{12}),\\
    \begin{array}{c}
      \tilde g_{123} - \tilde g_{132} + \tilde g_{312}\\
  \label{eq:Fouriercoeffrelationship4}
      - \tilde g_{321} + \tilde g_{231} - \tilde g_{213}
    \end{array}
    &\substack{=\\\mathcal F^{-1}} -i \hbar g_0 /2.
  \end{align}
  Expressions for the even coefficients of \(\tilde g(\bs \rho)\) in terms of the odd coefficients of \(g( \bs \xi)\) can be found by removing the tildes on the left-hand-side and adding them to the right-hand-side.

  This means that
\begin{equation}
  \mathcal F^{-1}: \text{Cl}_{3,0}^{[1]} (\mathbb R) \rightarrow i \text{Cl}_{3,0}^{[0]}(\mathbb R) .
\end{equation}
As a result, \(Cl_{3,0}(\mathbb R)\) can be shown to be isomorphic to \(\text{Cl}_{3,0}^{[0]}(\mathbb C)\) by using the inverse of the Grassmann Fourier transform \(\mathcal F\):
\begin{align}
  &\text{Cl}_{3,0}(\mathbb R) \equiv \text{Cl}_{3,0}^{[0]}(\mathbb R) \oplus \text{Cl}_{3,0}^{[1]}(\mathbb R) \\
  \cong& \text{Cl}^{[0]}_{3,0}(\mathbb C) \equiv \text{Cl}_{3,0}^{[0]}(\mathbb R) \oplus \mathcal F^{-1}(\text{Cl}_{3,0}^{[1]}(\mathbb R)) \cong \mathbb H \otimes \mathbb C.
\end{align}
Note that \(\text{Cl}_{3,0}^{[0]}(\mathbb C)\) forms a subalgebra while \(\text{Cl}_{3,0}^{[1]}(\mathbb C)\) does not. 

  In other words, the Grassmann Fourier transform captures how the complexified quaternion algebra, consisting of all products and sums of the Pauli matrices with coefficients in \(\mathbb C\), is isomorphic to \(\text{Cl}_{3,0}(\mathbb R)\). The real parts of the linear combinations of Pauli operators correspond to the odd monomials and the imaginary parts correspond to the even monomials in the isomorphic \(\text{Cl}_{3,0}(\mathbb R)\) algebra.

  This means that there is no loss of information from restricting \(g(\xi)\) to consist solely of even-dimensional monomials after complexification. %

  We proceed to define \(g(\xi) \in \text{Cl}_{3,0}^{[0]}(\mathbb C)\) instead of \(\text{Cl}_{3,0}(\mathbb R)\):
  \begin{align}
    \label{eq:Weylsymbol}
    g(\bs \xi) =& g_0 + \left(g_{12} \xi_1 \xi_2 + g_{23} \xi_2 \xi_3 + g_{13} \xi_1 \xi_3\right.\\
                & \left.+ g_{21} \xi_2 \xi_1 + g_{32} \xi_3 \xi_2 + g_{31} \xi_3 \xi_1\right)\nonumber,
  \end{align}
 Applying the Grassmann Fourier transform to this element now defines a redundant element consisting solely of odd monomials with complex coefficients, \(\tilde g(\bs \rho) \in \text{Cl}^{[1]}_{3,0}(\mathbb C)\):
  \begin{align}
    \label{eq:dualWeylsymbol}
    \tilde g(\bs \rho) =& \tilde g_1 \rho_1 + \tilde g_2 \rho_2 + \tilde g_3 \rho_3\nonumber\\
                        & + \left( \tilde g_{123} \rho_1 \rho_2 \rho_3 + \tilde g_{132} \rho_1 \rho_3 \rho_2 + \tilde g_{312} \rho_3 \rho_1 \rho_2 \right.\\
                        & \left. + \tilde g_{321} \rho_3 \rho_2 \rho_1 + \tilde g_{231} \rho_2 \rho_3 \rho_1 + \tilde g_{213} \rho_2 \rho_1 \rho_3 \right), \nonumber
  \end{align}

  This can be used to define a further map from \(\text{Cl}^{[0]}_{3,0}(\mathbb C)\) to \(\text{Cl}_{3,0}(\mathbb C)\):
  \begin{equation}
    \text{Cl}_{3,0}^{[0]}(\mathbb C) \overset{\hbar>0}{\hookrightarrow} \text{Cl}_{3,0}(\mathbb C) \equiv \text{Cl}_{3,0}^{[0]}(\mathbb C) \oplus \mathcal F( \text{Cl}_{3,0}^{[0]}(\mathbb C)).
  \end{equation}
  Since \(\tilde g \in \text{Cl}_{3,0}^{[1]}(\mathbb C)\) is redundant, this is an injective map, i.e.~a monomorphism. We note that since \(\mathcal F\) is \(\hbar\)-dependent, this monomorphism is \(\hbar\)-dependent.

  Even elements are sometimes called Weyl elements in other literature due to the historical development of this algebra in the Wigner-Weyl-Moyal formalism, where it is related to a Wigner formulation of quantum mechanics~\cite{Weyl27,Wigner32,Moyal49}. In the Fourier-sense, their dual elements can be interpreted as the characteristic function of the Weyl symbol, but are also the basis elements of the Clifford algebra \(\text{Cl}_{3,0}(\mathbb C)\). It is for this reason that this is called a superalgebra generalization of a CCR/CAR algebra in other literature~\cite{Slawny72} and can be related to \(C^*\)-algebras~\cite{Kadison97,Bratteli12} an operator-algebraic formulation~\cite{Wald94}.

  Yet another way to refer to the even elements \(g(\bs \xi) \in \text{Cl}^{[0]}_{3,0}(\mathbb C)\) and the odd elements \(\tilde g(\bs \rho) \in \text{Cl}^{[1]}_{3,0}(\mathbb C)\), is as the Bosonic and Fermionic superpartners with respect to a supersymmetry algebra. \(\text{Cl}_{3,0}(\mathbb C)\) is a Lie superalgebra and therefore defines such an algebra. The even elements commute with each other and the odd elements anti-commute. As we shall see, this turns out to be a supersymmetric embedding space~\cite{Mckeon04,Mckeon05,Mckeon13}.

  Note that the supersymmetry present in the quaternionic algebra does not suggest that superpartners are observable as meaningfully different particles when interpreted in quantum mechanics. As we have seen in Eqs.~\ref{eq:Fouriercoeffrelationship1}-\ref{eq:Fouriercoeffrelationship4}, the superpartners simply correspond to antisymmetrized Wick rotations of the complexified quaternion algebra that makes up the (generally non-pure) qubit Hilbert space. There is no real sense that these are separate observable particles.

\subsubsection{Symplectic Dynamics and Action}
  \label{sec:symplectic}

Any (unitary or non-unitary) qubit operation can be decomposed into Paulis: \(U_C = \sum_{i=0}^3 c_i \hat \xi_i\) where \(c_i \in \mathbb C\). Thus
\begin{align}
  \hat \xi_j \rightarrow& \hat U_C \hat \xi_j \hat U^\dagger_C\\
  =& \sum_{k,l=0}^3 c_k c^*_l \hat \xi_k \hat \xi_j \hat \xi_l\\
  \label{eq:hbarevolution}
  =& \sum_{k,l=0}^3 c_k c^*_l \hbar \epsilon_{jln} \epsilon_{knm} \hat \xi_m /2.
\end{align}

Since dequantization of an operator (not written as a product of operators) simply involves removing the hats from the \(\hat \xi_i\) in any expression, it follows that the Grassmann version of Hamilton's equation must be
  \begin{equation}
    \label{eq:GrassmannHamilton}
    \der{\xi_k}{\lambda} = \frac{\hbar}{2} \{H, \xi_k\}_{\text{P.B.}} = \frac{i\hbar}{2} H \frac{\cev \partial}{\partial \xi_k}.
  \end{equation}
  where the most general form for a one-particle Hamiltonian in this Grassmann algebra is
  \begin{equation}
    \label{eq:generalGrassmannHamiltonian}
    H(\bs \xi) = -\frac{i}{2} \sum_{k,l,m} \epsilon_{klm} b_k \xi_l \xi_m.
  \end{equation}
  Given our definition of involution from before, Hamiltonians are real-valued for unitaries, as expected.

This means that the equations of motion are
  \begin{equation}
    \label{eq:GrassmannHamiltonseq}
    \der{\xi_k}{\lambda} = \frac{\hbar}{2} \sum_{l,m} \epsilon_{klm} b_l \xi_m,
  \end{equation}
  which matches Eq.~\ref{eq:hbarevolution} when solved for a particular \(\lambda\). Note that we cannot strictly interpret the evolution parameter \(\lambda\) to be time, since like \(\hbar\) it is dimensionless.

  The presence of a non-zero \(\hbar\) coefficient, when interpreted as Planck's constant later, may be strange to see in an equation that purports to be a Grassmann version of Hamilton's equations. However, as we have seen, it is imposed by our quantization rule. We will soon see that this is intimately related to why the classically efficiently simulable unitary qubit Clifford stabilizer subtheory is contextual.
  
  Nevertheless, the \(\hbar\) can be factored out when the evolution is solved for periods \(\lambda\) that produce only single terms on the right-hand-side of Eq.~\ref{eq:GrassmannHamiltonseq} (i.e.~evolutions that take the Grassmann generators to themselves~\cite{Kocia17_2,Kocia19}). In this case, the \(\hbar\) factor can be effectively ignored as they are not relative constants, and the evolution becomes \(\mathcal O(\hbar^0)\). These evolutions turn out to be unitary Clifford gates, which as we shall see are non-contextual~\cite{Kocia17_2} in Appendix~\ref{sec:Cliffordhbar1}.

  Hamilton's equation is conventionally written in terms of two conjugate degrees of freedom, \(x^\mu\) and \(\pi^\mu\):
  \begin{equation}
    \der{x^\mu}{\lambda} = \partder{H}{\pi_\mu},\quad     \der{\pi_\mu}{\lambda} = -\partder{H}{x^\mu}.
  \end{equation}

  This implies that for the Clifford algebra \(\text{Cl}_{3,0}(\mathbb C)\), the Grassmann elements are conjugate under inversion: \(\xi_k \equiv -i x^k = i\pi_k\). %

  The Lagrangian, \(\mathcal L\), is related to the Hamiltonian,
  \begin{equation}
    \mathcal L \equiv \sum_\mu \pi_\mu \der{x^\mu}{\lambda} - H.
  \end{equation}
  This means that for \(\text{Cl}_{3,0}(\mathbb C)\),
  \begin{equation}
    \mathcal L = \sum_k \xi_k \left( H \frac{\cev \partial}{\partial \xi_k}\right) - H.
  \end{equation}
  Given the form of a general Hamiltonian (Eq.~\ref{eq:generalGrassmannHamiltonian}), this implies that the Lagrangian is just the negative of the Hamiltonian:
  \begin{equation}
    \label{eq:generalGrassmannLagrangian}
    \mathcal L = -i \sum_{k,l,m} \epsilon_{klm} b_l \xi_k \xi_m - H = \frac{i}{2} \sum_{k,l,m} \epsilon_{klm} b_k \xi_l \xi_m = -H.
  \end{equation}

  This means that \(\text{Cl}_{3,0}(\mathbb C)\) exhibits symplectic dynamics, which means that it preserves the ``area'' defined by any pair of the \emph{three} conjugate degrees of freedom. The two-form that defines this symplecticity is implicit in the anti-commutation of the Grassmann generators or their Poisson bracket, which as we saw, generates the Pauli anti-commutation relation:
  \begin{equation}
    \label{eq:symplecticity}
    \der{}{\lambda} \{\xi_i, \xi_j\}_\text{P.B.} = \{\dot \xi_i, \xi_j\}_\text{P.B.} + \{\xi_i, \dot \xi_j\}_\text{P.B.} = 0.
  \end{equation}

  See Appendix~\ref{sec:action} for a derivation of the associated generating action of these Lagrangians.

\subsubsection{Inner Products in \(\text{Cl}_{3,0}(\mathbb C)\) Define the Qubit Hilbert Space}%
\label{sec:G3andHilbertspace}

In Section~\ref{sec:hbarquantization}, we defined how to isomorphically map between elements in \(\text{Cl}_{3,0}(\mathbb R)\) and the complexified quaternion algebra \(\mathbb H \otimes \mathbb C\). We did this by quantizing the algebraic elements by \(\hbar\) and reconstructing an equivalent procedure using the \(\hbar\)-dependent Grassmann Fourier transform, which used Berezin calculus on the Grassmann generators of the Clifford algebra. %

In this section we will find that we can also impose a Hilbert space inner product on the Clifford algebra \(\text{Cl}_{3,0}(\mathbb R)\) (and thus its complexification) by similarly using the Berezin calculus. Instead of reconstructing an equivalent procedure to the quantized Poisson bracket, we will reconstruct an equivalent procedure to the trace in Hilbert space using a free commutative product of the quaternion algebra and the Clifford algebra, %
  \begin{equation}
    \label{eq:translationop}
    \hat T(\rho) = \exp \left( i \sum_{j=1}^3 \hat \xi_j \rho_j \right) \in \mathbb H \otimes \mathbb C \ast \text{Cl}_{3,0}(\mathbb R).
  \end{equation}
  We can use this as a map from \(\text{Cl}_{3,0}(\mathbb C)\) to the quaternion algebra by integrating out the Grassmann part of the free commutative product. 
  
  This allows us to concisely define the quantized operator \(\hat g\) (given by Eq.~\ref{eq:quantizedWeylsymbol}) to simply be the quantized Grassmann Fourier transform of the element in \(\text{Cl}_{3,0}(\mathbb C)\):
  \begin{equation}
    \label{eq:gop}
    \hat g = \int \hat T(\bs \rho) \tilde g(\bs \rho) \text d^3 \rho.
  \end{equation}

  This may seem like a lot more work than simply taking an element \(g\in \text{Cl}_{3,0}(\mathbb C)\) and adding hats to all of Grassmann elements, but it turns out that \(\hat T(\rho)\) is group-theoretically very useful for evaluating products of operators and concisely deriving many operations on \(\text{Cl}_{3,0}(\mathbb C)\) from their Hilbert space expressions.

  This is because \(\hat T(\rho)\) satisfies the properties of a translation operator:
  \begin{equation}
    \label{eq:translationgroup_1}
    \hat T(\bs \rho') \hat T(\bs \rho'') = \exp \left( \sum_{j=1}^3 \rho'_j \rho''_j \right) \hat T(\bs \rho' + \bs \rho''),
  \end{equation}
  and
  \begin{equation}
    \label{eq:translationgroup_2}
    \Tr \hat T(\bs \rho) = 2(1 + i \delta(\bs \rho)).
  \end{equation}

  Note that if \(g\) is defined to be in \(\text{Cl}^{[0]}_{3,0}(\mathbb R)\) and \(\tilde g\) in \(\text{Cl}^{[1]}_{3,0}(\mathbb R)\), then the trace of \(\hat T(\bs \rho)\) produces
  \begin{equation}
    \Tr \left( \hat T(- \bs \rho) \hat g \right) = 2( i \tilde g(\bs \rho) + g(i \bs \rho)).
  \end{equation}
This reflects the fact stated earlier that \(g(i\bs \rho)\) contains the real coefficients while \(\tilde g(\bs \rho)\) contains their antisymmetrized Wick rotation. The same identity holds after complexification. %

We can now see how the Hilbert space inner product (i.e.~traces) are expressed in terms of Grassmann integrals. Using Eq.~\ref{eq:gop} and \(\hat T\)'s translation identities, expectation values can be found to be~\cite{Kocia17_2}

[CHECK 2 pi hbar PREFACTORS BELOW!]
  \begin{align}
    \label{eq:WWMinnerproduct}
    \Tr(\hat O \hat \rho) &= \frac{i}{2 \pi \hbar} \int \int \rho(\bs \xi) e^{-i \sum_k \xi_k \rho_k} O(\bs \rho) \text d^3 \xi \text d^3 \rho\\
                          &= (2 \pi \hbar)^{-1} \int \rho(\bs \xi) \tilde O(\bs \xi) \text d^3 \xi \\
                          &= (2 \pi \hbar)^{-1} \int O(\bs \xi) \tilde \rho(\bs \xi) \text d^3 \xi.
  \end{align}
  
  Commonly, the structure of an inner product or Hilbert space is imposed on a Grassmann algebra by defining an even number $2k$ generators $\{\xi_i\}_{i=1}^k$ and $\{\xi^*_i\}_{i=1}^k$, and setting the inner product to be~\cite{Peskin18,Schwartz13}
\begin{equation}
  \label{eq:Grassmanninnerproduct}
  \int \int \xi_1 \ldots \xi_k \xi^*_1 \ldots \xi^*_k \text{d}^k \bs \xi \text{d}^k \bs \xi^*= 1.
\end{equation}
Since we are primarily considering the algebra \(\text{Cl}_{3,0}(\mathbb C)\) consisting of even and odd products of real-valued Grassmanns (i.e.~invariant under involution, whereas \(\xi_i^*\) is \(\xi_i\) under involution in Eq.~\ref{eq:Grassmanninnerproduct}), we see that Eq.~\ref{eq:WWMinnerproduct} involves both even elements and odd elements (i.e.~Grassmann Fourier transformed elements) in order to satisfy Eq.~\ref{eq:Grassmanninnerproduct}. %
In particular, Eq.~\ref{eq:WWMinnerproduct} satisfies the form of Eq.~\ref{eq:Grassmanninnerproduct} if the substitution \(i \rho_k \rightarrow \xi_k^*\) is made, which as we saw in Section~\ref{sec:hbarquantization} is also how the Grassmann Fourier transform maps between the even \(\text{Cl}^{[0]}_{3,0}(\mathbb C)\) elements and the odd \(\text{Cl}^{[1]}_{3,0}(\mathbb C)\) elements. Therefore, this Hilbert space inner product involves complex conjugation and an overall double integral over \(2k\) variables (where \(k=3\)) as expected.

  So for instance a density matrix corresponding to the \(\ketbra{0}{0}\) computation basis state, \(\hat \rho = \frac{1}{2}(\hat I + \hat Z)\), has associated element \(\rho(\bs \xi) = \frac{1}{2}(\xi_0 + i \xi_1 \xi_3) \in \text{Cl}^{[0]}_{3,0}(\mathbb C) \cong \text{Cl}_{3,0}(\mathbb R)\). Its expectation value with the Pauli matrix \(\hat X\) is \(\Tr(\hat X \hat \rho) = \int \rho(\bs \xi) \tilde X(\bs \xi) \text d^3 \bs \xi\), where \(\tilde X(\bs \xi) = \xi_1 \in \text{Cl}^{[1]}_{3,0}(\mathbb C)\).

  Again, using Eq.~\ref{eq:gop} and \(\hat T\)'s translation identities, the element in \(\text{Cl}^{[0]}_{3,0}(\mathbb C)\) from the product of two operators \(\hat O = \hat O_1 \hat O_2\) can be found to be~\cite{Kocia17_2}
  \begin{equation}
    O(\bs \xi) = \int O_1(\bs \xi') O_2(\bs \xi'') \exp\left(2\Delta_3(\bs \xi, \bs \xi', \bs \xi'')/\hbar\right) \text d^3 \bs \xi' \text d^3 \bs \xi'',
  \end{equation}
  where
  \begin{equation}
    \label{eq:symplecticarea}
    \Delta_3(\bs \xi, \bs \xi', \bs \xi'') \equiv \sum_{j=1}^3 (\xi'_j \xi''_j + \xi''_j \xi_j + \xi_j \xi'_j).
  \end{equation}
  This will prove useful later on for when we semiclassically expand this in orders of \(\hbar\) to develop the operator product expansion of products of elements from the Clifford even subalgebra.

  It also proves useful for defining correlation functions:
  \begin{equation}
    \Tr(\hat O_1 \hat O_2 \hat \rho) = \int O(\bs \xi) \tilde \rho(\bs \xi) \text d^3 \bs \xi.
  \end{equation}

\subsection{Operator Product Expansion in \(\text{Cl}_{3,0}(\mathbb C)\)}
\label{sec:hbarCliffordtheory}

In this section we will derive an operator product expansion (OPE). The OPE relates products of elements in \(\text{Cl}_{3,0}(\mathbb C)\) to an exact truncating series~\cite{Groenewold46} and will be defined so that it is isomorphic to the \(\hbar\)-quantized product in the quaternion algebra. We will show later in Section~\ref{sec:witt} that this is a basis-independent formulation of OPEs in continuous conformal field theories, which are usually presented for holomorphic functions but here will directly apply to the \(\text{Cl}_{3,0}(\mathbb C)\) algebra elements itself. %

Note that in the language of the Wigner-Weyl-Moyal CAR algebra, the OPE is called the Groenewold rule~\cite{Almeida98}.

\subsubsection{Qubit \(\text{Cl}_{3,0}(\mathbb C)\) Operator Product Expansion}

We will indicate the the product of two elements in \(\text{Cl}_{3,0}(\mathbb C)\) isomorphic to the product in the complexified quaternion algebra by \(\star\). It is clear right away that this will likely involve the Berezin calculus and powers of \(\hbar\) through Grassmann Fourier transforms; without supplementing with the Berezin calculus, the Grassmann product in \(\text{Cl}_{3,0}(\mathbb C)\) by itself will otherwise set all quadratic products to zero.

  Generally, for even elements
  \begin{align}
    &W_1 \star W_2(\bs \xi)  = \nonumber\\
    \label{eq:arealaw}
    & \int e^{\frac{2}{\hbar}(\bs \xi\cdot \bs \xi_1 + \bs \xi_2 \cdot \bs \xi + \bs \xi_1 \cdot \bs \xi_2)} W_1(\bs \xi_1) W_2(\bs \xi_2) \text d^3 \bs \xi_1 \text d^3 \bs \xi_2.
  \end{align}
  This is the integral representation of this identity and can be rewritten instead in terms of its odd elements:
  \begin{align}
    &W_1 \star W_2(\bs \xi)  = \nonumber\\
    & \int e^{\frac{2i}{\hbar}(\bs \rho_1\cdot \bs \xi + \bs \rho_2 \cdot \bs \xi + \bs \rho_1 \cdot \bs \rho_2)} \widetilde W_1(\bs \rho_1) \widetilde W_2(\bs \rho_2) \text d^3 \bs \rho_1 \text d^3 \bs \rho_2.
  \end{align}
  Now, following the similar treatment in~\cite{Almeida98} for the two-generator formalism, we can define
  \begin{align}
    &'{W_1 \star W_2}'(\bs \xi_1, \bs \xi_2) \equiv \nonumber\\
    & \int e^{\frac{2i}{\hbar}(\bs \rho_1\cdot \bs \xi_1 + \bs \rho_2 \cdot \bs \xi_2 + \bs \rho_1 \cdot \bs \rho_2)} \widetilde W_1(\bs \rho_1) \widetilde W_2(\bs \rho_2) \text d^3 \bs \rho_1 \text d^3 \bs \rho_2,
  \end{align}
  such that \('{W_1 \star W_2}'(\bs \xi, \bs \xi) = W_1 \star W_2(\bs \xi)\).
  
This analytically continued function allows us to obtain an expansion in orders of \(\hbar\) by expanding \(\exp(\frac{2i}{\hbar} \rho_1 \rho_2)\) and then setting \(\xi_1=\xi_2=\xi\):
  \begin{align}
    & '{W_1 \star W_2}'(\bs \xi_1, \bs \xi_2) \nonumber\\
    =& \int \left(1 + \frac{2i}{\hbar} \bs \rho_1 \cdot \bs \rho_2 - \frac{4}{\hbar^2} (\bs \rho_1 \cdot \bs \rho_2)^2\right)\\
    & \times e^{\frac{2i}{\hbar} \bs \rho_1 \cdot \bs \xi_1} \widetilde W_1(\bs \rho_1) e^{\frac{2i}{\hbar} \bs \rho_2 \cdot \bs \xi_2} \widetilde W_2(\bs \rho_2) \text d^3 \bs \rho_1 \text d^3 \bs \rho_2\nonumber\\
    =& \int \left(1 - \frac{\hbar i}{2} \vec \partial_{\xi_1} \cdot \vec \partial_{\xi_2} - \frac{\hbar^2}{4} (\vec \partial_{\xi_1} \cdot \vec \partial_{\xi_2})^2 \right)\\
    & \times e^{\frac{2i}{\hbar} \bs \rho_1 \cdot \bs \xi_1} \widetilde W_1(\bs \rho_1) e^{\frac{2i}{\hbar} \bs \rho_2 \cdot \bs \xi_2} \widetilde W_2(\bs \rho_2) \text d^3 \bs \rho_1 \text d^3 \bs \rho_2\nonumber\\
    \label{eq:analyticallycontinuedGroenewold}
    =& \exp\left(-i \frac{\hbar}{2} \vec \partial_{\xi_1} \cdot \vec \partial_{\xi_2} \right) W_1(\bs \xi_1) W_2(\bs \xi_2).
  \end{align}
  Therefore,
  \begin{align}
    & W_1 \star W_2 (\bs \xi) = \nonumber\\
    \label{eq:WeylSymbolGroenewoldRule}
    &= \exp\left(-i \frac{\hbar}{2} \vec \partial_{\xi_1} \cdot \vec \partial_{\xi_2} \right) W_1(\bs \xi_1) W_2(\bs \xi_2) \Big|_{\bs\xi_1=\bs\xi_2=\bs\xi}.
  \end{align}
  It is instructive to write out all three terms of the \(\hbar\) expansion:
  \begin{align}
    & W_1 \star W_2 (\bs \xi) \nonumber\\
    \label{eq:WeylSymbolGroenewoldRuleLeadingTerms}
    \equiv& \mathcal O(\hbar^0) + \mathcal O(\hbar^1) + \mathcal O(\hbar^2)\\
    =& W_1(\bs \xi) W_2(\bs \xi) \nonumber\\
    & - \frac{\hbar i}{2} \sum_i \vec \partial_{{\xi_1}_i} W_1(\bs \xi) \vec \partial_{{\xi_2}_i} W_2(\bs \xi) \\
    & - \frac{\hbar^2}{4} \sum_{i,j} \vec \partial^2_{{\xi_1}_i,{\xi_1}_j} W_1(\bs \xi) \vec \partial^2_{{\xi_1}_i,{\xi_1}_j} W_2(\bs \xi). \nonumber
  \end{align}

  Therefore, the \(\mathcal O(\hbar^0)\) term handles Pauli times identity, the \(\mathcal O(\hbar^1)\) term handles Pauli times Pauli (not including the identity), and the \(\mathcal O(\hbar^2)\) term handles Pauli squared operations.%

  The \(\mathcal O(\hbar^1)\) nature of Pauli times Pauli operation for even \(\text{Cl}^{[0]}_{3,0}(\mathbb C)\) elements can also be seen by noting that
  \begin{equation}
    \hat \xi_i \hat \xi_j \hat \xi_i \hat \xi_k = \hbar \hat \xi_k \hat \xi_j/2.
  \end{equation}
  Similarly, the \(\mathcal O(\hbar^2)\) nature of Pauli squared operations for even \(\text{Cl}^{[0]}_{3,0}(\mathbb C)\) elements can also be seen by noting that
  \begin{equation}
    \hat \xi_i \hat \xi_j \hat \xi_i \hat \xi_j = \hbar^2/4.
  \end{equation}

A similar expansion can be derived for the odd elements that make up \(\text{Cl}_{3,0}^{[1]}(\mathbb C)\):
\begin{align}
  &\reallywidetilde{W_1 \star W_2}(\bs \rho) \nonumber\\
  \equiv& \int W_1 \star W_2(\bs \xi) e^{-i \xi \rho} \text d^3 \bs \xi\\
  =& \int e^{\frac{2}{\hbar} (\bs \xi' \cdot \bs \xi'' + \bs \xi'' \cdot \bs \xi + \bs \xi \cdot \bs \xi' -i \bs \xi \cdot \bs \rho)} W_1(\bs \xi') W_2(\bs \xi'') \text d^3 \bs \xi' \text d^3 \bs \xi'' \text d^3 \bs \xi\\
  =& \int \left(1 + \frac{2}{\hbar} \bs \xi' \cdot \bs \xi'' + \frac{4}{\hbar^2} (\bs \xi' \cdot \bs \xi'')^2 + \ldots\right) \\
  & \times W_1(\bs \xi') W_2(\bs \xi'') e^{\frac{2}{\hbar}(\bs \xi'' \cdot \bs \xi_2 + \bs \xi_1 \cdot \bs \xi' - i \bs \xi \cdot \bs \rho)}  \text d^3 \bs \xi' \text d^3 \bs \xi'' \text d^3 \bs \xi\nonumber\\
  =& \int \left(1 - i \frac{\hbar}{2} \vec \partial_{\xi_1'} \cdot \vec \partial_{\xi_2} - \frac{\hbar^2}{4} (\vec \partial_{\xi_1} \cdot \vec \partial_{\xi_2})^2 + \ldots\right) \\
  & \times W_1(\bs \xi') W_2(\bs \xi'') e^{\frac{2}{\hbar}(\bs \xi'' \cdot \bs \xi_2 + \bs \xi_1 \cdot \bs \xi' - i \bs \xi \cdot \bs \rho)}  \text d^3 \bs \xi' \text d^3 \bs \xi'' \text d^3 \bs \xi \Big|_{\bs \xi_1 = \bs \xi_2 = \bs \xi}\nonumber\\
  =& \exp\left(-i \frac{\hbar}{2} \vec \partial_{\xi_1} \cdot \vec\partial_{\xi_2}\right) \reallywidetilde{\widetilde W_1(i \bs \rho_1) \widetilde W_2(i \bs \rho_2)} \Big|_{\bs \rho_1 = \bs \rho_2 = \bs \rho}.
  \end{align}
  Again, it is instructive to write out all three terms:
  \begin{align}
    & \reallywidetilde{W_1 \star W_2} (\bs \rho) \nonumber\\
    \label{eq:dualWeylSymbolGroenewoldRuleLeadingTerms}
    \equiv& \mathcal O(\hbar^0) + \mathcal O(\hbar^1) + \mathcal O(\hbar^2)\\
    =& \reallywidetilde{\widetilde W_1(i \bs \rho) \widetilde W_2(i \bs \rho)} \nonumber\\
    & - \frac{\hbar i}{2} \sum_i \reallywidetilde{\vec \partial_{{\rho_1}_i} \widetilde W_1(\bs \rho) \vec \partial_{{\rho_2}_i} \widetilde W_2(\bs \rho)} \\
    & - \frac{\hbar^2}{4} \sum_{i,j} \reallywidetilde{\vec \partial^2_{{\rho_1}_i,{\rho_2}_j} \widetilde W_1(\bs \rho) \vec \partial^2_{{\rho_1}_i, {\rho_2}_j}\widetilde W_2(\bs \rho)}. \nonumber
  \end{align}

This shows how for the odd \(\text{Cl}^{[1]}_{3,0}(\mathbb C)\) elements, the \(\mathcal O(\hbar^0)\) term still handles the Pauli times identity operations, but the two higher order terms are flipped in their role compared to the OPE for the even \(\text{Cl}^{[0]}_{3,0}(\mathbb C)\) elements: the \(\mathcal O(\hbar^1)\) term handles Pauli squared, and the \(\mathcal O(\hbar^2)\) term handles Pauli times Pauli (not including the identity). %

  The \(\mathcal O(\hbar)\) nature of Pauli squared operations for odd elements can also be seen by noting that
  \begin{equation}
    \hat \xi_i \hat \xi_i = \frac{\hbar}{2}.
  \end{equation}

  Note how the OPE expression for the even elements of \(\text{Cl}_{3,0}(\mathbb C)\) (Eq.~\ref{eq:WeylSymbolGroenewoldRuleLeadingTerms}) show how to replace the insertion of two ``nearby'' operators by a series of single ``local'' operators, where by this we mean that two operators defined on two separate algebras spanned by \(\xi_1\) and \(\xi_2\) are expressed by a series of single operators evaluated on a single algebra spanned by \(\xi\). This interpretation does not strictly hold for the odd expansion (Eq.~\ref{eq:dualWeylSymbolGroenewoldRuleLeadingTerms}) since the overall Fourier transform on the terms in that series require a ``global definite'' integration. This is another manifestation of the subalgebra property of the even elements compared to the odd elements. Strictly speaking, this means that only the expansion for \(\text{Cl}^{[0]}_{3,0}(\mathbb C)\) is a proper OPE in the conformal theory sense. We formally show this more precisely in Section~\ref{sec:conformal}.

  This shows how restricting to the isomorphic complexified even subalgebra of \(\text{Cl}_{3,0}(\mathbb R)\), \(\text{Cl}^{[0]}_{3,0}(\mathbb C)\), than the corresponding OPE that is \(\mathcal O(\hbar^2)\) with two terms supplementing the \(\mathcal O(\hbar^0)\) term. However, if we perform the monomorphism to the larger algebra \(\text{Cl}_{3,0}(\mathbb C)\), then we can lower our algebra's OPE dependence by one order of \(\hbar\) by using the odd term \(\mathcal O(\hbar^1)\) term for Pauli squared instead of the even algebra's \(\mathcal O(\hbar^2)\) term.

  Written in terms of the even subalgebra, this will remove one of its overall Grassmann Fourier transforms. If expressing everything within the even subalgebra is not important (i.e.~if you consider both the even and odd elements of \(\text{Cl}_{3,0}(\mathbb C)\) to be local terms), then the even OPE can be written fully at \(\mathcal O(\hbar^1)\) by replacing its \(\mathcal O(\hbar^2)\) term by the Fourier transform of the \(\mathcal O(\hbar^1)\) odd expansion:
  \begin{align}
    & W_1 \star W_2 (\bs \xi) \nonumber\\
    \label{eq:WeylSymbolGroenewoldRuleLeadingTerms_combined}
    \equiv& \mathcal O(\hbar^0) + \mathcal O(\hbar^1) + \mathcal O(\hbar^1)\\
    =& W_1(\bs \xi) W_2(\bs \xi) \nonumber\\
    & - \frac{\hbar i}{2} \sum_i \vec \partial_{{\xi_1}_i} W_1(\bs \xi) \vec \partial_{{\xi_2}_i} W_2(\bs \xi) \\
    & - \frac{\hbar i}{2} \sum_i \vec \partial_{{\rho_1}_i} \widetilde W_1(\bs \rho) \vec \partial_{{\rho_2}_i} \widetilde W_2(\bs \rho). \nonumber
  \end{align}

  This means that the \(\text{Cl}_{3,0}(\mathbb C)\) OPE of all single-qubit operations (decomposable into unitary gates, non-unitary channels, and single-qubit projections) are expressed in the Clifford algebra \(\text{Cl}_{3,0}(\mathbb C)\) as linear combinations of \(\mathcal O(\hbar^0)\) and \(\mathcal O(\hbar^1)\) terms involving products of either even or odd elements. The number of such terms generally proliferate exponentially if BQP$\ne$BPP.

  Note that in quantum field theory there is often an implicit renormalization performed of the fields \(\xi_i \rightarrow \xi_i/\sqrt{\hbar}\)~\cite{Brodsky11}. This means that their even-dimensional superpartners should be rescaled as \(\xi_i \xi_j \rightarrow \xi_i \xi_j/\hbar\). This transforms the orders in Eq.~\ref{eq:WeylSymbolGroenewoldRuleLeadingTerms_combined} from \(\mathcal O(\hbar^0) + \mathcal O(\hbar^1)\) to \(\mathcal O(\hbar^{-1}) + \mathcal O(\hbar^0)\).

  As can be seen in Appendix~\ref{sec:Cliffordhbar1}, the Clifford algebra \(\text{Cl}_{3,0}(\mathbb C)\) is actually able to capture the unitary single-qubit unitary Clifford stabilizer subtheory with a single \(\mathcal O(\hbar^1)\) term (and generally a polynomial number of \(\mathcal O(\hbar^1)\) terms for \(n\) qubits, though we do not show this here).

  This OPE written in terms of both even an odd elements of \(\text{Cl}_{3,0}(\mathbb C)\) will be enormously important in the upcoming sections. Its use necessitates using the full \(\text{Cl}_{3,0}(\mathbb C)\) algebra in the mapping from complexified quaternions, instead of just the even elements \(\text{Cl}^{[0]}_{3,0}(\mathbb C)\) of \(\text{Cl}_{3,0}(\mathbb C)\), which is otherwise sufficient. By incorporating the full \(\text{Cl}_{3,0}(\mathbb C)\) algebra we are able to lower the OPE to first order in \(\mathcal O(\hbar^1)\), which we will see in the next section makes it equivalent to the conformal Virasoro algebra. %
  This form of the OPE will be crucial in later allowing us to produce entanglement entropies that agree with the Ryu-Takayanagi formula for AdS spacetime.

  \subsection{Conformal Symmetry}
\label{sec:conformal}

In this section we will show how the OPE found in Section~\ref{sec:hbarCliffordtheory} %
is a conformally symmetry OPE. By this we mean that we will find the OPE is equivalent to the Witt algebra commutation relations at \(\mathcal O(\hbar^0)\) %
and that the \(\mathcal O(\hbar^1)\) terms are generated by a central charge anomaly and extend the Witt algebra to the infinite-dimensional Virasoro algebra. But first, we begin with a general discussion of conformal symmetry in \(\text{Cl}_{3,0}(\mathbb R)\), when expressed in \(\text{Cl}_{3,0}(\mathbb C)\), and its Casimir operator.

We present a similar parallel presentation in Appendix~\ref{sec:Cliffordsubtheory} for how expressing the Clifford stabilizer subtheory within \(\text{Cl}_{3,0}(\mathbb C)\) respects the dimensional-dependence of contextuality for qubits and qutrits. Contextuality (defined therein) is equivalent to the conformal or central charge anomaly.

\subsubsection{Action Invariance}

The complexification of \(\text{Cl}_{3,0}(\mathbb R)\) that allows for the Clifford algebra \(\text{Cl}_{3,0}(\mathbb C)\) to exhibit conformal symmetry after the monomorphism from \(\text{Cl}_{3,0}(\mathbb R)\). One way to see this is that the inclusion of a third generator \(i \xi_3\), to complement \(\xi_1\) and \(\xi_2\) defines a two-dimensional special orthogonal group that is a conformal symmetry group (as discussed in Section~\ref{sec:SLinoctonions}) and \(\text{Cl}_{3,0}(\mathbb C)\) contains all three permutations of such bases.%

Consider a Lagrangian of the general form given by Eq.~\ref{eq:generalGrassmannLagrangian} with only one term, \(\mathcal L = \frac{i}{2} \xi_l \xi_m\). We define a new Lagrangian \(\tilde {\mathcal L}\) related to \(\mathcal L\) by a conformal factor containing the Grassmann element that it is independent of:
  \begin{equation}
    \tilde {\mathcal L} \equiv e^{\alpha \xi_k} \left(\frac{i\hbar}{4} \xi_l \xi_m\right)  = e^{\alpha \xi_k} \mathcal L,
  \end{equation}
  where \(\alpha \in \mathbb R\).
  The Euler-Lagrange equations for this conformal coordinate \(\xi_k\) are then
  \begin{equation}
    \partder{\xi_k}{\lambda} = -i e^{\alpha \xi_k} \mathcal L \frac{\cev \partial}{\partial \xi_k} = - \alpha \tilde {\mathcal L},
  \end{equation}
  indicating that this Lagrangian's corresponding equations of motion are related to the old one's by just an overall rescaling. Therefore, \(\tilde {\mathcal L}\) (and trivially \(\mathcal L\) by setting \(\alpha = 0\)) exhibit conformal symmetry.

  Similarly, considering a general generating action within the \(\text{Cl}^{[0]}_{3,0}(\mathbb C)\) algebra,
  \begin{equation}
    \label{eq:Grassmannaction}
    S(\bs \xi; t) {=} \sum_{l,m} \epsilon_{klm} \tan(b_kt/2) \xi_l \xi_m, 
  \end{equation}
  it follows that any transformation \(\xi_k \rightarrow \xi_k + \epsilon(\xi_k)\) leaves this action unchanged.

  Notice that this conformal symmetry requires a third generator to be defined; given a Lagrangian or generating action written in terms of two generators, \(\xi_l\) and \(\xi_m\), the conformal coordinate will always be the third \(\xi_k\). A two-generator formalism with generators momentum \(p\) and position \(q\), as is common for finite odd-dimensional systems, cannot generally exhibit such conformal symmetry.

  Also notice that any rotation between the three Grassmann generators simply redefines the third conformally invariant element. It is always possible to actively (rigidly) rotate the generators in a Lagrangian \(\mathcal L = \frac{i}{2} \xi_l \xi_m\) to a general quadratic form,
  \begin{equation}
    \label{eq:generalLagrangian}
    \mathcal L = \frac{i}{2} (b_k \xi_l \xi_m + b_l \xi_m \xi_k + b_m \xi_k \xi_l),
  \end{equation}
  where \(b_k^2 + b_l^2 + b_m^2 = 1\), by evolving them under the dynamics of a Hamiltonian
  \begin{equation}
  \label{eq:threegenHam}
  H = \sum_{i, j} c_{i j} \xi_i \xi_j/2,
\end{equation}
where \(c_{i j}\) is antisymmetric, and \(\sum_{i < j} c_{i j} = 1\).
After evolution for $t=\pi$, \(\xi_k \rightarrow c_l \xi_m - c_m \xi_l\), \(\xi_l \rightarrow -c_k \xi_m + c_l \xi_k\), and \(\xi_m \rightarrow c_k \xi_l - c_l \xi_k\) and so we can identify $b_k = c_k^2$, $b_l = c_k c_l$, and $b_m = c_k c_l$.

This means that in terms of these new rotated three-generators the Lagrangian is conformally related to \(\tilde {\mathcal L} \equiv e^{\xi'_k} {\mathcal L}\) with conformal coordinate \(\xi'_k \equiv c_l \xi_m - c_m \xi_l\).

\subsubsection{Central Charge Anomaly}
\label{sec:centralcharge}

We found in Section~\ref{sec:SLinoctonions} that the Casimir operator for the conformal group defined by the \(\text{Cl}_{3,0}(\mathbb C)\) is 
\begin{align}
  \hat C_2 =& \frac{1}{2}\left(\frac{1}{4}\left(\hat \xi_2 + i \hat \xi_1\right)\left(\hat \xi_2 - i \hat \xi_1\right) \right. \nonumber\\
           & \left. + \frac{1}{4}\left(\hat \xi_2 - i \hat \xi_1\right)\left(\hat \xi_2 + i \hat \xi_1\right)\right) + \frac{1}{4}\hat \xi_3^2.
\end{align}
Performing the above products using the OPE of the Clifford subalgebra \(\text{Cl}^{[0]}_{3,0}(\mathbb C)\) to \(\mathcal O(\hbar^1)\) produces
\begin{equation}
  C_2 = \frac{1}{4}(\xi_3^2) = \frac{\hbar}{8}.
\end{equation}
It follows that the central charge (using the usual rescaling in physics) is \(c = \frac{\hbar}{2} \frac{1}{2}\).

Therefore, for \(\hbar = 0\), the central charge is \(c = 0\). In condensed matter, this is typically associated with critical systems where second order phase transitions occur with quenched disorder~\cite{Gurarie05}. At this value, the central charge is non-informative about solutions of any conformal field theory. This makes sense, since the Clifford algebra \(\text{Cl}_{3,0}(\mathbb C)\) becomes divergently disassociated with the qubit quantum theory at this value.

Typically, for \(\hbar > 0\), the value is set to \(\hbar=2\) to reproduce the Pauli matrices in physics. This sets the central charge to be \(c=1/2\), which is the central charge associated with the free Majorana fermion representation. Note that the Majorana mapping that produces the Clifford algebra \(\text{Cl}_{3,0}(\mathbb C)\) does not correspond to this free Majorana representation, since it produces bound Hamiltonians for non-zero \(\hbar\).

(We note that \(\hbar > 0\) and the semiclassical limit \(\hbar \rightarrow 0\) is typically not meant to be interpreted as actually changing \(\hbar\), but rather the set of equations under consideration that depend on it. For example, in OPEs it refers to when its series expansions in terms of powers of \(\hbar\) is dominated by the lower order terms because the symplectic area \(\Delta_3\) (see Eq.~\ref{eq:symplecticarea}) is far larger than \(\hbar\) so that \(\hbar\) is effectively a ``small'' but still fixed constant.)

\subsubsection{Witt Algebra Centrally Extended to the Virasoro Algebra}
\label{sec:witt}

Let us choose a preferred element of the \(\text{Cl}^{[0]}_{3,0}(\mathbb C)\) that corresponds to a projector onto a (generally) mixed state,
\begin{equation}
  \label{eq:stressenergytensor}
 T(\bs \xi) \equiv T_0 \xi_0 + T_{3} \xi_1 \xi_2 + T_{2} \xi_1 \xi_3 + T_{1} \xi_2 \xi_3,
\end{equation}
where \(T_0 \ne 0\) and at least one \(T_i \ne 0\).

Borrowing the language of conformal field theory, this is a ``secondary field'' since the OPE between \(T\) and some other element \(V\) given in Eq.~\ref{eq:WeylSymbolGroenewoldRule} can be rewritten as
\begin{align}
  \label{eq:OPEofenergymomentumtensor}
  'T \star V'(\bs \xi_1, \bs \xi_2) =& T_0 \left(V(0) + \sum_{j,k,l=1}^3 \epsilon_{ijk} \vec \partial_{\bs \xi_{2_j}} \tilde V(\bs \xi_2) \xi_{2_k} \xi_{2_l}\right)\\
                             & - \frac{i \hbar}{2} \bs T \cdot \vec \partial_{\bs \xi_{2}} \left( V(\bs \xi_2) + \tilde V(\bs \xi_2)\right), \nonumber
\end{align}
where \(\bs T \equiv (T_1, T_2, T_3)\). %
Thus, Eq.~\ref{eq:OPEofenergymomentumtensor} defines a local field \(T(\bs \xi_1)\) that does not depend on \(\bs \xi_2\) and in general (with other secondary fields) it will have on non-zero term in its final line proportional to the central charge. Elements \(V\) that produce a zero term proportional to the central charge are called ``primary fields'' according to the convention of conformal field theory.

Note that according to Eq.~\ref{eq:OPEofenergymomentumtensor} a non-zero \(\hbar\) is necessary to both construct the conformal Witt algebra, since the second term's \(\tilde V(\bs \xi_2)\) in the first line requires the \(\hbar\)-dependent Grassmann Fourier transform, \emph{and} the (quantum) conformal anomaly defined by the final term. This effect of both defining the conformal algebra and a deviation from it can be attributed to \(\text{Cl}_{3,0}(\mathbb C)\)'s feature of being a cover of \(\text{SL}(2, \mathbb R)\) instead of an isomorphism, as discussed in Sections~\ref{sec:summary}-\ref{sec:SLinoctonions}.

The OPE defined by Eq.~\ref{eq:OPEofenergymomentumtensor} is equivalent to a basis-independent formulation or representation of the commutation relations of the Virasoro algebra, which is commonly defined in a basis on the ring \(\bs C[z,z^{-1}]\):
\begin{equation}
  \label{eq:Virasarorelations}
  [L_j, L_k] = (j-k) L_{j+k} + \frac{c}{12} (j^3 - j) \delta_{j+k, 0},
\end{equation}
for
\begin{equation}
L_j \equiv - z^{j+1} \partial_{z}.
\end{equation}
This is satisfied by setting \(L_{\pm 1} \equiv -(\hat \xi_1 \pm i \hat \xi_3)/2\) and \(L_0 = \hat \xi_2/2\) and the algebra multiplication \([\cdot,\cdot]\) to the commutator (or, equivalently, by setting \(L_i\) to the Killing vectors \(V^\mu\) in Appendix~\ref{sec:ads3blackholes} and the algebra multiplication to their Lie bracket).

Assuming that fields \(V(z)\) are dependent on their position through \(L_{-1} V(z)\), this implies that the stress-energy field \(T(y)\) can be defined holomorphically as \(T(y)V(z) = \sum_{n \in \mathbb Z} \frac{L_n V(z)}{(y-z)^{n+2}}\) and implies that the OPE of the stress-energy tensor with itself is
\begin{equation}
  T(y)T(z) = \frac{c/2}{(y-z)^4} + \frac{2 T(z)}{(y-z)^2} + \frac{\partial T(z)}{(y-z)} + \mathcal O(1),
\end{equation}
which can be compared to the basis-independent Eq.~\ref{eq:OPEofenergymomentumtensor}.

\subsubsection{Trace Anomaly}
\label{sec:traceanomaly}

Since we have identified the Grassmann Fourier transform is an \(\mathcal O(\hbar)\) operation (Eq.~\ref{eq:GrassmannFouriertransform}) it follows that its trace (Eq.~\ref{eq:WWMinnerproduct}) of the stress-energy tensor \(T(\bs \xi)\) (Eq.~\ref{eq:stressenergytensor}) exhibits a \emph{trace anomaly}~\cite{}:
\begin{equation}
  i \int T(\bs \xi) \mathcal F(\xi_0) \text d^3 \bs \xi = 2/\hbar \int \xi_0 \xi_1 \xi_2 \xi_3 \text d^3 \bs \xi = 2/\hbar
\end{equation}

Similarly to \(\hbar\) being proportional the the central charge anomaly for the OPE with the stress-energy tensor, this identifies \(\hbar^{-1}>0\) as proportional for the trace anomaly for the stress-energy tensor.

\subsection{Bi-Quantized Seven-Generator Grassmanns and Octonions}
\label{sec:octonions}

We established that the complexified quaternion algebra is isomorphic to \(\text{Cl}^{[0]}_{3,0}(\mathbb C)\) in Section~\ref{sec:hbarCliffordtheory} and defined an OPE equivalent to the Virasoro algebra through Eq.~\ref{eq:WeylSymbolGroenewoldRuleLeadingTerms_combined} in the larger \(\text{Cl}_{3,0}(\mathbb C)\) algebra. %

We have so far used two Clifford algebras that are isomorphic to the complexified quaternion algebra, \(\text{Cl}_{3,0}(\mathbb R)\) and \(\text{Cl}^{[0]}_{3,0}(\mathbb C)\), and expressed them after an \(\hbar\)-dependent monomorphism in the larger \(\text{Cl}_{3,0}(\mathbb C)\) algebra.

This larger \(\text{Cl}_{3,0}(\mathbb C)\) algebra is itself also isomorphic to another Clifford algebra: \(\text{Cl}^{[0]}_{4,0}(\mathbb C)\). However, this isomorphism is ``non-canonical'', i.e.~the isomorphism is generally shown with respect to the orthogonal idempotents of their isomorphic matrix algebras over \(\mathbb C\) (using the Wedderburn-Artin theorem~\cite{Beachy99}), which requires arbitrarily choosing a basis or representation, instead of by a ``basis-independent'' element-to-element morphism.

This non-canonical aspect of the isomorphism suggests we can express one physical theory in terms of another, or mathematically-speaking, in terms of a dual algebra that produces an extension of the original algebra.

Luckily, there is another Clifford algebra that is isomorphic to the complexified quaternion algebra, \(\text{Cl}_{2,0}(\mathbb C)\), from which it is particularly straight-forward to define a monomorphism to \(\text{Cl}^{[0]}_{4,0}(\mathbb C)\), as we saw in Section~\ref{sec:SLinoctonions}.%

Following this intuition, we would like to create an extension of the quaternion algebra by the algebra \(\text{Cl}_{2,0}(\mathbb C)\) monomorphically mapped to \(\text{Cl}^{[0]}_{4,0}(\mathbb C)\). Our intuition seems to be justified by the fact that the only non-trivial normed division algebra that can be an extension of the quaternion algebra must contain twice the number of elements and be eight-dimensional~\cite{Hurwitz98,Baez02}. This means that if we are committed to only dealing with Clifford algebras, we \emph{must} express \(\text{Cl}_{2,0}(\mathbb C)\) as a subalgebra in \(\text{Cl}^{[0]}_{4,0}(\mathbb C) \cong \text{Cl}_{3,0}(\mathbb C)\) to double the dimension appropriately and produce a normed division algebra.

This mapping from \(\text{Cl}_{2,0}(\mathbb C)\) will require a new monomorphism, and we choose for it to depend on a new dimensionless quantization constant, $G$. We will find that this produces the octonion algebra as discussed in Section~\ref{sec:SLinoctonions}.

We define this new \(G\)-dependent monomorphism in a similar manner to the \(\hbar\)-dependent one by first defining a quantization of the Poisson bracket of Clifford algebra elements to the complexified quaternion algebra. However, note that here the Poisson bracket we quantize will actually be in \(\text{Cl}_{4,0}(\mathbb C)\), instead of \(\text{Cl}_{4,0}^{[0]}(\mathbb C) \substack{\cong\\\text{n.c.}} \text{Cl}_{3,0}(\mathbb C)\), though we will subsequently focus on the latter subalgebra of the former. %
We quantize the Poisson bracket within the (even larger) \(\text{Cl}_{4,0}(\mathbb C)\) algebra because we want to quantize four generators so that they become the four new elements to extend our normed division algebra, not three.

\subsubsection{Canonical Quantization}

For \(\mu\), \(\nu \in \{4,5,6,7\}\), we define
\begin{equation}
  \{\xi_\mu, \xi_\nu\}_{\text{P.B.}} \rightarrow \{\hat \xi_\mu, \hat \xi_\nu\} = G \delta_{\mu \nu},
\end{equation}
At this point we do not mean to imply anything physical by selecting our quantization variables to be \(G\), and this constant should just be treated similarly to \(\hbar\): a dimensionless parameter. We will not justify its interpretation as a dimensionless analog to the gravitational constant until Section~\ref{sec:holographicprinciple}.

This means that
\begin{equation}
  \hat \xi_\mu \hat \xi_\nu = \begin{cases} \hat \xi_\nu & \text{if $\mu = 0$}\\ \hat \xi_\mu & \text{if $\nu=0$}.\end{cases}
\end{equation}

When we quantized the Poisson bracket in \(\text{Cl}_{3,0}(\mathbb R)\) we produced Pauli operators and the quaternion algebra, which is an algebra that is considerable less esoteric in the physics community compared to the octonion algebra. Since the non-quaternion quantized operators are likely less familiar to many readers, we will explore their relations here in more detail.

Since \(\text{Cl}_{3,0}^{[0]}(\mathbb C) \cong \text{Cl}_{2,0}(\mathbb C)\), it follows that the three quaternion \(\hbar\)-quantized generators \(\hat \xi_i\) and the four non-quaternion \(G\)-quantized generators \(\hat \xi_\mu\) satisfy the following multiplication rules
\begin{equation}
 \hat \xi_\mu \hat \xi_\nu = \begin{cases} \delta_{\mu \nu} \hat \xi_0 - \epsilon_{\mu\nu\rho} \hat \xi_\rho & \text{if $\mu, \nu \ne 0$}, \end{cases}
\end{equation}
where \(\epsilon_{\mu\nu\rho}\) is a completely antisymmetric tensor with value \(1\) when \(\mu\nu\rho=123\), \(145\), \(176\), \(246\), \(257\), \(347\), \(365\) (an arbitrary choice of convention), and we have temporarily set \(\hbar = G = 2\).

We can find the appropriate proportionality factor by noting that since \(\hat \xi_\mu \hat \xi_\nu\) is proportional to \(\hat \xi_i\), %
\begin{align}
  (\hat \xi_{\mu} \hat \xi_{\nu})^2 =& -\hat \xi_\mu ^2 \hat \xi_\nu^2 = -G^2/4 \\
  \implies& \hat \xi_{\mu} \hat \xi_{\nu} = \pm i \sqrt{G^2 /2\hbar} \hat \xi_i.
\end{align}

For future reference, we rewrite the full octonion multiplication rules with \(\hbar\) and \(G\) shown:
\begin{equation}
  \label{eq:octonionmultiplicationrules}
  \hat \xi_\mu \hat \xi_\nu = \begin{cases} \hat \xi_\nu & \text{if $\mu = 0$}\\ \hat \xi_\mu & \text{if $\nu=0$}\\ \delta_{\mu \nu} \hat \xi_0 - \epsilon_{\mu\nu\rho} \frac{G}{\sqrt{2 \hbar}} \hat \xi_\rho & \text{otherwise}, \end{cases}
\end{equation}
where \(\epsilon_{\mu\nu\rho}\) is a completely antisymmetric tensor with value \(1\) when \(\mu\nu\rho=123\), \(145\), \(176\), \(246\), \(257\), \(347\), \(365\).

\begin{figure}[t]
\includegraphics[]{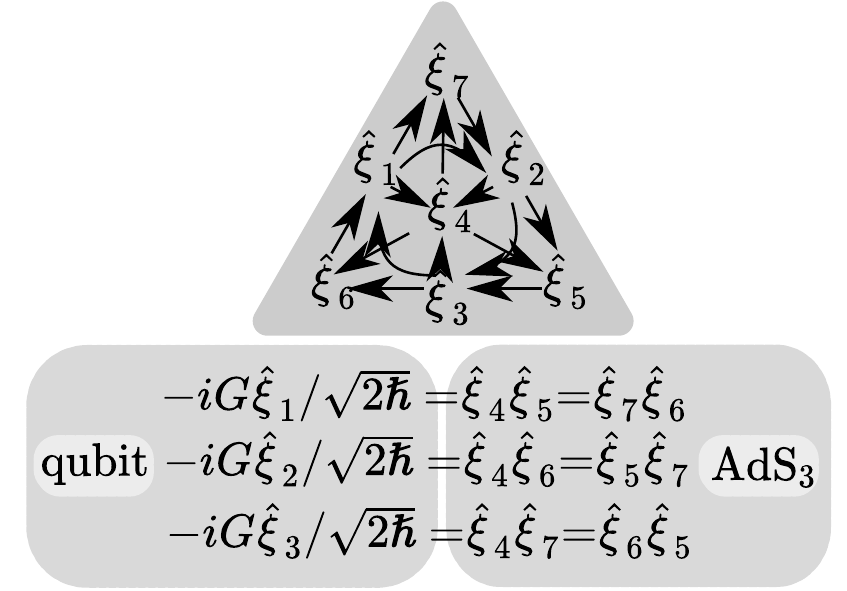}
\caption{The octonion multiplication rules are shown at the top by arrows (this is called an octonion Fano plane mnemonic~\cite{Hirschfeld98}). Below that are listed one order of the corresponding octonion multiplication identities and how they split up in the AdS/qubit duality.} %
\label{fig:octonions}
\end{figure}

The corresponding octonion relations are shown in Figure~\ref{fig:octonions}. 

For completeness, we can also note that since \(\hat \xi_i \hat \xi_\mu\) is proportional to a \(\hat \xi_\nu\), %
it follows that%
\begin{align}
  (\hat \xi_i \hat \xi_{\mu})^2 =& - \hat \xi_i^2 \hat \xi_{\mu}^2 = -\hbar G/4\\
  \implies& \hat \xi_i \hat \xi_{\mu} = \pm i \sqrt{\hbar /2} \hat \xi_\nu.
\end{align}

Finally, since \(\hat \xi_\mu \hat \xi_\nu\) is proportional to \(\hat \xi_\rho \hat \xi_\sigma\) for \(\mu \ne \nu \ne \rho \ne \sigma\). %
\begin{align}
  (\hat \xi_\mu \hat \xi_\nu)^2 =& - \hat \xi_\mu^2 \hat \xi_\nu^2 = - G^2 /4\\
  \label{eq:fourgeneratorGidentity}
  \implies& \hat \xi_\mu \hat \xi_\nu = \pm \hat \xi_{\rho} \hat \xi_\sigma.
\end{align}

Along with \(\epsilon_{ijk} \hat \xi_i \hat \xi_j = -i \sqrt{\hbar/2} \hat \xi_k\), this completes all possible multiplication rules with their explicit constants.

\(\hat \xi_\mu\) extend the algebra to a non-associative algebra. This means that there is no matrix representation of \(\hat \xi_\mu\), unlike the case for \(\hat \xi_i\), which we showed could be represented by the Pauli matrices in two-dimensions.

We saw in Section~\ref{sec:SLinoctonions} that the even products of the non-quaternion operators, \(\hat \xi_\mu \hat \xi_\nu\), correspond to AdS$_3$ isometry group operations. To see what the single non-quaternion operators correspond to, consider \((1+1)\times(1+1)\) tensored Minkowski spacetime with coordinates \((t, x, t', x')\). These are the coordinates of the four-dimensional embedding spacetime with Minkowski metric.

The following identities, which mix \(\hat \xi_i\) and \(\hat \xi_\mu\), hold from the octonion algebra:
\begin{align}
  \label{eq:mixedoctonionidentities1}
  i \sqrt{G^2/2 \hbar} \hat \xi_4 =& \hat \xi_1 \hat \xi_5 = \hat \xi_2 \hat \xi_6 = \hat \xi_3 \hat \xi_7,\\
  i \sqrt{G^2/2 \hbar} \hat \xi_5 =& \hat \xi_4 \hat \xi_1 = \hat \xi_2 \hat \xi_7 = \hat \xi_6 \hat \xi_3,\\
  i \sqrt{G^2/2 \hbar} \hat \xi_6 =& \hat \xi_7 \hat \xi_1 = \hat \xi_4 \hat \xi_2 = \hat \xi_3 \hat \xi_5,\\
  \label{eq:mixedoctonionidentities4}
  i \sqrt{G^2/2 \hbar} \hat \xi_7 =& \hat \xi_1 \hat \xi_6 = \hat \xi_5 \hat \xi_2 = \hat \xi_4 \hat \xi_3.
\end{align}

\begin{figure}[t]
\includegraphics[]{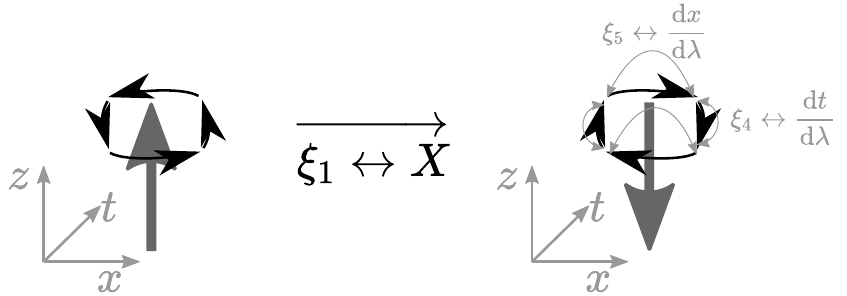}
\caption{Flipping a spin in the \(z\)-direction (i.e.~the operation \(\xi_1 \leftrightarrow X\)) is equivalent to flipping its angular momentum arrow component-wise in the \(x\)-direction and \(t\)-direction, which ends up reversing the angular momentum and thus flipping the associated spin to its opposite value. Discretized translations in the \(x\)-direction and \(t\)-direction correspond to such flips in the case of a discrete modulus \(2\) spacetime.}
\label{fig:embeddingspace}
\end{figure}

As found in Section~\ref{sec:SLinoctonions}, taking into account the imaginary constants in Eqs.~\ref{eq:AdSirreducibleelements1}-\ref{eq:AdSirreducibleelements4} that relate these to the spacetime, it follows that \(-(i\hat \xi_4)^2 = -(i \hat \xi_6)^2 = \hat \xi_5^2 = \hat \xi_7^2 = 1\) and this defines the metric signature \(({-}{+}{-}{+})\) of the embedding space.

Thus, \((\hat \xi_4, \hat \xi_5, \hat \xi_6, \hat \xi_7)\) can be interpreted as the quantization of translations along \((t, x, y, z)\),
\begin{equation}
  (\hat \xi_4, \hat \xi_5, \hat \xi_6, \hat \xi_7) \equiv \left( \der{t}{\lambda}, \der{x}{\lambda}, \der{t'}{\lambda}, \der{x'}{\lambda} \right).
\end{equation}

These identities can be interpreted as a relationship between spin degrees of freedom and time boosts and space shifts in special relativity. For example, consider the spin-flip operation \(\hat X\) associated with \(\xi_1\), when acting on a spin in the \(z\)-direction. The first line (Eq.~\ref{eq:mixedoctonionidentities1}) of the identities above imply that boosting a spin in \(t-x\), which we found in the prior subsection is given by the Lorentz element \(\xi_4 \xi_5\) (the embedded space inherits the isometries of the embedding space), is equivalent to flipping its spin, i.e. \(\xi_1\equiv X\). See Figure~\ref{fig:embeddingspace} for a depiction of this relationship and how a shift in time \(t\) and a shift in space \(x\) reverses the perpendicular components of a \(z\)-spin.

Similarly, rotating in the \(t-t'\) plane or boosting along \(t-x'\) is equivalent to \(\xi_2\equiv Z\) and \(\xi_3\equiv Y\), respectively.

Flipping the order of \(\xi_i \xi_\nu\) above introduces a minus sign and corresponds to time reversal for \(\xi_4\) and \(\xi_6\) and space parity reversal for \(\xi_5\) and \(\xi_7\).

\subsubsection{Monomorphism from \(\text{Cl}_{4,0}^{[0]}(\mathbb C)\) to \(\text{Cl}_{2,0}(\mathbb C) \subseteq \text{Cl}_{4,0}^{[0]}(\mathbb C)\)}
\label{sec:fourgenWWM}%

We examine the identity from the octonion algebra that relates the non-quaternion generators to the quaternion generators:
\begin{align}
  \label{eq:sigmaoddPaulimap}
  \frac{1}{\sqrt{2 \hbar}} \hat \xi_i = -i \frac{1}{G} \hat \xi_\mu \hat \xi_\nu.
\end{align}

Similarly, the even elements are related by
\begin{align}
  \label{eq:sigmaevenPaulimap}
  \frac{1}{\hbar} \hat \xi_i \hat \xi_j = \frac{1}{G} \hat \xi_\rho \hat \xi_\mu.
\end{align}

This is a convenient identity since as mentioned in Section~\ref{sec:hbarCliffordtheory}, in quantum field theory these fields are renormalized \(\xi_i \xi_j \rightarrow \xi_i \xi_j/\hbar\) and in string theory a similar renormalization is frequently performed: \(\xi_\mu \xi_\nu \rightarrow \xi_\mu \xi_\nu/ G\).

With this in mind, we define an isomorphism \(\sigma\) between \(\text{Cl}^{[0]}_{3,0}(\mathbb C)\) and a subalgebra of \(\text{Cl}^{[0]}_{4,0}(\mathbb C)\) isomorphic to \(\text{Cl}_{2,0}(\mathbb C) \cong \text{Cl}^{[0]}_{3,0}(\mathbb C) \cong \text{Cl}_{3,0}(\mathbb R)\), by identifying the corresponding elements:
\begin{align}
  \label{eq:sigmaisomorphism1}
  \xi_0 &\overset{\sigma}{\leftrightarrow} \bar \xi_0,\\
  \label{eq:sigmaisomorphism2}
 \sqrt{2 \hbar} i \xi_1,\, \xi_2 \xi_3  &\overset{\sigma}{\leftrightarrow} \bar \xi_4 \bar \xi_5 = \bar \xi_7 \bar \xi_6\\
  \label{eq:sigmaisomorphism3}
  \sqrt{2 \hbar} i \xi_2,\, \xi_1 \xi_3  &\overset{\sigma}{\leftrightarrow} \bar \xi_4 \bar \xi_6 = \bar \xi_5 \bar \xi_7\\
  \label{eq:sigmaisomorphism4}
  \sqrt{2 \hbar} i \xi_3,\, \xi_1 \xi_2  &\overset{\sigma}{\leftrightarrow} \bar \xi_4 \bar \xi_7 = \bar \xi_6 \bar \xi_5.
\end{align}
Notice that even though there are two even non-quaternion terms that are equal to every quaternion even term, they are meant to be defined on different subspaces so this is still a bijective map and thus an isomorphism.

We denote the elements of this subalgebra of \(\text{Cl}^{[0]}_{4,0}(\mathbb C)\) by bars, instead of without any overhead modifier, to take into account that they are dependent on \emph{both} \(\hbar\) and \(G\), as we shall shortly see.

This map can be trivially extended to apply to the complexified larger algebra, \(\text{Cl}_{3,0}(\mathbb C)\), %
and thereby define a monomorphism from the \(\text{Cl}_{2,0}(\mathbb C)\) subalgebra of \(\text{Cl}^{[0]}_{4,0}(\mathbb C)\) to \(\text{Cl}^{[0]}_{4,0}(\mathbb C)\) using the monomorphism from \(\text{Cl}^{[0]}_{3,0}(\mathbb C)\) to \(\text{Cl}_{3,0}(\mathbb C)\) defined in Section~\ref{sec:hbarquantization}:
\begin{equation}
  \label{eq:Gmonomorphismthroughhmonomorphism}
  \text{Cl}^{[0]}_{4,0}(\mathbb C) \supseteq \text{Cl}_{2,0}(\mathbb C) \overset{\sigma}{\leftrightarrow} \text{Cl}^{[0]}_{3,0}(\mathbb C) \overset{\hbar>0}{\hookrightarrow} \text{Cl}_{3,0}(\mathbb C) \overset{\sigma}{\leftrightarrow} \text{Cl}^{[0]}_{4,0}(\mathbb C). %
\end{equation}

If we define a four-generator element in the \(\text{Cl}^{[0]}_{4,0}(\mathbb C)\) algebra to be
\begin{equation}
  \bar g(\bar{\bs \xi}) = \bar g_0 \bar \xi_0 + \sum_{\mu,\nu=4}^7 \bar g_{\mu \nu} \bar \xi_\mu \bar \xi_\nu + \sum_{\mu,\nu,\rho,\sigma=4}^7 \bar g_{\mu \nu \rho \sigma} \bar \xi_\mu \bar \xi_\nu \bar \xi_\rho \bar \xi_\sigma,
\end{equation}
we can similarly define a dual to this non-quaternionic \(\bar g (\bar {\bs \xi})\) through the Grassmann Fourier transform restricted to these four generators:
  \begin{equation}
    \label{eq:nonquaternionGrassmannFouriertransform}
    {\bar g}(\bar {\bs \xi}) = \bar{\mathcal F}(\bar{\tilde g}(\bs \rho)) \equiv \frac{2}{G} \int e^{i\sqrt{G/2} \sum_{k=4}^7 \bar \xi_k \bar \rho_k} \tilde{\bar g}(\bar \rho) \text d \bar \rho_7 \text d \bar \rho_6 \text d \bar \rho_5 \text d \bar \rho_4.
  \end{equation}
  Therefore, the inverse Grassmann Fourier transform on these four generators must be,
  \begin{align}
    \tilde{\bar  g}(\bar{\bs \rho}) &= \bar{\mathcal F}^{-1}(\tilde g(\bar{\bs \xi}))\\
    \label{eq:nonquaternionGrassmanninverseFouriertransform}
    &\equiv \frac{2}{G} \int e^{-i\sqrt{G/2} \sum_{k=4}^7 \bar \xi_k \bar \rho_k} \bar g(\bar \xi) \text d \bar \xi_7 \text d \bar \xi_6 \text d \bar \xi_5 \text d \bar \xi_4.
  \end{align}

  In Section~\ref{sec:hbarquantization} the Grassmann Fourier transform allowed us to express the monomorphism \(\text{Cl}^{[0]}_{3,0}(\mathbb C) \overset{\hbar>0}{\hookrightarrow} \text{Cl}_{3,0}(\mathbb C)\) without referring to the quantization of the Poisson bracket. Similarly, here we will be able to use the Grassmann Fourier transform restricted to \(\text{Cl}^{[0]}_{4,0}(\mathbb C)\) to not only define the monomorphism given by Eq.~\ref{eq:Gmonomorphismthroughhmonomorphism} without referring to the quantization of its Poisson bracket, but also without referring to the intermediate map \(\sigma\) through \(\text{Cl}^{[0]}_{3,0}(\mathbb C)\).

  As we have defined it so far, \(\tilde{\bar g}(\bar{\bs \rho})\) has the same form as \(\bar g(\bar{\bs \xi})\), namely
  \begin{equation}
    \tilde{\bar g}(\bar{\bs \rho}) = \tilde{\bar g}_0 \bar \rho_0 + \sum_{\mu,\nu=4}^7 \tilde{\bar  g}_{\mu \nu} \bar \rho_\mu \bar \rho_\nu + \sum_{\mu,\nu,\rho,\sigma=4}^7 \tilde{\bar g}_{\mu \nu \rho \sigma} \bar \rho_\mu \bar \rho_\nu \bar \rho_\rho \bar \rho_\sigma.
  \end{equation}
When defined over three generators, we found that the Grassmann Fourier transform related odd and even coefficients. Here, we can find that the Grassmann Fourier transform relates same-grade coefficients of \(\tilde{\bar g}(\bar{\bs \rho})\) to the those of \(\bar g(\bar{\bs \xi})\) in the following manner:
  \begin{align}
    \tilde{\bar g}_{45} - \tilde{\bar g}_{54} &= \bar g_{67} - \bar g_{76},\\
    \tilde{\bar g}_{47} - \tilde{\bar g}_{74} &= \bar g_{56} - \bar g_{65},\\
    \tilde{\bar g}_{46} - \tilde{\bar g}_{64} &= \bar g_{57} - \bar g_{75},\\
    \sum_{\mu,\nu,\rho,\sigma=4}^7 \epsilon_{\mu \nu \rho \sigma}\tilde{\bar g}_{\mu\nu\rho\sigma}
    &= \bar g_0,
  \end{align}
  along with the corresponding equations found from exchanging the tilde'd and untilde'd coefficients above.

We choose to redefine
  \begin{align}
  \bar g(\bar{\bs \xi}) =& \bar g_0 + \bar g_{45} \bar \xi_4 \bar \xi_5 + \bar g_{46} \bar \xi_4 \bar \xi_6 + \bar g_{56} \bar \xi_5 \bar \xi_6\\
                         & + \bar g_{54} \bar \xi_5 \bar \xi_4 + \bar g_{64} \bar \xi_6 \bar \xi_4 + \bar g_{65} \bar \xi_6 \bar \xi_5, \nonumber
\end{align}
and
\begin{align}
  \tilde{\bar g}(\bar{\bs \rho}) =& \tilde{\bar g}_{67} \bar \rho_6 \bar \rho_7 + \tilde{\bar g}_{57} \bar \rho_5 \bar \rho_7 + \tilde{\bar g}_{47} \bar \xi_4 \bar \xi_7\\
                                  & + \tilde{\bar g}_{76} \bar \rho_7 \bar \rho_6 + \tilde{\bar g}_{75} \bar \rho_7 \bar \rho_5 + \tilde{\bar g}_{74} \bar \xi_7 \bar \xi_4\\
                                  & + \sum_{\substack{\mu,\nu,\rho,\sigma=4\\\mu\ne\nu\ne\rho\ne\sigma}}^7\tilde{\bar g}_{\mu \nu \rho \sigma} \bar \rho_\mu \bar \rho_\nu \bar \rho_\rho \bar \rho_\sigma,
\end{align}
where all coefficients are defined to be in \(\mathbb C\).

We will refer to \(\bar g(\bar{\bs \xi})\) as the \emph{AdS dual} element and \(\tilde{ \bar g}(\bar{\bs \xi})\) as its Grassmann Fourier transform. %
As we have defined it, the AdS dual element is related to quadratic products of only three of the four generators. Thus, it is clearly isomorphic to \(\text{Cl}^{[0]}_{3,0}(\mathbb C) \in \text{Cl}_{2,0}(\mathbb C)\). %

It follows that we can equivalently define the monomorphism given in Eq.~\ref{eq:Gmonomorphismthroughhmonomorphism} by
\begin{equation}
  \label{eq:Gmonomorphism}
\text{Cl}_{2,0}(\mathbb C) \overset{G>0}{\hookrightarrow} \text{Cl}^{[0]}_{4,0}(\mathbb C) \equiv \text{Cl}_{2,0}(\mathbb C) \oplus \mathcal F(\text{Cl}_{2,0}(\mathbb C)),
\end{equation}
where by \(\text{Cl}_{2,0}(\mathbb C)\) in Eq.~\ref{eq:Gmonomorphism} we mean to imply the \(\text{Cl}_{2,0}(\mathbb C)\) subalgebra expressed within \(\text{Cl}^{[0]}_{4,0}(\mathbb C)\). We note that the monomorphism defined by Eq.~\ref{eq:Gmonomorphism} can only be applied to \(\bar g(\bar{\bs \xi}) \in \text{Cl}_{2,0}(\mathbb C) \subseteq \text{Cl}^{[0]}_{4,0}(\mathbb C)\) %
much like the case for the monomorphism defined from \(\text{Cl}^{[0]}_{3,0}(\mathbb C)\) to \(\text{Cl}_{3,0}(\mathbb C)\), which only applied to \(g(\bs \xi) \in \text{Cl}^{[0]}_{3,0}(\mathbb C)\), not their Grassmann Fourier transforms.

Since we have defined the AdS dual element to be quadratic products of only three of the four generators, it follows that we need to only minimally change the Hilbert space inner product that we defined in Section~\ref{sec:G3andHilbertspace} to define a Hilbert space for \(\text{Cl}_{3,0}(\mathbb C)\) to work for \(\text{Cl}_{4,0}(\mathbb C)\):
\begin{align}
  \Tr(\hat O \hat \rho) =& G^{-1} \int \bar O(\bar{\bs \xi}) \tilde{\bar \rho}(\bar{\bs \xi}) \text d^4 \bs \xi\\
  =& G^{-1} \int \tilde{\bar O}(\bar{\bs \xi}) {\bar \rho}(\bar{\bs \xi}) \text d^4 \bs \xi
\end{align}
where \(\text d^4 \bs \xi \equiv \text d \xi_4\text d \xi_5\text d \xi_6\text d \xi_7\). Namely, the inner product must have an additional Grassmann integral to run over the fourth \(\xi_7\) Grassmann variable that is spanned by the AdS dual elements \(\tilde{\bar g}\).

\subsubsection{Dual AdS OPE}

Performing an expansion in orders of \(G\) for products of AdS dual elements in \(\text{Cl}_{2,0}(\mathbb C) \in \text{Cl}^{[0]}_{4,0}(\mathbb C)\) trivially produces the same expansion as found for \(\text{Cl}^{[0]}_{3,0}(\mathbb C)\):
  \begin{align}
    & \bar W_1 \star \bar W_2 (\bar{\bs \xi}) \nonumber\\
    \label{eq:AdSWeylSymbolGroenewoldRule}
    \equiv& \mathcal O(G^0) + \mathcal O(G^1) + \mathcal O(G^2)\\
    =& \bar W_1(\bar{\bs \xi}) \bar W_2(\bar{\bs \xi_2}) \nonumber\\
    & - \frac{G i}{2} \sum_\mu \vec \partial_{\bar \xi_\mu} \bar W_1(\bar{\bs \xi}) \vec \partial_{\bar \xi_\mu} \bar W_2(\bar{\bs \xi}) \\
    & - \frac{G^2}{4} \sum_{\mu,\nu} \vec \partial^2_{\bar \xi_\mu,\bar \xi_\nu} \bar W_1(\bar{\bs \xi}) \vec \partial^2_{\bar \xi_i, \bar \xi_j} \bar W_2(\bar{\bs \xi}). \nonumber
  \end{align}

  It is easy to verify that %
  the \(\mathcal O(G^0)\) term captures Pauli times identity, the \(\mathcal O(G^1)\) term captures Pauli times Pauli, and the \(\mathcal O(G^2)\) term captures Pauli squared terms.

  These identifications can also be seen by examining their quantized products:
  \begin{align}
    \hat \xi_\mu \hat \xi_\nu \hat \xi_0 &= \hat \xi_\mu \hat \xi_\nu \in \mathcal O(G^0),\\
    \label{eq:mixedAdSWeylSymbolGroenewoldRule}
    \hat \xi_\mu \hat \xi_\nu \hat \xi_\mu \hat \xi_\rho &= -i\frac{G}{2}\hat \xi_\rho \hat \xi_\nu \in \mathcal O(G^1),\\
    (\hat \xi_\mu \hat \xi_\nu)^2 &= \frac{G^2}{4}\hat \xi_0 \in \mathcal O(G^2).
  \end{align}

   However, taking the full \(\text{Cl}^{[0]}_{4,0}(\mathbb C)\) algebra into account to reduce the order of the last term can be reduced to \(\mathcal O(G^1)\) produces a different form:
  \begin{align}
    & \bar W_1 \star \bar W_2 (\bar{\bs \xi}) \nonumber\\
    \label{eq:AdSWeylSymbolGroenewoldRule_combined}
    \equiv& \mathcal O(G^0) + \mathcal O(G^1) + \mathcal O(G^1)\\
    =& \bar W_1(\bar{\bs \xi}) \bar W_2(\bar{\bs \xi}) \nonumber\\
    & - \frac{G i}{2} \sum_\mu \vec \partial_{\bar \xi_\mu} \bar W_1(\bar{\bs \xi}) \vec \partial_{\bar \xi_\mu} \bar W_2(\bar{\bs \xi}) \\
    & - \frac{G}{2} \reallywidetilde{{\bar W}_1(\bar{\bs \xi}) \widetilde{\bar W}_2(\bar{\bs \xi})}. \nonumber
  \end{align}

  Notice that the second \(\mathcal O(G^1)\) term involves a global Grassmann Fourier transform unlike the OPE in \(\text{Cl}_{3,0}(\mathbb C)\). %
  This means it is not strictly an OPE in the conformal sense, i.e.~it is not equivalent to the Virasoro algebra. This can be explained with the reminder that \(\text{Cl}_{4,0}^{[0]}(\mathbb C) \substack{\cong\\\text{n-c}} \text{Cl}_{3,0}(\mathbb R)\), where \(\substack{\cong\\\text{n-c}}\) denotes a non-canonical isomorphism. Therefore, term-by-term, this non-canonical isomorphism maps the OPE in \(\text{Cl}_{3,0}(\mathbb C)\) to a non-conformal form in \(\text{Cl}^{[0]}_{4,0}(\mathbb C)\).

    Finally, note again that in string theory, much like in quantum field theory, there is often an implicit renormalization performed of the fields \(\xi_\mu \xi_\nu \rightarrow \xi_\mu \xi_\nu/G\). This transforms the orders in Eq.~\ref{eq:AdSWeylSymbolGroenewoldRule_combined} from \(\mathcal O(G^0) + \mathcal O(G^1)\) to \(\mathcal O(G^{-1}) + \mathcal O(G^0)\).

  \subsubsection{Symplectic Structure}
  \label{sec:Gsymplecticstructure}

Here we will find that we can define a Lagrangian as a quadratic function of \(\bar \xi_\mu\) in \(\text{Cl}^{[0]}_{4,0}(\mathbb C)\) similar to Lagrangians in \(\text{Cl}_{3,0}(\mathbb C)\):
\begin{equation}
  \mathcal L = -\sum_{\mu, \nu, \rho=4}^7 \epsilon_{\rho \mu \nu} g_{\mu \nu} \bar \xi_\mu \bar \xi_\nu.
\end{equation}

To see that this is justified, consider the general quaternion Lagrangian,
\begin{equation}
  \label{eq:Lagrangian}
  -\mathcal L = \frac{i}{2} \xi_j \xi_k.
\end{equation}
The corresponding qubit Hamilton equations, captured fully at \(\mathcal O(\hbar^1)\), are
\begin{align}
  \der{\xi_i}{\lambda} =& 0,\\
  \der{\xi_j}{\lambda} =& i \hbar \xi_k/2,\\
  \der{\xi_k}{\lambda} =& -i \hbar \xi_j/2.
\end{align}
The dual AdS Euler-Lagrange equations, captured at \(\mathcal O(G^1)\) can be found from these to be
\begin{align}
  \label{eq:dualAdSELeq1}
  \der{\bar \xi_\mu \bar \xi_\nu}{\lambda} &= 0,\\
  \label{eq:dualAdSELeq2}
  \der{\bar \xi_\sigma \bar \xi_\rho}{\lambda} &= 0,\\
  \label{eq:dualAdSELeq3}
  \der{\bar \xi_\mu \bar \xi_\rho}{\lambda}  &= \der{\bar \xi_\nu \bar \xi_\sigma}{\lambda} = iG(\bar \xi_\mu \bar \xi_\sigma + \bar \xi_\rho \bar \xi_\nu),\\
  \label{eq:dualAdSELeq4}
  \der{\bar \xi_\mu \bar \xi_\sigma}{\lambda} &= \der{\bar \xi_\rho \bar \xi_\nu}{\lambda} = -iG(\bar \xi_\mu \bar \xi_\rho + \bar \xi_\nu \bar \xi_\sigma),
\end{align}

Setting Hamilton's equation of motion to be
  \begin{align}
    \label{eq:Killing}
    \der{\bar \xi_\mu \bar \xi_\nu}{\lambda} &= \der{\bar \xi_\mu}{\lambda} \bar \xi_\nu + \bar \xi_\mu \der{\bar \xi_\nu}{\lambda}\\
    =& \frac{G}{2} \left(\{H, \bar \xi_\mu\}_{\text{P.B.}}\bar \xi_\nu + \bar \xi_\mu \{H, \bar \xi_\nu\}_{\text{P.B.}}\right)\\
    \label{eq:GrassmannHamilton_dual}
    =& \frac{iG}{2} \left(H \frac{\cev \partial}{\partial \bar \xi_\mu} \bar \xi_\nu + \bar \xi_\mu H \frac{\cev \partial}{\partial \bar \xi_\nu}\right).
  \end{align}
reveals that the Lagrangian \(\mathcal L = -H\) in terms of the dual non-quaternion generators is
\begin{equation}
  \label{eq:dualLagrangian}
  -\mathcal L \equiv -(\bar {\mathcal L} + \tilde{\bar {\mathcal L}})=  i(\bar \xi_\rho \bar \xi_\mu + \bar \xi_\sigma \bar \xi_\nu).
\end{equation}
This can also be found by performing the isomorphism from the prior Lagrangian in \(\text{Cl}_{3,0}(\mathbb C)\) and including both equal non-quaternion terms in a uniformly weighted linear combination. %

As we motivated in Section~\ref{sec:SLinoctonions} and show in Appendix~\ref{sec:ads3blackholes}, \(\bar \xi_\mu \bar \xi_\nu\) are equal to the Killing vectors, \(\{K^\rho_{(\mu,\nu)}\}_{\substack{\mu,\nu\\\mu\ne\nu}}\), of the Minkowski embedding spacetime, \(K_{(\mu,\nu)} = \bar \xi_\mu \bar \xi_\nu \equiv \epsilon_{\mu\nu} \xi_\nu \partial_\mu = x_\mu \partial_{x_\nu} - x_\nu \partial_{x_\mu}\), when expressed in local coordinates \(\bs x\) (the subscript \((\mu,\nu)\) indexes the Killing vectors and is in parentheses so as not to be mistaken for a tensor index).

Therefore, while we can associate the \(\mathcal L\) in \(\text{Cl}_{3,0}(\mathbb C)\) as the Lagrangian of spin equations of motion, the isomorphic \(\mathcal L\) in \(\text{Cl}^{[0]}_{4,0}(\mathbb C)\) can be equivalently associated with the Lagrangian of isometry generator equations of motion.

\subsection{The holographic principle}
\label{sec:holographicprinciple}

The quantization of three (of seven) generators by \(\hbar\) and the remaining four by \(G\) effectively defined two different monomorphisms, the first from \(\text{Cl}_{3,0}(\mathbb R)\) to \(\text{Cl}_{3,0}(\mathbb C)\) and the second from \(\text{Cl}_{2,0}(\mathbb C) \subseteq \text{Cl}^{[0]}_{4,0}(\mathbb C)\) to \(\text{Cl}^{[0]}_{4,0}(\mathbb C)\). %
We identify \(\hbar\) with a dimensionless analog to Planck's constant in Appendix~\ref{sec:hbar0qutritCliffordtheory}. %
Here we will also justify identifying the quantization variable of the extension of the quaternion algebra, \(G\), with a dimensionless analog to the gravitational constant (technically \(G_N^{(3)}\), the Newton constant in \(2+1\) spacetime).

The full algebra can be defined by the triply-nested anticommutator
\begin{align}
  \label{eq:fullLiebracket}
  \Bigg\{ \Big\{ \hat\xi_\mu \{\hat \xi_i, \hat \xi_j\}\Big\},\hat \xi_{\nu}\Bigg\} =& G \hbar \delta_{ij} \delta_{\mu \nu}.
\end{align}
This defines the three different products possible between the two quantized groups of generators.

Eq.~\ref{eq:fullLiebracket} reveals that \(G \hbar\) is a conserved area in the sense that the full quantum gravity theory can be formulated semiclassically in terms of linear combinations of \(\mathcal O(\hbar^1 G^1)\) stationary phase points each corresponding to the symplectic dynamics of the seven-generator Grassmann algebra, which individually preserve area \(G \hbar\). This is the same as the quantum subtheory, which can be formulated semiclassically in terms of \(\mathcal O(\hbar^1)\) stationary phase points corresponding to the symplectic dynamics of the three-generator Grassmann algebra, which individually preserve the area \(\hbar\)~\cite{Kocia17_2}.

For a two-dimension CFT the Planck length, \(l_P = \left(\frac{\hbar G}{c^3}\right)^{1/2}\) and Planck time, \(t_P = \left(\frac{\hbar G}{c^5}\right)^{1/2}\), is constant. Here, for a \(2+1\) bulk spacetime, area corresponds to length times time, which is thus proportional to \(\hbar G\). This means that this quantization into \(\hbar\) and \(G\) is really just enforcing the area law and holographic principle, as expected from a \(2D\) conformal field theory, and thereby defines the \(G\) quantization variable as the gravitational constant.

As a result, here we can finally formally associate \(G\) to be a dimensionless analog to the gravitational constant and we do so henceforth.

\subsection{Dictionary}
\label{sec:dictionary}

We would like to summarize the findings from Sections~\ref{sec:octonions}-\ref{sec:holographicprinciple} here and formally establish the ``dictionary'' of the AdS$_3$/qubit correspondence presented. To do this it is instructive to review three steps in the quantization procedure above, and how they each produce a specific field:
\begin{enumerate}
\item Quantization of \(\xi_i\) by \(\hbar\) produces fields corresponding to generators of single-qubit spin degrees of freedom (represented by the Pauli matrices):
\(\hat \xi_i \leftrightarrow \hat \sigma_i\)

\item Quantization of \(\xi_\mu\) by \(G\) produces fields corresponding to generators of translation along the degrees of freedom of an embedding spacetime with a \((2,2)\) signature:
\(\hat \xi_\mu \leftrightarrow \der{x^\mu}{\lambda}\)

\item Imposing the octonion algebraic multiplication rules upon the non-quaternion elements produces \(\mathcal O(G^0)\) fields \(\bar \xi_\mu \bar \xi_\nu\) corresponding to generators of spacetime isometries (i.e.~Killing vectors) of the embedded AdS$_3$ spacetime with \((1,2)\) signature:
  \(\bar \xi_\mu \bar \xi_\nu \leftrightarrow x^\mu \partial_{x^\nu} \pm x^\nu \partial_{x^\mu}\)
\end{enumerate}

We proceed to highlight how the usual consequences of this duality in the continuous AdS/CFT correspondence appear here in terms of a conformal/isometry duality and a strong/weak duality.

\subsubsection{Conformal Theory/\(\text{SO}^+(2,2)\) Isometries of AdS$_3$}

We found in Section~\ref{sec:conformal} that \(\text{Cl}_{3,0}(\mathbb C)\) is able to capture the multiplication rules of the \(\text{Cl}_{3,0}^{[0]}(\mathbb C)\) enveloping algebra of the \(SO^+(2,2)/SO^+(1,2)\) isometry group in terms of an operator product expansion that is equivalent to the Virasoro algebra. This is similar to the CFT/AdS duality of the eponymous correspondence.

\subsubsection{Strong/Weak Duality}

We found in Section~\ref{sec:fourgenWWM} (Eq.~\ref{eq:sigmaevenPaulimap}) that
\begin{equation}
  \frac{\hbar}{G} \hat \xi_\mu \hat \xi_\nu = \hat \xi_i \hat \xi_j.
\end{equation}

Therefore, taking \(\hbar \rightarrow 0\) and fixing \(G>0\) is equivalent to taking \(G \rightarrow \infty\) and fixing \(\hbar>0\). Similarly, taking \(\hbar \rightarrow \infty\) and fixing \(G>0\) is equivalent to taking \(G \rightarrow 0\) and fixing \(\hbar>0\).

Consider \(\mathcal L = \xi_i \xi_j\) and the isomorphic \(\mathcal L = \bar \xi_\mu \bar \xi_\nu + \bar \xi_\rho \bar \xi_\sigma\).
It follows that taking \(\hbar \rightarrow 0\) and fixing \(G > 0\) produces
\begin{align}
  \der{\xi_k}{\lambda} &= 0 \quad \left(\substack{\text{uncoupled}\\\text{Majoranas}}\right),\\
  \der{\xi_i \xi_j}{\lambda} = \der{\bar \xi_\mu \bar \xi_\nu}{\lambda} &= G (\bar \xi_\mu \bar \xi_\rho + \bar \xi_\sigma \bar \xi_\nu) \quad \left(\substack{\text{coupled}\\\text{isometries}}\right),
\end{align}
while taking \(\hbar \rightarrow \infty\) and fixing \(G > 0\) produces
\begin{align}
  \der{\xi_i}{\lambda} &= \hbar \xi_j \quad \left(\substack{\text{coupled}\\\text{Majoranas}}\right),\\
  \der{\xi_j \xi_k}{\lambda} = \der{\bar \xi_\mu \bar \xi_\sigma}{\lambda} &= 0 \quad \left(\substack{\text{uncoupled}\\\text{isometries}}\right). %
\end{align}

Therefore, in the limit that \(\hbar \rightarrow \infty\) the \(\hbar\) coupling between the Majoranas given by \(H = \frac{i}{2} \sum_{ijk} \epsilon_{ijk} b_i \xi_i \xi_j\) becomes strong and in this limit the equations of motion for \(\bar \xi_\mu \bar \xi_\nu\) correspond to classical equations of motion for isometry generators in AdS$_3$. %

On the other hand, in the limit that \(\hbar \rightarrow 0\) the \(\hbar\) coupling between the Majoranas in the qubit Hamiltonians \(H = \frac{i}{2} \sum_{ijk} \epsilon_{ijk} b_i \xi_i \xi_j\) becomes weak (since \(\der{\xi_i}{\lambda} = (\hbar/2) H \cev \partial_{\xi_i}\)) and in this limit the equations of motion for \(\bar \xi_\mu \bar \xi_\nu\) correspond to non-classical equations of motion for coupled isometry generators in AdS$_3$. %

Note that \(\hbar \rightarrow \infty\) does not correspond the the equations become more quantum, as this is a rescaling that also changes the effective Planck's constant (action area). Indeed, the simplification that occurs in the dual algebra reveals that the effective Planck's constant actually vanishes in this limit. In fact, its equations of motion can be associated with the trivial dynamics of a maximally mixed thermal state at the center of the Bloch sphere.

\subsubsection{Bulk/Boundary Correspondence}

The AdS$_3$ algebra elements \(\hat \xi_\mu\) produce a larger-dimensional space corresponding to the Minkowski \(2+2\) embedding space. They define a lower-dimensional space spanned by \(\hat \xi_i\) wherein the \(SO(2,2)\) isometry conformal group is defined. In this sense, there is a bulk/boundary correspondence. However, as discussed in Figure~\ref{fig:AdS/qubit}, in the AdS/CFT correspondence the AdS bulk only corresponds to a CFT at conformal infinity where it becomes a Minkowski spacetime. Here we are associating the bulk with a global Minkowski spacetime. %

\subsubsection{Supersymmetry}

In the original AdS/CFT conjecture~\cite{Maldacena99}, \(\mathcal N = 4\) SYM theory, which is maximally symmetric, was conjectured to be dual to AdS$_5 \times \mathbb S^5$ space, which is a maximally supersymmetric solution of ten-dimensional supergravity. The \(32\) Killing spinors corresponded to the supercharges of \(\mathcal N = 4\) SYM generated by complex Majoranas, along with similar correspondences among the other symmetries and fields. As discussed in Section~\ref{sec:SLinoctonions}, Section~\ref{sec:octonions} and Section~\ref{sec:dictionary}, the \(\xi_i\) generators expressed in \(\text{Cl}_{3,0}(\mathbb C)\) are similarly supersymmetric Majorana fields that are isomorphic to Killing vectors \(\bar \xi_\mu \bar \xi_\nu\) expressed in \(\text{Cl}^{[0]}_{4,0}(\mathbb C)\).

\subsection{Perturbed AdS$_3$ Spacetime}
\label{sec:perturbedAdS3}

AdS$_d$ is a maximally symmetric spacetimes and so contain \(n(n+1)/2\) isometries (i.e.~Killing vectors). Therefore, AdS$_3$ has six isometries. By expressing AdS$_3$ as a homogeneous symmetric space, the six isometries of AdS$_3$ are paired into three equivalent isometries (corresponding to clockwise and counter-clockwise isometry transformations found in Appendix~\ref{sec:ads3blackholes}).

We would like to explore non-maximally symmetric spacetimes that are perturbatively close to AdS$_3$. One way to accomplish this is by introducing a perturbation that removes one of the isometries. Here, we will accomplish this through a \(\mathcal O(G)\) transformation that corresponds to an infinitesimal symmetry-breaking boost between \(x\) and \(x'\) (in terms of the coordinates defined in Section~\ref{sec:octonions}). %

As discussed before, the Paulis \(\xi_i = \xi_\mu \xi_\nu\) and the identity element \(\xi_0\), together constituting the quaternion algebra, generate the isometries of the AdS$_3$ spacetime. Rigid rotations take these isometries to linear combinations of the original isometries. %

For instance, as we show in Appendix~\ref{sec:quantummagicstates}, the rotation effected by evolution under the Hamiltonian \(H = \xi_1 \xi_3\) for time \(\pi/2\)
produces the \(T\) gate magic state, \(\ketbra{T}{T} = (I + (X + Y)/\sqrt{2})/2\). These are characterized by consisting of two \(\mathcal O(\hbar^1)\) terms. Supplementing the stabilizer states and including Pauli measurements, this produces a universal set of states for measurement-based quantum computation~\cite{Bravyi05}.

\begin{figure}[h]
\includegraphics[]{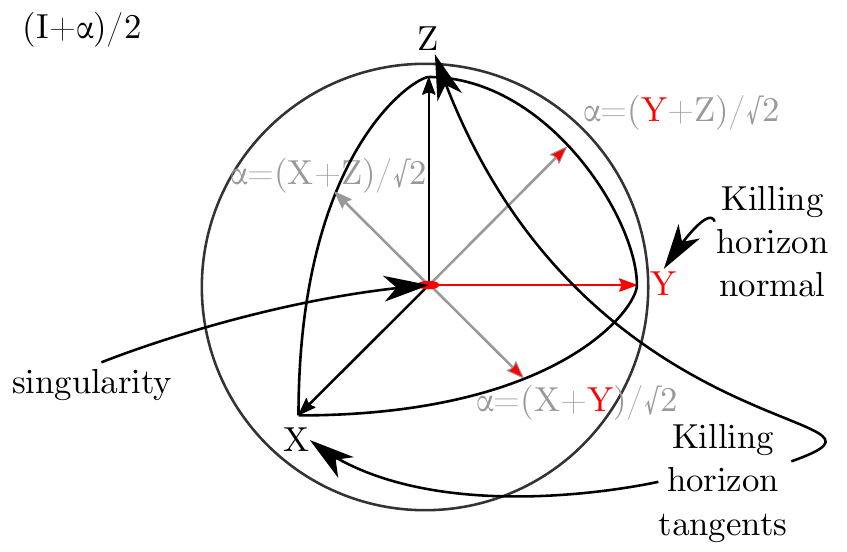}
\caption{Isometries of AdS space without magic (black), and with magic states (grey). In either case, there exists at least one Killing horizon (red) with a black hole corresponding to the Bloch sphere origin.}
\label{fig:bloch_sphere}
\end{figure}

This set, along with equivalent axial magic states, is sketched in Figure~\ref{fig:bloch_sphere}.

Notice that the inclusion of a quantum magic state such as \(\ketbra{T}{T}\) through a rotation of one of the generators by Eq.~\ref{eq:threegenHam} does not supplement the stabilizer states with the remaining generators, as they are also rigidly rotated so that they maintain their mutual relations (see Eqs.~\ref{eq:tgatemagicstate1}-\ref{eq:tgatemagicstate3}).
Such a transformation does not change the number of isometries; the group of the spacetime remains the same, and so the spacetime remains AdS$_3$ after the transformation.

A transformation that produces isometries corresponding to \emph{both} stabilizer states and magic states requires a Hamiltonian (Eq.~\ref{eq:threegenHam}) over three generators that are not restricted to \(\xi_1\), \(\xi_2\) and \(\xi_3\). This cannot be effected by a transformation to any order in \(\hbar\) without higher order than \(\mathcal(G^0)\), as such quantum transformations are still restricted to be conformal (i.e. respect the Virasoro algebra commutation relations). However, as we shall see, this is possible with a \(\mathcal O(G)\) transformation.

Consider the transformation
\begin{align}
  \label{eq:LorentzGboost}
  \hat \xi_5 \rightarrow& \cosh(\delta \tau) \hat \xi_5 + \sinh(\delta \tau) \hat \xi_7\\
  \implies \hat \xi_5 =& \hat \xi_5 + \delta \tau \hbar \hat \xi_7 + \mathcal O(\delta \tau^2),\\
  \hat \xi_7 \rightarrow& \cosh(\delta \tau) \hat \xi_7 + \sinh(\delta \tau) \hat \xi_5\\
  \implies \hat \xi_7 =& \hat \xi_7 + \delta \tau \hbar \hat \xi_5 + \mathcal O(\delta \tau^2).
\end{align}
This is not a \(\mathcal O(G^0)\) symplectic transformation and so cannot be associated with a Hamiltonian in \(\text{Cl}_{3,0}(\mathbb C)\). Since it involves the octonion algebra addition, it is a \(\mathcal O(G^1)\) transformation and can be considered to be the result of a linear combination of two Hamiltonians evolved for \(\tau = \pi\)---\(H_1 = i \xi_0\) and \(H_2 = \frac{i}{\sqrt{2}} (\xi_5 \xi_6 + \xi_6 \xi_4 + \xi_4 \xi_5 + \xi_7 \xi_4)\)---the identity that leaves \(\hat \xi_5\) and \(\hat \xi_7\) unchanged and a rotation that exchanges them. Note that these two evolutions together form a Lorentz boost in this non-quaternion sub-algebra, which is not a symplectic (area-preserving) transformation for usual two-forms relating conjugate degrees of freedom.

Notice that this differs from the \(\mathcal O(\hbar^1)\) transformation that produces the T gate. The T gate can be viewed as a symplectic transformation to a linear combination of \(\mathcal O(\hbar^1)\) terms in \(\text{Cl}_{3,0}(\mathbb C)\). On the other hand, Eq.~\ref{eq:LorentzGboost} is an \(\mathcal O(G)\) non-symplectic transformation to a linear combination of \(\mathcal O(G^1)\) terms in the non-isomorphic larger algebra \(\text{Cl}_{4,0}(\mathbb C)\). We will see that this its \(\hbar\)-dependence will be \(\mathcal O(\hbar^2)\).

Restricting to the subalgebra \(\text{Cl}^{[0]}_{4,0} \substack{\cong\\\text{n.c.}} \text{Cl}_{3,0}(\mathbb C)\), this \(\mathcal O(G)\) transformation has the effect of splitting the two-fold correspondence between \(\hat \xi_\mu \hat \xi_\nu\), \(\hat \xi_\rho \hat \xi_\tau\) and \(\hat \xi_i\):
\begin{align}
  \label{eq:gravitymagicstate1}
  \hat \xi_4 \hat \xi_5 &\rightarrow \hat \xi_4 \hat \xi_5 + \delta \tau \hat \xi_4 \hat \xi_7 + \mathcal O(\delta \tau^2),\\
  \hat \xi_7 \hat \xi_6 &\rightarrow \hat \xi_7 \hat \xi_6 + \delta \tau \hat \xi_5 \hat \xi_6 + \mathcal O(\delta \tau^2),\\
  \hat \xi_4 \hat \xi_6 &\rightarrow \hat \xi_4 \hat \xi_6,\\
  \hat \xi_5 \hat \xi_7 &\rightarrow \hat \xi_5 \hat \xi_7 + \delta \tau G \hbar^2 \hat \xi_0 + \mathcal O(\delta \tau^2),\\%
  \hat \xi_4 \hat \xi_7 &\rightarrow \hat \xi_4 \hat \xi_7 + \delta \tau \hat \xi_4 \hat \xi_5 + \mathcal O(\delta \tau^2),\\
  \label{eq:gravitymagicstate6}
  \hat \xi_6 \hat \xi_5 &\rightarrow \hat \xi_6 \hat \xi_5 + \delta \tau \hat \xi_6 \hat \xi_7 + \mathcal O(\delta \tau^2),
\end{align}

The former \(\hat \xi_5 \hat \xi_7\) isometry obtains a grade-zero component under this transformation and thus ceases to be an isometry.

Notice that this transformation between \(\xi_5\) and \(\xi_7\) (corresponding to \(x\) and \(x'\)) produces an algebra non-canonically isomorphic to \(\mathfrak{sl}(2, \mathbb R)\). \(\mathfrak{sl}(2, \mathbb R)\) is generated by \(\hat \xi_0\), \(-\hat \xi_1 = K^{01} \oplus K^{32}\), \(-\hat \xi_2 = K^{21} \oplus K^{03}\), and \(-i \xi_3 = J^{02} \oplus J^{13}\), where \(\hat \xi_\mu \hat \xi_\nu = -i G \hat \xi_j /\sqrt{2 \hbar}\) . This means that only two of the isometries are equal to \(\hat \xi_\mu \hat \xi_\nu\) with a real-valued constant: \(\hat \xi_5 \hat \xi_7 = J^{02}\) and \(\hat \xi_4 \hat \xi_7 = J^{13}\). This means that the real-valued \(\mathcal O(G)\) term above for \(\hat \xi_5 \hat \xi_7\) remains real-valued. A similar transformation between \(\xi_4\) and \(\xi_6\) (corresponding to \(t\) and \(t'\)) would similarly preserve \(\mathfrak{sl}(2, \mathbb R)\). On the other hand, non-hyperbolic rotations between the other non-quaternion generators (which would now correspond to simply rotations instead of boosts) would introduce an imaginary-valued \(\mathcal O(G)\) term because they are equal to \(i K^{\mu}\) instead of \(K^{\mu}\). Such transformations would no longer preserve \(\mathfrak{sl}(2, \mathbb R)\).

Eqs.~\ref{eq:gravitymagicstate1}-\ref{eq:gravitymagicstate6} can be more succinctly written just in terms of the quaternions:
\begin{align}
  \label{eq:dS3isometries1}
  \hat \xi'_1 &\equiv \hat \xi_1 + \delta \tau \hat \xi_2 + \mathcal O(\delta \tau^2),\\ 
  \hat \xi''_1 &\equiv \hat \xi_1 - \delta \tau \hat \xi_2 + \mathcal O(\delta \tau^2),\\ 
  \hat \xi'_3 &\equiv \hat \xi_3 + \mathcal O(\delta \tau^2),\\ 
  \hat \xi''_3 &\equiv \hat \xi_3 + \delta \tau G \hbar^2 \hat \xi_0 + \mathcal O(\delta \tau^2),\\ 
  \hat \xi'_2 &\equiv \hat \xi_2 + \delta \tau \hat \xi_1 + \mathcal O(\delta \tau^2),\\ 
  \label{eq:dS3isometries6}
  \hat \xi''_2 &\equiv \hat \xi_2 - \delta \tau \hat \xi_1 + \mathcal O(\delta \tau^2).
\end{align}

Notice that \(\hat \xi_i \ne \hat \xi'_i \ne \hat \xi''_i\). This means that this \(\mathcal O(G)\) transformation splits every isometry into two isometries except for \(\xi_3\). The latter is split into itself and a linear combination that includes a term proportional to the identity, which is no longer an isometry. Therefore, the five new isometries are no longer Clifford-isomorphic to the Paulis. Recall that quantum magic state transformations preserved the three isometries Clifford-isomorphic to the Paulis due to the restriction that the theory is conformal when expressed in terms of three generators. %

This means that the \(\hat \xi'_i\) and \(\hat \xi''_i\) no longer define a conformally invariant field theory. This can also be seen by finding that there is no way to express a Lagrangian that is independent of all the new isometries, since they all sequentially overlap with each other. %

Taking a closer look at Eqs.~\ref{eq:dS3isometries1}-\ref{eq:dS3isometries6}, the tangent to the Killing horizon \(\xi_1\) is now split into \(\xi'_1\) and \(\xi''_1\) which have opposite components along the Killing horizon normal \(\xi_3\). These correspond to trapping and escaping isometries as they lie on either side of the Killing horizon. \(\xi_2\) has been accordingly split as well into \(\xi'_2\) and \(\xi''_2\). %

As already discussed, the normal to the Killing horizon, \(\xi_3\), has been split into itself, \(\xi'_3\), and \(\xi''_3\). This is a non-normalized linear combination for any non-zero \(\hbar\) since \(\xi''_3\) has a component on \(\xi_0\) with coefficient proportional to \(G \hbar^2\). This is reflective of the non-conformality of this field theory after the \(\mathcal O(G)\) transformation and can be interpreted as indicating that this is a higher order \(G \hbar^2\) non-unitary effect. %

As we shall see, the \(\xi'_3\) term corresponds to a product of the Killing horizon with an even linear combination of escaping and trapping isometries. We will find that it is responsible for Hawking radiation from pair production before the Page curve turning point. Similarly, we will see that the \(\xi''_3\) term corresponds to a product of the Killing horizon with a linear combination of these isometries along with a term proportional to the identity. We will see that this last term allows \(\xi''_3\) to be interpreted as possessing a ``wormhole'' term, which is an \(\mathcal O(\hbar^2 G^1)\) ``uniformizing'' term, wherein the two isometries interact with each other through the identity or origin term.

These associations can be made explicit by noticing that the usual Pauli product rule \(\hat \xi_3 = i \sqrt{\hbar/2} \hat \xi_1 \hat \xi_2\) can be rewritten in terms of these new isometries,
\begin{equation}
  \label{eq:prePageturningptisometry}
  \hat \xi'_3 = i \sqrt{\hbar/2} (\hat \xi'_1 + \hat \xi''_1) \hat \xi_2 + \mathcal O(\delta \tau^2),
\end{equation}
\begin{equation}
  \label{eq:postPageturningptisometry}
  \hat \xi''_3 = i \sqrt{\hbar/2} (\hat \xi'_1 + \hat \xi''_1) \hat \xi_2 + \delta \tau G \hbar^2 \hat \xi_0 + \mathcal O(\delta \tau^2).
\end{equation}
Eq.~\ref{eq:prePageturningptisometry} is similar to the solution of the Airy function when the area subtended by two stationary phase points is large compared to \(\hbar\) given by Eq.~\ref{eq:preuniformizationAiry} (see Appendix~\ref{sec:Airy}). Eq.~\ref{eq:postPageturningptisometry} is similar to the uniformized solution when the area subtended by the two stationary phase points become on the order of \(\hbar G\) given by Eq.~\ref{eq:postuniformizationAiry}.

This means that both \(\xi'_3\) and \(\xi'_3\) isometries are a product of the escaping and trapping isometries. However, as explained in Appendix~\ref{sec:Airy}, instead of ``interacting'' with the origin, for \(\xi''_3\) these two isometries interact with each other.

Moreover, the \(\hat \xi'_i\) and \(\xi''_i\) isometries can be related to linear combinations of particle/anti-particle isometries by noting that
\begin{align}
  \hat \xi_i =& (\hat u_i + \hat u_i^*)/2,\\
  \hat \xi_\mu =& -(\hat u_i - \hat u_i^*)/2,
\end{align}
where \(\hat u_i\) and \(\hat u^*_i\) are the Fermi annihilation and creation operators introduced in Appendix~\ref{sec:hbar0qutritCliffordtheory} (Eqs.~\ref{eq:splitoctonions1}-\ref{eq:splitoctonions2}). Therefore, you can associate these isometries with probabilistic quantum operations on both particles and anti-particles. This means that some of the particles can be inside the black hole with probability \(p\) and outside the black hole with probability \(1-p\) (and vice-versa for the anti-particles) depending on the linear combination (i.e. magnitude of \(\delta \tau\) or of the Lorentz boost generally).

A general real octonion \(\hat G\) can be written as a direct product~\cite{Gunaydin74}
\begin{equation}
  \hat G = \hat G_L + \hat G_T,
\end{equation}
where
\begin{align}
  \hat G_L \equiv& G_0 \hat u_0 + G_0^* \hat u^*_0,\\
  \hat G_T \equiv& \sum_{i=1}^3 (G_i \hat u_i + G^*_i \hat u^*_i),
\end{align}
such that
\begin{equation}
  \hat u_0 \hat G = \sum_{i=0}^3 G_i \hat u_i, \qquad \hat u^*_0 \hat G = \sum_{i=0}^3 G^*_i \hat u^*_i.
\end{equation}

This implies that creating a particle from vacuum in the \(j\)th state is the same as creating an antiparticle from vacuum in the \(j\)th state and reveals that such a Fermi particle is entangled with its anti-particle. This entanglement property is what leads to the information paradox given that these isometries lie on opposite sides of the Killing horizon~\cite{Hawking76,Harlow16}.

\begin{figure}[h]
\includegraphics[]{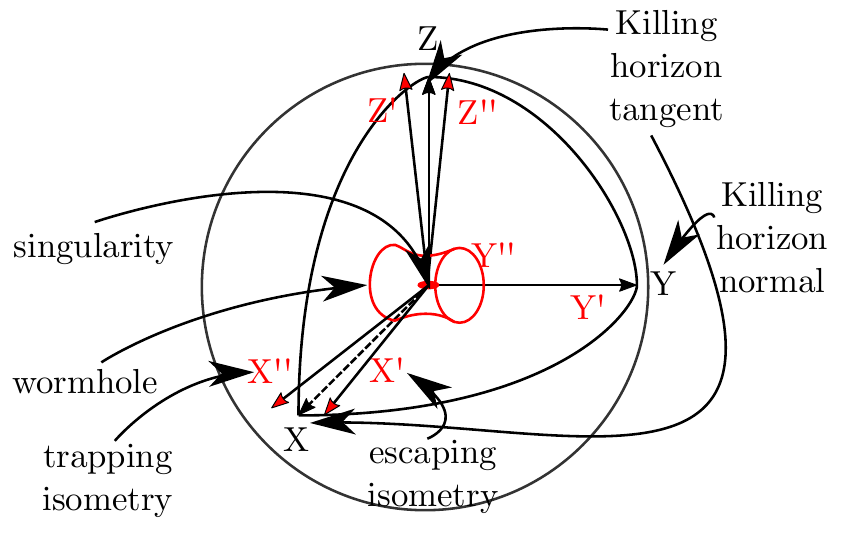}
\caption{Isometries of perturbed AdS$_3$ space after \(\mathcal O(G)\) transformation given by Eq.~\ref{eq:LorentzGboost}.}
\label{fig:bloch_sphere_dS3}
\end{figure}

The isometries are sketched in Fig.~\ref{fig:bloch_sphere_dS3}.

Notice that performing the \(\mathcal O(G)\) transformation given by Eq.~\ref{eq:LorentzGboost} in any other rotation direction will still produce the same qualitative isometries above. Only the coefficients will be changed.

It is important to note that any \(\mathcal O(G)\) transformation that splits an isometry from being tangential to the Killing horizon to two isometries that lie on opposite sides will necessarily introduce a ``wormhole''-containing \(\xi''_3\) term to the normal isometry. This is because the normal isometry must contain one \(\xi_\mu \xi_\nu\) that is unaffected by the transformation and another \(\xi_\rho \xi_\tau\) that is. Since the two must lie at equal angles to the original isometry and \(\xi_\mu \xi_\nu\) remains the same, it follows that \(\xi_\mu \xi_\nu\) must remain colinear and can only develop a new component along \(\xi_0\).

Therefore, this \(\mathcal O(G)\) transformation's production of escaping and trapping isometries (which produces a thermodynamically stable evaporative black hole and perturbed AdS$_3$ spacetime) necessarily produces a ``wormhole'' term. The two phenomena are inseparable in this formalism.

\subsection{The RT Formula for AdS$_3$}
\label{sec:RT}

\subsubsection{General Formula}

Given the density matrix of a state \(\hat \rho\), the entanglement entropy of a subsystem \(A\) is defined as
\begin{align}
  \label{eq:entanglemententropy}
  S(\rho_A) = -\Tr(\rho_A \log \rho_A) = {-}\int (\rho_A \star \log \rho_A)(\bs \xi) \xi_3 \xi_2 \xi_1 \text d^3 \bs \xi,
\end{align}
where \(\rho_A \equiv \Tr_B \rho\).

As it is written in the \(\text{Cl}_{3,0}(\mathbb C)\) algebra in Eq.~\ref{eq:entanglemententropy}, this is an integral over a product of two operators. %
We would like to express this product as a dual AdS OPE in the \(\text{Cl}^{[0]}_{4,0}(\mathbb C)\) subalgebra using Eq.~\ref{eq:AdSWeylSymbolGroenewoldRule_combined}. Since we are taking a Grassmann four-generator integral of \((\rho_A \star \log \rho_A)(\bs \xi)\) with \(\xi_7 \xi_6 \xi_5 \xi_4\), it follows that the integral is only non zero for the \(\mathcal O(\hbar^0 G^0)\) identity squared term and the second \(\mathcal O(G^1 \hbar^1)\) Pauli squared term of the OPE: %
\begin{align}
  \label{eq:Weylentanglemententropydual}%
  S(\rho_A) =& -\int \bar \rho_A(\bs \xi) \overline{\log \rho}_A(\bs \xi) \bar \xi_4 \bar \xi_5 \bar \xi_6 \bar \xi_7 \text d^4 \bs \xi\\
  & + \frac{G}{2} \int \reallywidetilde{{\bar \rho}_A(\bar{\bs \xi}) \widetilde{ \overline{\log  \rho}}_A(\bar{\bs \xi})} \bar \xi_4 \bar \xi_5 \bar \xi_6 \bar \xi_7 \text d^4 \bs \xi. \nonumber
\end{align}

If \(\rho_A(\bs \xi)\) is \(\mathcal O(\hbar^0 G^0)\), the first term in Eq.~\ref{eq:Weylentanglemententropydual} is \(\mathcal O(\hbar^0 G^0)\) and the second is \(\mathcal O(\hbar^1 G^1)\). We shall see that this is the case for quantum theory within AdS$_3$ spacetime.

The Ryu-Takayanagi formula~\cite{Ryu06,Ryu06_2} for the entropy of \(\rho_A\) is
\begin{equation}
  \label{eq:RyuTakayanagi}
  S(\rho_A) = \Tr(\rho \mathcal L_A) + S_\text{bulk}(\rho_{\epsilon_A}) + \mathcal O(\hbar^0),
\end{equation}
where \(\rho_A\) is a boundary CFT state on boundary \(A\) of a larger \(\rho\) CFT state b, and \(\mathcal L_A\) is a local operator in the bulk integrated over \(\gamma_A\), and \(S_\text{bulk}(\rho_{\epsilon_A})\) is the bulk von Neumann entropy in \(\epsilon_A\).
At leading order in \(G\), \(\mathcal L_A = \text{Area}(\gamma_A)/4G\). Accounting for \(\hbar^{-1} G^{-1}\) implicit normalization, the first term in Eq.~\ref{eq:RyuTakayanagi} is \(\mathcal O(G^0 \hbar^{0})\) and the second is \(\mathcal O(G^1 \hbar^1)\)~\cite{Engelhardt15}.

With these order expansions, we can relate the corresponding terms in Eq.~\ref{eq:Weylentanglemententropydual} to the terms in Eq.~\ref{eq:RyuTakayanagi}, assuming \(\rho_A(\bs \xi)\) is \(\mathcal O(\hbar^0)\):
\begin{align}
  \label{eq:Weylentanglemententropydualterm1}
  \Tr(\rho \mathcal L_A) =& -\int \bar \rho_A(\bar{\bs \xi}) \overline{\log \rho_A}(\bar{\bs \xi}) \bar \xi_4 \bar \xi_5 \bar \xi_6 \bar \xi_7 \text d^3 \bar{\bs \xi}
\end{align}
\begin{align}
  \label{eq:Weylentanglemententropydualterm2}
S_\text{bulk}(\rho_{\epsilon_A}) =& \frac{G}{2} \int \reallywidetilde{\bar {\rho}_A(\bar{\bs \xi}) \widetilde{\overline{\log  \rho_A}}(\bar{\bs \xi})} \bar \xi_4 \bar \xi_5 \bar \xi_6 \bar \xi_7 \text d^3 \bar{\bs \xi}
\end{align}

Entanglement between subsystem A and traced out degrees of freedom will produce a mixed state after the trace. Any mixed state on a qubit can be written as a convex combination of pure states corresponding to stabilizer states and (quantum) magic states. We will now proceed to examine these two cases of pure states obtained by tracing out to subsystem A, with the expectation that they correspond to a single term in such a convex decomposition. We will also examine the case of a ``gravitational'' magic pure state corresponding to the \(\mathcal O(G^1)\) transformation in Section~\ref{sec:perturbedAdS3}.

\subsubsection{RT Applied to Stabilizer States}

Consider a stabilizer state \(\hat \rho_A = (\hat \xi_0 + i \hat \xi_i \hat \xi_j)/2\) (see Appendix~\ref{sec:Cliffordstabilizersubtheory} for their derivation). This has AdS dual element
\begin{align}
{\bar \rho}_A(\bar {\bs \xi}) = (\bar \xi_0 + \bar \xi_\mu \bar \xi_\nu)/2 &\equiv \bar \xi_0/2 + i \bar {\mathcal L}_A(\bar{\bs \xi})/2\\ &= \bar \rho_A(0) e^{i \bar {\mathcal L}_A(\bar{\bs \xi})} .
\end{align}

The \(\mathcal O(G^0)\) term of the RT formula is
\begin{align}
  & -\int \bar{ \rho}_A(\bar{\bs \xi}) \log \bar{ \rho}_A(\bar{\bs \xi}) \bar \xi_4 \bar \xi_5 \bar \xi_6 \bar \xi_7 \text d^4 \bar{\bs \xi}\\
  =& -\int \bar{\rho}_A(0) e^{-i {\mathcal L}_A(\bar{\bs \xi})} (\log\bar{\rho}_A(0) + i {\mathcal L}_A(\bar{\bs \xi})) \bar \xi_4 \bar \xi_5 \bar \xi_6 \bar \xi_7 \text d^4 \bar{\bs \xi}\\
  =& -\log(\bar{\rho}_A(0)) \bar{\rho}_A(0) \int \bar \xi_4 \bar \xi_5 \bar \xi_6 \bar \xi_7 \text d^4 \bar{\bs \xi}\\
  =& -\log(\bar{\rho}_A(0)) \bar{\rho}_A(0).
\end{align}

The \(\mathcal O(G^1)\) ``bulk'' correction of the RT formula is
\begin{align}
  & \frac{G}{2} \int \reallywidetilde{\bar{\rho}_A(\bar{\bs \xi}) \widetilde{\overline{\log \rho}}_A(\bar{\bs \xi})} \bar \xi_4 \bar \xi_5 \bar \xi_6 \bar \xi_7 \text d^4 \bar{\bs \xi}\\
  =& i \frac{G}{2} \bar{\rho}_A(0) \int \reallywidetilde{e^{-i \bar {\mathcal L}_A(\bar{\bs \xi})} (\tilde{\bar L}_A(\bar{\bs \xi}) - \widetilde{\overline{\log \rho}}_A(0))} \nonumber\\
  & \quad \qquad \qquad \times \bar \xi_4 \bar \xi_5 \bar \xi_6 \bar \xi_7 \text d^4 \bar{\bs \xi} \\
  =& i \frac{G}{2} \bar{\rho}_A(0) \int (\bar {\mathcal L}_A(\bar{\bs \xi}) - i \reallywidetilde{\bar {\mathcal L}_A(\bar{\bs \xi}) \tilde{\bar L}}_A(\bar{\bs \xi}) - \overline{\log \rho_A(0)} ) \nonumber\\
  & \quad \qquad \qquad \times \bar \xi_4 \bar \xi_5 \bar \xi_6 \bar \xi_7 \text d^4 \bar{\bs \xi} \\
  =& \frac{G}{2} \bar{\rho}_A(0) |\bar {\mathcal L}_A((\bar \xi_\mu=1,\bar \xi_\nu=1))|^2 \nonumber\\
\end{align}

Together these produce the entanglement entropy
\begin{align}
  S(\rho_A) =& -\bar{\rho}_A(0) \left(\log \bar{\rho}_A(0) - \frac{G}{2}\sum_{\mu,\nu \in \mathcal S} |\bar {\mathcal L}_A((\bar \xi_\mu{=}1,\bar \xi_\nu{=}1))|^2\right) \nonumber\\
  & + \mathcal O(G^2).
\end{align}
As seen from the previous section with Eqs.~\ref{eq:Weylentanglemententropydualterm1}-\ref{eq:Weylentanglemententropydualterm2}, %
the first term is \(\mathcal O(G^{0})\) and the second is \(\mathcal O(G^1)\), which become \(\mathcal O(G^{-1})\) and the second is \(\mathcal O(G^0)\) in the usual renormalized convention used in CFT and string theory. %

We take the holographic area to be proportional to \(\hbar G\) as discussed in Section~\ref{sec:holographicprinciple} and adapt the argument motivated in Appendix~\ref{sec:Airy} for the Airy function to this situation: smaller holographic areas maker higher-order terms dominate semiclassical expansions. This suggests that we interpret the \(\mathcal O(G^0) \equiv \mathcal O(\hbar^{0} G^0)\) term (where we have included the redundant \(\hbar\) dependence) as corresponding to the entropy of the young black hole in AdS when it is large, and the \(\mathcal O(G^1) \equiv \mathcal O(\hbar^1 G^1)\) term as corresponding to the entropy of the intermediate-age black hole after it has evaporated off some of its initial area.

Notice that, unlike for the Airy function, there is no higher \(\mathcal O (\hbar^2)\) or \(\mathcal O(G^2)\) term here to perform a uniformization. This means that the area of the black hole does not change after intermediate evolution. For AdS$_3$, thermodynamically stable large (BTZ) black holes do not evaporate away, so this makes sense. (They still radiate and initially shrink, but the open boundary conditions also bathe them in their own radiation leading them to a thermodynamically eternally radiating state~\cite{Almheiri19,Penington20}.) Moreover, this result captures the correct behavior of small black holes in AdS, which evaporate away but whose entanglement entropy never reaches the later decreasing part of the Page curve.

This suggests that though stabilizer state contextuality produces black holes, thermodynamically stable \emph{radiating} black holes require something more to evaporate away. %
We will see that this is still true with \(\mathcal O(\hbar^1)\) quantum magic states since they do not affect the AdS$_3$ symmetry of spacetime. 

\subsubsection{RT Applied to Quantum Magic States}
\label{sec:RTquantummagicstates}

Consider the quantum magic state \(\rho_A = (\bar \xi_0 + (\bar \xi_i \bar \xi_j + \bar \xi_j \bar \xi_k)/\sqrt{2})/2\) (see Appendix~\ref{sec:quantummagicstates} for their derivation). This has the dual AdS element
\begin{align}
  \bar {\rho}_A =& (\bar \xi_0 + (\bar \xi_\mu \bar \xi_\nu + \bar \xi_\mu \bar \xi_\rho )/\sqrt{2})/2.\\
  \equiv& \bar \xi_0/2 + i \bar {\mathcal L}_A(\bar{\bs \xi})/2.
\end{align}

Given the similar form, the entanglement entropy of this state can be expressed similarly to the one found for the stabilizer state in the previous section,
\begin{align}
  S(\rho_A) =& -\bar{\rho}_A(0) \left(\log\bar{\rho}_A(0) - \frac{G}{2} \sum_{\mu,\nu \in \mathcal S} |{\mathcal L}_A((\bar \xi_\mu{=}1,\bar \xi_\nu{=}1))|^2\right) \nonumber\\
  & + \mathcal O(G^2).
\end{align}

The value of this entanglement entropy is equal to the one found for the stabilizer state. However, it contains two \(\mathcal O(G^1)\) terms compared to the one for the stabilizer state. %
(In general, as long as \(BQP \ne BPP\), the number of these terms will increase exponentially with the number of qubits.)

This reveals that while the dual of this state is a black hole with more terms required to express its isometries, it is still eternal (i.e.~its area does not decrease after intermediate times) since it does not possess higher order \(\mathcal O(\hbar^2)\) or \(\mathcal O(G^2)\) terms.

\subsubsection{RT Applied to Gravity Magic States in Perturbed AdS$_3$}

Now we consider the state corresponding to the fourth isometry \(\xi''_2\) in Eq.~\ref{eq:dS3isometries1}-\ref{eq:dS3isometries6} after the \(\mathcal O(G)\) transformation defined in Section~\ref{sec:perturbedAdS3} and write it in the even Clifford algebra \(\text{Cl}_{3,0}(\mathbb C)\) representation:
\begin{align}
  \hat \xi''_1 \hat \xi''_3 =& \hat \xi_1 \hat \xi_3 + G \delta \tau \hbar^2 \hat \xi_0 + \mathcal O(\delta \tau^2).
\end{align}
The corresponding ``gravitational'' magic state is
\begin{align}
  \bar \rho_A(\bar{\bs \xi}) =& (\bar \xi_0 + \bar \xi_4 \bar \xi_6 + G \delta \tau \hbar^2 \bar \xi_0)/2 + \mathcal O(\delta \tau^2),
\end{align}

It follows that
\begin{equation}
\bar \rho_A(0) = \left(1 + G \delta \tau \hbar^2 \right)/2 + \mathcal O(\delta \tau^2).
\end{equation}
but otherwise the state is the same as a stabilizer state or quantum magic state: \(\bar \rho(\bar{\bs \xi}) = \bar \rho_A(0) e^{i {\mathcal L}_A(\bar{\bs \xi})}\).

Given that the \(\mathcal O(G)\) transformation is performed in the larger algebra \(\text{Cl}_{4,0}(\mathbb C)\) (though we are restricting ourselves to its \(\text{Cl}_{4,0}^{[0]}(\mathbb C)\) subalgebra, as discussed in Section~\ref{sec:perturbedAdS3}), the terms in the OPE depend on both the \(\hbar\) and \(G\) parameters (i.e.~the OPE is no longer conformal and they are no longer equivalent expansion parameters). Hence, the terms that were formerly found to be \(\mathcal O(\hbar^j)\) or \(\mathcal O(G^j)\) now become \(\mathcal O(\hbar^j G^j)\).

Given the dual AdS \(\bar \rho(\bar{\bs \xi})\) state's similarity to a stabilizer state or magic state, its entanglement entropy will contain a term that scales as \(\mathcal O(\hbar^{0} G^0)\) and \(\mathcal O(\hbar^1 G^1)\) like before for the stabilizer and magic states. However, there will also now be a term that scales as \(\mathcal O(\hbar^{2}G^1)\) due to the additional \(\hbar^{2} G\)-dependent term in \(\bar \rho_A(0)\). %

This changes the former leading order term for the entanglement entropy \(S(\rho_A)\) to be
\begin{align}
  &-\int \bar \rho_A(\bar{\bs \xi}) \log \bar \rho_A(\bar{\bs \xi}) \bar \xi_4 \bar \xi_5 \bar \xi_6 \bar \xi_7 \text d^4 \bar{\bs \xi} \nonumber\\
  =& -\log_2 (\bar \rho_A(0))\bar \rho_A(0)\\
  =& -\log_2 ((1 + G\delta \tau \hbar^2)/2)(1 + G\delta \tau \hbar^2)/2\\
  =& 1/2 + G (1 - \log^{-1}(2)) \delta \tau \hbar^2 /2 + \mathcal O(\delta \tau^2),
\end{align}
where \(\log^{-1}(2)\) is an (inverse) natural logarithm (to base \(e\)). 
This means that the leading order term now contains \emph{both} a \(\mathcal O(\hbar^{0} G^0)\) and \(\mathcal O(\hbar^{2} G^1)\) term for any finite rotation \(\delta \tau\) that produces a gravitational magic state. Note that \(1-\log^{-1}(2) < 0\), and so the \(\mathcal O(\hbar^{2} G^1)\) term has opposite sign compared to \(\mathcal O(\hbar^{0} G^0)\) term.

The remaining \(\mathcal O(\hbar^0 G^0)\) sub-leading term is
\begin{equation}
\frac{G}{2} \sum_{\mu,\nu \in \mathcal S}|{\mathcal L}_A((\bar \xi_\mu=1, \bar \xi_\nu=1))|^2 = \frac{G}{2}.
\end{equation}

Adding everything together produces
\begin{align}
  S(\bar \rho_A) =& 1/2 + G (1 - \log^{-1}(2)) \delta \tau\hbar^2 /2 \sqrt{2}\\
                  & + \frac{G}{2} \sum_{\mu,\nu \in \mathcal S}|{\mathcal L}_A((\bar \xi_\mu=1, \bar \xi_\nu=1))|^2 \nonumber\\
                  & + \mathcal O(\delta \tau^2) \nonumber\\
  \equiv& \mathcal O(\hbar^{0}G^0) + \mathcal O(\hbar^{2}G^1) + \mathcal O(\hbar^{1}G^1) \\
                  & + \mathcal O(\delta \tau^2). \nonumber
\end{align}

As the black hole area---proportional to \(\hbar G\) in three-dimensional spacetime---reduces due to evaporation, its entropy \(S(\bar \rho_A)\) is dominated by the first \(\mathcal O(\hbar^{0} G^0)\) term when the black hole is young and the \(\mathcal O(\hbar^{0} G^0) + \mathcal O(\hbar^{2}G^1) + \mathcal O(\hbar^{1} G^1)\) terms when it is old. %
Since the second term has opposite sign to the other two, if the third term preceeds the second term in dominance, this produces values for the entanglement entropy that follow the Page curve, sketched in Figure~\ref{fig:PagecurvedS3}. We discuss how this is similar to the quantum extremal surface prescription in Appendix~\ref{sec:quantumextremalsurfaces}.

\begin{figure}[ht]
\includegraphics[]{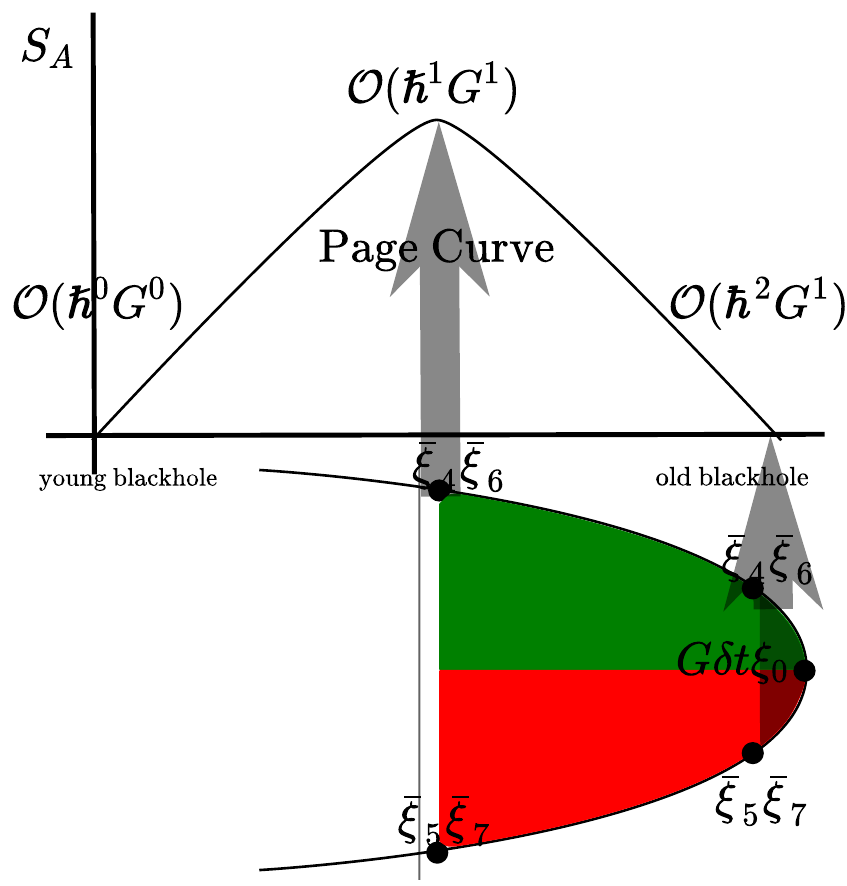}
\caption{Page curve for (radiating) black holes in perturbed AdS$_3$ if the \(\mathcal O(\hbar^2 G^1)\) term is treated as the sub-sub-leading term. See Appendix~\ref{sec:Airy} for an explanation of what the bottom part refers to (i.e.~semiclassical uniformization). A reminder that (as found in Section~\ref{sec:hbarCliffordtheory}) in quantum field theory bosonic fields are often implicitly normalized by \(\hbar\) and in string theory fields are often implicitly normalized by \(G\), and so the three terms above become \(\mathcal O(\hbar^{-1} G^{-1})\), \(\mathcal O(\hbar^0 G^0)\), and \(\mathcal O(\hbar^{1}G^0)\), from left-to-right.}
\label{fig:PagecurvedS3}
\end{figure}

\subsection{Irreversibility of Evolution}
\label{sec:irreversibility}

Consider the reversal of dynamics w.r.t.~\(\lambda\),
  \begin{equation}
    - \der{(i\xi_k)}{\lambda} = - \left(-i\frac{\hbar}{2} H \frac{\cev \partial}{\partial (-i\xi_i)}\right) = -i H \frac{\cev \partial}{\partial (i \xi_i)},
  \end{equation}
  where reversal is performed by flipping the sign of the momentum analog \(-i \xi_j\) of the real Grassmann degree of freedom (as we found in Section~\ref{sec:symplectic}) and complex conjugating. This is clearly distinguishable from the forward evolution, \(\der{(i\xi_k)}{\lambda} = H \frac{\cev \partial}{\partial (i \xi_i)}\), but \emph{a priori} there is no reason that dynamics should run with increasing \(\lambda\) instead of decreasing \(\lambda\), or vice-versa; dynamics in the \(\lambda\) degree of freedom is reversible.

We established the central charge operator of the Majorana mapping in Section~\ref{sec:centralcharge}. We have also related this mapping to a representation of a double cover of \(\text{SO}^+(2,2)/\text{SO}^+(1,2)\), which is the symmetry group of two-dimensional conformal transformations. Thus, we might expect that the property of an irreversible renormalization group of two-dimensional conformal field theories holds here as well.

Specifically, for coupling constant \(\hbar\), per the $c$-theorem~\cite{Zamolodchikov86}, we search for a function \(c(\alpha) \ge 0\) with coupling constant \(\alpha\) such that
\begin{equation}
  \der{c(\alpha)}{\lambda} \le 0,
\end{equation}
where critical fixed points (stationary points) saturate the above equality and are found at values of the central charge \(c\) that correspond to the (scaled) coefficient of the second term in the Virasoro algebra relations in Eq.~\ref{eq:Virasarorelations}~\cite{Zamolodchikov86}.

Since we know that we have two critical values for the central charge---\(c=0\) when we get the Clifford algebra \(\text{Cl}_{3,0}(\mathbb C)\) without a well-defined Fourier association between its even and odd terms, and \(c=1/2\) when we do have a well-defined \(\hbar=2\) Fourier association whose associated multiplication rules we can use to extend the quaternion algebra to the octonion algebra \(\mathbb O\)---we can identify our coupling constant \(\alpha \equiv \hbar\) and find
\begin{align}
  \der{c(\hbar)}{\lambda} &= \hbar(\hbar-1/2)\der{f(\lambda)}{\lambda} + \mathcal O(\hbar^2),\, c(0) = 0,\, c(2) = 1/2, \nonumber\\
  \label{eq:Cformula}
  \implies& c(\hbar) = \hbar(\hbar-1/2)f(\lambda) + \hbar/4 + \mathcal O(\hbar^2),
\end{align}
where \(f(\lambda)\) is some function of \(\lambda\) independent of \(\hbar\). Expressing this equation to \(\mathcal O(\hbar^2)\) is sufficient for our purposes here. %

(As we noted in Section~\ref{sec:centralcharge}, though we frequently have not set \(\hbar =2\) in prior Sections and simply leave it as \(\hbar > 0\), this is merely to allow for easier explanation when lower \(\mathcal O(\hbar)\) terms (such as in OPE series expansions) dominate because the symplectic area \(\Delta_3\) (see Eq.~\ref{eq:symplecticarea}) is far larger than \(\hbar\) so that \(\hbar\) is effectively a ``small'' but still a fixed constant, or vice-versa when higher \(\mathcal O(\hbar)\) terms dominate.)

The c-theorem states that evolution w.r.t~\(\hbar\) is irreversible and must occur in the direction of increasing \(\hbar\), since it is the coupling constant of (an irreversible) two-dimensional RG flow. Increasing \(\hbar\) corresponds to reverse RG flow. Forward evolution must be associated with the \emph{reverse} RG flow since the forward process is dissipative~\cite{Zamolodchikov86} since the number of degrees of freedom decreases; evolution in the opposite direction would mean that information would actually be lost as the physical properties from prior value of \(\hbar\) would be unrecoverable from the present \(\hbar\) value's fewer degrees of freedom.

This theorem agrees with our construction: for \(\hbar=0\) our monomorphism from \(\text{Cl}_{3,0}^{[0]}(\mathbb C)\) to \(\text{Cl}_{3,0}(\mathbb C)\) is undefined and our theory is thus defined over the full unconstrained \(\text{Cl}_{3,0}(\mathbb C)\) algebra. For \(\hbar=2\), the monomorphism means that we are effectively dealing with fewer degrees of freedom encapsulated by the smaller \(\text{Cl}_{3,0}^{[0]}(\mathbb C)\) subalgebra of \(\text{Cl}_{3,0}(\mathbb C)\).

In other words, for the case of black hole evolution in AdS$_3$, the \(\mathcal O(\hbar^0)\) term (or equivalently the dual \(\mathcal O(G^0)\) term) of the RT formula is associated with \(\text{Cl}_{3,0}(\mathbb C)\) while the \(\mathcal O(\hbar^1)\) term (or equivalently the dual \(\mathcal O(G^1)\) term) is associated with \(\text{Cl}^{[0]}_{3,0}(\mathbb C)\).

  Therefore, even if evolution in \(\lambda\) reverses for a given value of \(\hbar\), it is not necessarily related to evolution at a ``prior'' smaller value of \(\hbar\) by a sign, since it is now potentially dealing with a higher number of \(\mathcal O(\hbar)\) terms. For instance, in the case of evaporating black holes, we see that the dynamics of evaporation must progress along increasing \(\hbar\).

  This suggests that evolution is irreversible when \(\lambda\) and \(\hbar\)---which become dimensionful conjugates of each other in any continuous theory (corresponding to time and action, respectively)---are considered together as a parameterization of evolution. The c-function given by Eq.~\ref{eq:Cformula} captures how they should perhaps be considered together. Heuristically, this suggests that the direction of the ``arrow of time'' must lie along \(\hbar\), and therefore is not sign agnostic.

  The situation changes for perturbed AdS$_3$. The \(\mathcal O(G)\) transformation acts on the largest \(\text{Cl}_{4,0}(\mathbb C)\) algebra and increases the size of relevant algebra from the \(\text{Cl}_{2,0}(\mathbb C)\) subalgebra of \(\text{Cl}_{4,0}^{[0]}(\mathbb C)\) to the full \(\text{Cl}_{4,0}^{[0]}(\mathbb C)\) algebra, which is no longer canonically isomorphic to a subgroup of \(\text{Cl}_{3,0}(\mathbb C)\). This means that its resultant OPE is not equivalent to the Virasoro algebra and as a result the c-theorem no longer holds.  %

  However, its \(\mathcal O(\hbar^0 G^0)\) and \(\mathcal O(\hbar^1 G^1)\) terms (multiplying together the dual \(\hbar\) and \(G\) orders) that are associated with the increasing Page curve are still fully supported in \(\text{Cl}_{3,0}(\mathbb C)\), while the \(\mathcal O(\hbar^2 G^1)\) term associated with the decreasing Page curve is not. This interestingly parallels the derivation of the Page curve (see Appendix~\ref{sec:infoparadox}), wherein the increasing Page curve is due to quantum unitarity while the decreasing Page curve is due to a thermodynamic argument. This suggests that the \(\mathcal O(\hbar^2 G^1)\) term associated with information escape through a wormhole mechanism is probabilistic and can occur at any point in the evolution after the formation of the black hole (the \(\mathcal O(hbar^0 G^0)\) term).
  
%
  
%

%

%
%


%

%
%

%

%


%

%

%

\subsection{Discussion}
\label{sec:discussion}

Conformal field theories have often been applied to holographic systems corresponding to continuum quantum field theories without a clear notion of qubits (or qudits) by taking a continuum limit over algebraic generators (or modes): \(\xi_i \rightarrow \xi_x = \xi(x)\). As we have seen, keeping \(i\) supported over a finite discrete set \(\{\xi_i\}_i\) through a Majorana mapping proved key to produce operator product expansion that is equivalent to the Virasoro algebra commutation relations and thereby satisfy the holographic principle.%

While many finite \(N\) agreements have been found with the AdS/CFT correspondence~\cite{Maldacena99,Gubser98,Witten98,Aharony08}, they are mostly defined by string (or M-)theories that produce compactified spacetimes with corresponding boundary CFTs with continuous degrees of freedom. The correspondence we demonstrate here requires no such dependence on string theory and is based on a finite number of algebraic generators. Other efforts have also used models that are not directly related to string theories to explain gravitational phenomena~\cite{Piroli20,Jian21,Liu21}. Perhaps the results here will help unify these efforts. %

Nevertheless, there is a notion of compactified degrees of freedom in the theory presented here related to the larger dimensional embedding spacetime expressed by the non-quaternion elements of the octonion elements. We found that \(\mathcal O(G^1)\) transformations in these ``hidden'' degrees of freedom determine whether the embedded spacetime is AdS$_3$ or perturbed AdS$_3$. %

We have found that the AdS$_3$ isometry/qubit correspondence at the heart of the AdS$_3$/CFT$_2$ correspondence inherently defines the octonion algebra when it is treated as a homogeneous symmetric space. This octonionic theory treats gravity in AdS spacetime as an emergent phenomenon exhibiting a conformal anomaly (similar to contextuality or non-classicality). Similar relationships to non-classical properties have been postulated elsewhere~\cite{Maldacena13,Susskind16_2}. %

Therefore, a fundamental relationship we have established here is that the ubiquity of the central charge anomaly in quantum mechanics %
can be directly related to the ubiquity of singularities within general curved spacetimes. This means that, in the dual sense, the double cover property of qubits in \(\text{SU}(2)\) is equivalent to the ubiquitous presence of black holes in general relativity spacetimes. %

In terms of practical application, we note that the strong/weak duality between the \(\hbar\)-dependent formulation in \(\text{Cl}_{3,0}(\mathbb C)\) and the \(G\)-dependent formulation in \(\text{Cl}^{[0]}_{4,0}(\mathbb C)\) does not seem to seem to lead to a practically useful easier solution in the strong-coupling limit using the \(G\)-dependent formulation, as has been found in cases of the continuous correspondence in condensed matter~\cite{Johnson10} and high-energy~\cite{Kovtun05}. %
This is because the strong ``coupling'' means that \(\hbar \rightarrow \infty\), which is not meaningfully related to coupling between qubits but rather coupling between Majoranas. 
As we found in Section~\ref{sec:dictionary}, the quantum strong coupling limit of \(\hbar \rightarrow \infty\) is equivalent to \(G \rightarrow 0\) and makes the dual equations of motion become uncoupled. Such uncoupled equations necessarily correspond to maximally mixed state dynamics. %

In terms of predictive capability, this theory is as predictive as quantum mechanics. In particular, as discussed in Section~\ref{sec:holographicprinciple}, much like any quantum mechanical process can be written as a finite series of \(\mathcal O(\hbar^1)\) terms, so can any quantum gravitational process in the octonion theory be written in a finite series of \(\mathcal O(\hbar^1 G^1)\) terms.

In terms of novel predictions made in this paper, the most significant of course are the \(\mathcal O(G^1)\) mechanistic resolution to the black hole information paradox related to wormhole formation. %
There are also unexplored consequences from ``back-action'' effects when \(\mathcal O(G\hbar)\) transformations of the \(G\)-quantized generators act on the \(\hbar\)-quantized generators or vice-versa (i.e.~spacetime-matter interactions). We leave these for future work.

Perhaps the most pressing future direction of study is the extension of this theory to more than one qubit. Naively, adding more qubits adds exponentially more units of \(\hbar G\) in this phase space. Tracing out degrees of freedom in calculations of the entanglement entropy would produce the particular profile of the Page curve for a given subsystem.

\subsection{Conclusion}
\label{sec:conc}

There is much to be said about the AdS/CFT correspondence and its validity in various regimes as well as the approximative and conjectural nature of its ``dictionary''. Here we have provided an explicit non-approximative octonionic theory of quantum gravity that relates generally non-pure quantum states with gravitationally curved spacetime. We have shown how the presence of the ubiquitous central charge anomaly leads to the development of this theory as a superalgebra generalization of algebras defined through \(\hbar\)- and \(G\)-quantization. The result reveals how intimately linked the central charge anomaly is to the ubiquity of singularities in curved spacetime. We then offered a resolution of the information paradox from black hole radiation by showing that old radiative black holes in perturbed AdS$_3$ produce \(\mathcal O(\hbar^2 G^1)\) wormholes.

We summarize the concepts that we have found are very closely related in this formalism:
\begin{itemize}
  \item qubit contextuality, central charge anomaly, and supersymmetry,
  \item the qubit holographic principle, symplectic area conservation, and unitarity,
  \item RG fixed critical points, central charge critical points, and operator product expansion terms (semiclassical expansion orders),
  \item the two-dimensional renormalization group, central charge as a running coupling, and the irreversibility of evolution,
  \item uniformization and wormholes,
  \end{itemize}
We found that the quaternion subalgebra of the octonion algebra emergently defines an AdS$_3$ spacetime on an embedding space. %

As we discussed, the octonion algebra can be viewed as built up from progressively larger normed division algebras, wherein each step removes a property. (The complex algebra loses trivial involution, the quaternion algebra loses commutativity, the octonion algebra loses associativity.) %
Extending the octonion algebra further necessarily introduces zero divisors and produces a non-normed algebra. %
The fact that it is not possible to extend this theory further without losing normalizability suggests that it is in that sense complete. There are impressive results formulating the Standard Model in terms of the octonion algebra~\cite{Gunaydin73,Gunaydin74,Furey16,Furey18_1,Furey18_2,Singh20}, which if joined together with the quantum gravity theory presented here, suggests that it is an excellent candidate for a unified theory of physics.

\vskip5pt
\noindent--

L{.}~K{.}~Kovalsky~thanks M.~Sarovar, A.~Baczewski, E.~Knill, S.~Vardhan, S.~Pathak, and B.~Swingle for their help and correspondence during this study. %
This material is based upon work supported by the U.S. Department of Energy, Office of Science, Office of Advanced Scientific Computing Research, under the Accelerated Research in Quantum Computing (ARQC) programs.

Sandia National Laboratories is a multimission laboratory managed and operated by National Technology \& Engineering Solutions of Sandia, LLC, a wholly owned subsidiary of Honeywell International Inc., for the U.S. Department of Energy’s National Nuclear Security Administration under contract DE-NA0003525. 
This paper describes objective technical results and analysis. 
Any subjective views or opinions that might be expressed in the paper do not necessarily represent the views of the U.S. Department of Energy or the United States Government.

\bibliography{biblio}{}
\bibliographystyle{unsrt}

\appendix

\section{Algebraic Preliminaries of Pauli Algebra \(\text{Cl}_{3,0}(\mathbb R)\) as Classical Canonical Variables}
\label{sec:Paulialgebra}

  We define \(\xi_1\), \(\xi_2\) and \(\xi_3\) to be three real generators of the Clifford algebra \(\text{Cl}_{3,0}(\mathbb C)\) with identity \(\xi_0\), which is isomorphic to the Pauli algebra. %
  This means that the maximal monomial power of this algebra is three.

  Any element \(g \in \text{Cl}_{3,0}(\mathbb C)\) can be represented as a finite sum of homogeneous monomials consisting of these three generators:
  \begin{equation}
    \label{eq:gelement2}
    g(\bs \xi) = g_0 \xi_0 + \sum_{i=1}^3 g_i \xi_i + \sum_{i,j=1}^3 g_{ij} \xi_i \xi_j + \sum_{i,j,k=1}^3 g_{ijk} \xi_i \xi_j \xi_k.
  \end{equation}

  We define an involution operation that is analogous to complex conjugation:
  \begin{equation}
    (g^*)^* = g, \qquad (\alpha g)^* = \alpha^* g^*,
  \end{equation}
  where \(\alpha \in \mathbb C\).

  An element \(g \in \text{Cl}_{3,0}( \mathbb C)\) is real if \(g^* = g\). Since the generators are defined to be real (\(\xi_i^* = \xi_i\)), \(g\) is real if and only if the coefficients of its polynomials are real.

  Since this is an exterior algebra, polynomials of Grassmann generators serve as sufficient building blocks to define all operations on it. Moreover, since this algebra has an associated topology (discussed in Section~\ref{sec:SLinoctonions}) that possesses a compact subspace and connectedness, the operation of integration can be defined. Since \(\xi_i^2 = 0\), integration needs to be defined a little differently compared to other normed algebras, such as the real or complex numbers. Integration is defined to satisfy
  \begin{equation}
    \int \xi_0 \text d \xi_i = 0, \qquad    \int \xi_i \text d \xi_i = 1.
  \end{equation}
  
  This means that
  \begin{equation}
    \int \xi_{k_1} \xi_{k_2} \xi_{k_3} \text d \xi_3 \text d \xi_2 \text d \xi_1 = \epsilon_{k_1 k_2 k_3},
  \end{equation}
  and
  \begin{equation}
    \int g(\bs \xi) \text d^3 \bs \xi = \sum_{\bs k} \epsilon_{k_1 k_2 k_3} g_{k_1 k_2 k_3},
  \end{equation}
  where \(\epsilon_{k_1 k_2 k_3}\) is the Levi-Civita tensor.

  This definition is perhaps non-intuitive since we have framed it as an indefinite integration by not putting limits on the integral symbols, and from our intuition dealing with regular calculus on other algebras we expect indefinite integration to raise the power of polynomials instead of lowering them. Nevertheless, this definition is a topological generalization and can also be justified by its consequences~\cite{Walecka10}, which is to say that it could be considered to be an operation that we are calling ``integration'' because we will see that we can write a Grassmann version of Fourier transforms, exponential integrals, Gaussian integrals, etc{.}, in a manner equivalent to other algebras in terms of this integration operation. In fact, it is perhaps more intuitive to consider this definition of Grassmann integration as a \emph{definite} integral over the full (curiously unspecified) domain (like \(-\infty\) to \(\infty\) for the real algebra). 

  Due to the anti-commutation relation between Grassmann generators, derivatives can be defined as the following left- and right-handed operations:
  \begin{align}
    \vec \partial_{\xi_i} \xi_{k_1} \cdots \xi_{k_m} =& \delta_{k_1i} \cdots \xi_{k_m} - \delta_{k_2i} \xi_{k_1} \xi_{k_3} \cdots \xi_{k_m} \nonumber\\
    \label{eq:Grassmannleftderivative}
                                              & + \ldots + (-1)^m\delta_{k_mi} \xi_{k_1} \cdots \xi_{k_{m-1}},
  \end{align}
  and
  \begin{align}
    \xi_{k_1} \cdots \xi_{k_m} \cev \partial_{\xi_i} =& \delta_{k_mi} \xi_{k_1} \cdots \xi_{k_{m-1}} - \delta_{k_{m-1}i} \xi_{k_1} \cdots \xi_{k_{m-2}} \xi_{k_m} \nonumber\\
    \label{eq:Grassmannrightderivative}
                                                      & + \ldots + (-1)^m \delta_{k_1i} \xi_{k_2} \cdots \xi_{k_m}.
  \end{align}
  Again, this is chosen due to \emph{a posteriori} reasons, most notably that it allows for a similar Grassmann version of a Taylor series expansion of any element \(g \in \text{Cl}_{3,0}(\mathbb C)\):
  \begin{align}
    &g(\bs \xi) = g(0)\xi_0 + \sum_{i=1}^3 \vec \partial_{\xi_i} g(\bs \xi)|_{\bs \xi = 0} \xi_i\\
    & + \sum_{i,j=1}^3 \vec \partial^2_{\xi_i,\xi_j} g(\bs \xi)|_{\bs \xi = 0}\xi_i\xi_j + \sum_{i,j,k=1}^3 \vec \partial^3_{\xi_i,\xi_j,\xi_k} g(\bs \xi)|_{\bs \xi = 0}\xi_i\xi_j\xi_k. \nonumber
  \end{align}
  This is very useful, as it will allow for the definition of Grassmann versions of the exponential function
  \begin{equation}
    \exp(g(\bs \xi)) \equiv \xi_0 + \sum_{n=1}^\infty (n!)^{-1} g^n(\bs \xi),
  \end{equation}
  and the logarithm function,
  \begin{equation}
    \log(-\xi_0 + g(\bs x)) \equiv -\sum_{n=1}^\infty n^{-1} g^n(\bs \xi),
  \end{equation}
  which will turn out to satisfy the expected usual properties. The Grassmann exponential function can even be used to define a Grassmann version of the delta function when paired with Grassmann integration:
  \begin{equation}
    i \int \exp (i \sum_{j=1}^3 \rho_k \xi_k) \text d^3 \rho = \xi_1 \xi_2 \xi_3 \equiv \delta (\xi_1) \delta (\xi_2) \delta (\xi_3),
  \end{equation}
  since
  \begin{equation}
    \int g(\bs \xi) \xi_1 \xi_2 \xi_3 \text d^3 \xi = g(0).
  \end{equation}
  Note that \(\rho_j\) are also Grassmann elements, and should be considered to be an extension of the \(\xi_j\) Grassmann elements; \(\xi_j\) and \(\rho_j\) anti-commute.

  Notice that the definitions for Grassmann integration does not have an intuitive or formal ``anti-derivative'' property.

  As alluded to, these definitions allow us to define a Grassmann Gaussian integral as
  \begin{align}
    \int e^{\sum_{i,j=1}^3 a_{ij} \xi_i \xi_j} \text d^3 \bs \xi = \pf(2 a_{ij}) = \sqrt{ \det(2 a_{ij})},
  \end{align}
  where \(a_{ij} = -a_{ji}\) and \(\pf(A)\) is the Pfaffian of matrix \(A\) -- the determinant of a skew-symmetric matrix is equal to the square of a polynomial of its elements. Notice that this result produces a determinant in the numerator, instead of the denominator as for the real algebra.

  For \(A = \left(\begin{array}{ccc}0 &a_{12} & a_{23}\\ -a_{12}& 0& a_{23}\\ -a_{23} & -a_{13}& 0\end{array}\right)\), \(\pf(A) = 0\), and in general the Pfaffian is only non-vanishing for even-dimensional square matrices.

  These definitions for integration and derivative in the Grassmann \(\mathcal G_n\) algebra are sometimes called the \emph{Berezin calculus}~\cite{Berezin77}.

  The definition for differentiation given by Eqs.~\ref{eq:Grassmannleftderivative}-\ref{eq:Grassmannrightderivative} allows for a ``Poisson bracket'' to be defined:
  \begin{equation}
    i \sum_j \left( \xi_k \frac{\cev{\partial}}{\partial \xi_j} \right) \left( \frac{\vec{\partial}}{\partial \xi_j} \xi_l \right) \equiv \{\xi_k, \xi_l\}_{\text{P.B.}} = i \delta_{kl}.
  \end{equation}

  In this way, the Grassmann generators can be treated as conjugate classical canonical variables. However, to ascribe an evolution (from a Hamiltonian) to these canonical variables, we will want to ensure that any such evolution satisfies the expected quantum evolution under quantization.

  \section{Action}
  \label{sec:action}

    Given any real quadratic Hamiltonian \(H\), an associated symplectic matrix can be produced from exponentiation,
  \begin{equation}
    \label{eq:symplecticexpmatrix}
    \mathcal E_{ij}(\lambda) = \exp \left(i \frac{\hbar}{2} H \frac{\cev \partial^2}{\partial_{\xi_i}\partial_{\xi_j}} \lambda \right),
  \end{equation}
  which determines the evolution for all \(\lambda\):
  \begin{equation}
    \label{eq:Grassmannevolutionsymplecticmatrix}
    \bs \xi' = \bs{\mathcal E}(\lambda) \bs \xi.
  \end{equation}
  However, this is not a complete map between symmetric matrices (i.e.~quadratic Hamiltonians) and symplectic matrices~\cite{Almeida98}.

  A more useful definition can be made by associating a symplectic matrix \(\bs{\mathcal E}\) with an antisymmetric matrix \(\bs A\) through the Cayley parameterization,
  \begin{equation}
    \bs{\mathcal E} = (\bs A - i \bs I)(\bs A + i \bs I)^{-1}.
  \end{equation}
  For \(\bs{\mathcal E}\) given by Eq.~\ref{eq:symplecticexpmatrix},
  \begin{equation}
    \bs A_{jk} = \sum_{i} \epsilon_{ijk} \frac{b_k}{|b|} \tan (|b|t/2).
  \end{equation}

  We can identify this as the quadratic part of the usual \emph{generating action}
  \begin{equation}
    \label{eq:generatingaction}
    S(\bs \xi; t) = \sum_{jk} A_{jk} \xi_j \xi_k,
  \end{equation}
  which generates translations,
  \begin{equation}
    \xi'_i = \xi_i - \frac{\hbar}{4} S \cev \partial_{\xi_i}.
  \end{equation}
  This is easy to verify by substituting the definition of \(S\) (Eq.~\ref{eq:generatingaction}) into this equation and obtaining Eq.~\ref{eq:Grassmannevolutionsymplecticmatrix}. This is the action \(S\) that we will associate with a given Lagrangian \(L\).

  The action can also be seen in the general qubit unitary~\cite{Kocia17}:
\begin{equation}
  \hat U(\bs \xi, t) = \cos(bt/2) \exp \left( -\frac{2 i}{\hbar} \sum_{k,l,m} \epsilon_{klm} \frac{b_k}{b} \tan(bt/2) \hat \xi_l \hat \xi_m \right).
\end{equation}
We can identify the generating action (Eq.~\ref{eq:generatingaction}) in the exponent scaled by \(\hbar\) and its second derivative in the overall prefactor. This identifies this expression as a \(\mathcal O(\hbar^1)\) expansion of a full Feynman path integral~\cite{Kocia17}. This agrees with our finding that \emph{any} qubit quantum evolution can be represented by a quadratic Grassmann Hamiltonian \(H = -\frac{i}{2} \sum_{klm} \epsilon_{klm} b_k \xi_l \xi_m\) since the Hamilton's equations are generically \(\mathcal O(\hbar)\). Upon quantization, we must set \(\hbar = 2\) to produce the conventional relationship to Pauli matrices.

\section{The Clifford Stabilizer Subtheory, Contextuality, and Relationship to Central Charge}
\label{sec:Cliffordsubtheory}

\subsection{The Clifford Stabilizer Subtheory}
 \label{sec:Cliffordstabilizersubtheory}

 Note that the Clifford stabilizer subtheory is unrelated to the Clifford algebras.

We begin by defining the Pauli algebra in the traditional manner with respect to Pauli operators and then proceed to define the Clifford stabilizer subtheory. %

  We define Pauli operators to be the basis of \(\text{SU}(2)\), which is three-dimensional as a real manifold. One common representation is \(\hat \sigma_0 \equiv \left( \begin{array}{cc}1& 0\\ 0& 1\end{array}\right)\), \(\hat \sigma_1 \equiv \hat \sigma_x = \left( \begin{array}{cc}0& 1\\ 1& 0\end{array}\right)\), \(\hat \sigma_2 \equiv \hat \sigma_z = \left( \begin{array}{cc}1& 0\\ 0& -1\end{array}\right)\), and \(\hat \sigma_3 \equiv \hat \sigma_y = i \hat \sigma_x \hat \sigma_z\). Note that \(\{\hat \sigma_i, \hat \sigma_j\} = \delta_{ij} \sigma_0\). We define the Pauli group \(\mathcal P_n\) to consist of \(n\) tensor products of Paulis operators multiplied by \(\pm 1\). 

  Stabilizer states \(\ket \psi\) on \(n\)-qubits are defined by \(n\) (non-unique) generating commuting Pauli group elements \(\{\hat P_i\}_{i=1}^n\) such that \(\hat P_i \ket \psi = \ket \psi\). Thus, every \(n\)-qubit stabilizer state can be equivalently represented by \(n\) commuting Pauli operators. This is the operational representation of stabilizer states called the Heisenberg representation~\cite{Gottesman98}.

  The Clifford group \(\mathcal C_n\) consists of the Pauli group \(\mathcal P_n\) and unitary operators that take Pauli group elements to themselves. This means that Clifford unitaries take stabilizer states to other stabilizer states.

  Pauli group elements are also closed under themselves, and so are trivially a subgroup of the Clifford group. Since stabilizer states are unit eigenvectors of Pauli group elements, it follows that Pauli \emph{measurements} also take stabilizer states to other stabilizer states.
  
 Adding Pauli measurements to the Clifford group and restricting initial states to be stabilizer states produces a complete quantum subtheory consisting of initial states, a unitary gateset, and measurements. This subtheory of the full quantum theory is called the Clifford stabilizer subtheory.

 We have only discussed the Clifford stabilizer subtheory for two-dimensional systems, but a similar construction can be made in odd- and infinite-dimensions~\cite{Kocia17}. However, a difference exists between the two-dimensional theory and these others, which is that the qubit Clifford stabilizer subtheory exhibits quantum \emph{contextuality}.%

 \subsection{Contextuality}
 \label{sec:conxtuality}

Examining dimension-dependent properties of the Clifford stabilizer subtheory is a useful alternative way of characterizing the central charge anomaly in the complexified Clifford algebra \(\text{Cl}_{3,0}(\mathbb C)\) when mapped under the \(\hbar\)-dependent monomorphism to the Pauli algebra \(\text{Cl}_{3,0}(\mathbb R)\). In particular, qubits (two-dimensional Hilbert spaces) exhibit the unique property that they require non-classical anomalous terms in their operator product expansions in order to express the Clifford stabilizer subtheory of quantum theory. This phenomenon can be equivalently referred to as quantum contextuality. We will define this subtheory and contextuality here, and elaborate on this equivalence.

This is in contrast to the Clifford subtheory for odd-dimensional qudits and the harmonic Gaussian subtheory for infinite-dimensional systems, which are non-contextual. %
As we shall see here, contextuality generally indicates non-classicality, so this difference is often cited as peculiar since both subtheories are equally classically computationally simulable (as per the Gottesman-Knill theorem~\cite{Aaronson04}). %

  Contextuality is defined to exist if, given a set of measurements on a state, their measurement outcomes depend on which independent set is measured first~\cite{Kochen75,Redhead87,Mermin93}. More formally, this is the definition of state-independent measurement contextuality, but it will suffice for us. It is a uniquely quantum phenomenon, and is a necessary (though not sufficient) resource for universal quantum computation.

  Contextuality is perhaps best introduced through an example. We choose the Peres-Mermin square~\cite{Mermin93}, which consists of four Pauli measurements on two qubits: \(XI\), \(IZ\), \(IX\), and \(ZI\). The can be divided into two mutually anti-commuting sets consisting of commuting measurements. These are given by the rows and columns of Table~\ref{tab:PeresMermin}.
\begin{table}[h]
\begin{tabular}{|c|c||c|}
  \hline
  $XI$ & $IZ$ & $\boldsymbol{XZ}$\\
  \hline
  $IX$ & $ZI$ & $\boldsymbol{ZX}$\\
  \hline
  \hline
  $\boldsymbol{XX}$ & $\boldsymbol{ZZ}$ & $\pm\boldsymbol{YY}$\\
  \hline
\end{tabular}
\caption{The Peres-Mermin square.}
\label{tab:PeresMermin}
\end{table}

The outcomes of these measurements are \(\pm 1\). Regardless of the given state on the two qubits, the measurement outcomes of the measurements in any row or column are independent of each other since they commute. Furthermore, it follows that the outcome of the third element in the row or column is known given the outcomes of the first two, since they are separable and multiply together to form the third.

It is easy to show that given a measurement on \(XI\), whether a subsequent measurement on \(ZI\) produces outcomes \(\pm 1\) will depend on whether \(IZ\) or \(IX\) is measured first. In other words, the measurement outcome of \(ZI\) depends on which independent set is measured first: the top-most row or the left-most column. In this way, the context of the measurement on \(ZI\) affects its outcome. %

This contextuality can be made algebraically apparent in the Peres-Mermin square by noticing that the sign of the final \(YY\) product of the third elements depends on whether it is obtained from multiplying together the third column or row.

Since all these measurement operators are Paulis, which are in the Clifford stabilizer subtheory, the Peres-Mermin square construction demonstrates that the qubit Clifford stabilizer subtheory exhibits contextuality regardless of the state measured.

We will soon see that this state-independent contextual property of the qubit Clifford subtheory is related to the ubiquitous singularities that are present in curved spacetime when we extend a quantum formalism described by the Paulis (the quaternion algebra) to the full octonion algebra.

  \subsection{Clifford Subtheory Captured at \(\mathcal O(\hbar^1)\)}
  \label{sec:Cliffordhbar1}

  We have now established a well-defined mapping between the Majorana fermions that produce the Clifford algebra \(\text{Cl}_{3,0}(\mathbb C)\) and the Hilbert space of a (generally non-pure) qubit. In particular, the map can be expressed through the \(\hat T\) operator mapping given by Eq.~\ref{eq:gop}. This mapping is very useful because it allows us to formally introduce a convergent OPE finite-term series expansions in terms of a parameter \(\hbar\). We have also interspersed the last few sections with demonstrations that this is also technically possible to do strictly working in Hilbert space without the Majorana mapping, but is far less natural as the parameter \(\hbar\) is really only defined w.r.t.~the quantization of the Clifford algebra \(\text{Cl}_{3,0}(\mathbb C)\).

  As we discussed in Section~\ref{sec:Cliffordstabilizersubtheory}, the Clifford stabilizer subtheory is defined to contain the closed set of Clifford unitary gates with final Pauli measurements, both acting on initial stabilizer states. Clifford gates are defined to consist of all unitaries that take stabilizer states to themselves. As we mentioned in Section~\ref{sec:symplectic}, they have corresponding (quadratic) Hamiltonians in the \(\text{Cl}_{3,0}(\mathbb C)\) algebra and periods of evolution \(\lambda\) such that they take the elements of \(\text{Cl}_{3,0}^{[0]}(\mathbb C)\) to themselves. We found that this means that their overall \(\hbar\) dependence can be factored out and this procedure can be handled at order \(\hbar^0\).

  On the other hand, this ceases to be true for multi-qubit Pauli measurements. This is shown in detail for the Clifford algebra \(\text{Cl}^{[0]}_{3,0}(\mathbb C)\) in~\cite{Kocia17_2}, and we will only summarize the main point by saying that relative phases \(\pm i\) become important between commuting sets of Paulis in the Heisenberg representation. Since all Pauli multiplication rules can be performed at order \(\hbar^1\) if both elements in \(\text{Cl}^{[0]}_{3,0}(\mathbb C)\) and their Grassmann Fourier transform in \(\text{Cl}^{[1]}_{3,0}(\mathbb C)\) are calculated as in Eq.~\ref{eq:WeylSymbolGroenewoldRuleLeadingTerms_combined}, this means that the overall Clifford stabilizer subtheory can be captured at order \(\hbar^1\) within the Clifford algebra (with a single term for a single qubit).

  There is nothing fundamentally non-classical about \emph{tracking} a distribution or its characteristic Fourier transformation. %
  However, the necessity of using the Fourier transform after Pauli measurements to determine how to update the probability distribution \(\text{Cl}^{[0]}_{3,0}(\mathbb C)\) from its dual \(\text{Cl}^{[1]}_{3,0}(\mathbb C)\) (or vice-versa) \emph{is} non-classical as they are related by a global Fourier transformation. This is a non-classical update rule (incompatible with Kolmogorov's axioms for classical probability distribution update rules~\cite{Kocia19}, for instance), but it is still classically efficiently simulable.

  The importance of the Fourier transform's imaginary number can also be found in the well-known efficient tableau algorithm for classically simulating the qubit Clifford stabilizer subtheory. Therein, calculating the ``parity bit'' requires mod \(4\) arithmetic~\cite{Aaronson04} for rowsum's update of the phase, whereas for the odd-dimensional case, mod \(p\) arithmetic suffices~\cite{Kocia17}.

  This is in contrast to the odd- and infinite-dimensional case, where similar formalisms are able to accomplish this at \(\mathcal O(\hbar^0)\) with only the equivalent of the even subalgebra~\cite{Kocia17}.

  Note that this means that the Grassmann Fourier transform is itself an \(\mathcal O(\hbar)\) operation. We have thus almost established \(\hbar\) to be a dimensionless analogue to Planck's constant. But we refrain from declaring it so until we do a bit more work.

\begin{figure}[t]
\includegraphics[]{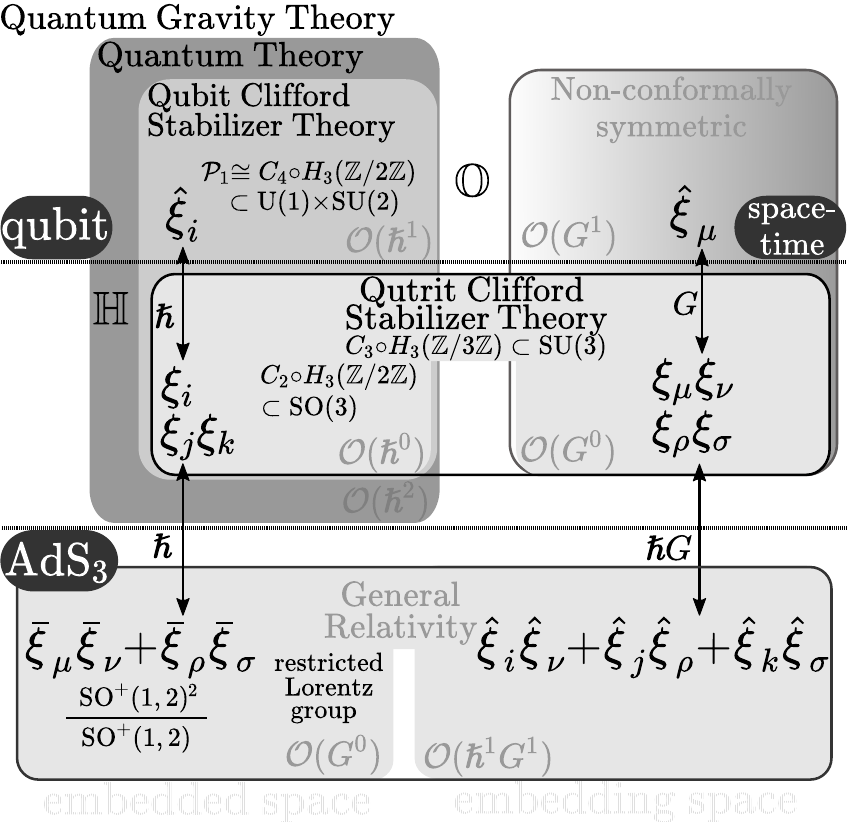}
\caption{Quantum gravity's relationship to all other physical theories. \(i=1,\, j=2,\, k=3 \) and \(\mu=4,\, \nu=5,\, \rho=6,\, \sigma=7\), or any cyclic permutation within these two respective sets. \(\mathcal P_1\) is the one-qubit Pauli group, \(C_p\) is the order-\(p\) cyclic group, and \(H_3(F)\) is the Heisenberg group over the field \(F\). %
}
\label{fig:dualities}
\end{figure}

A sketch of the dual elements of the full theory can be found in Figure~\ref{fig:dualities}. As we have seen, expressing everything in terms of the larger complexified Clifford algebras \(\text{Cl}_{3,0}(\mathbb C) \substack{\cong\\\text{n.c.}} \text{Cl}_{4,0}^{[0]}(\mathbb C)\) proved crucial to expressing everything conformally when possible, and identifying the algebra elements' role in the full theory.

  \subsection{Interpretation}

  We make a small digression in this section. In particular, we will examine how the qubit Clifford stabilizer subtheory's contextuality, apparent in the \(\text{Cl}_{3,0}(\mathbb C)\) algebra with its \(\Omega(\hbar^1)\) description, can be fundamentally attributed to the property that defining a classical probability space commensurate with qubit stabilizer states must be double-covered if they are to describe Pauli measurement outcomes properly. This double-covering manifests itself in the Clifford algebra through the superpartners that must both be tracked. %
This alternative perspective on the omnipresence of contextuality in quantum theory offers us the first hint of the AdS black hole geometry that we will soon formalize: doubly-covered spaces are particularly well-suited to minimally express particle and anti-particle creation or dynamics across null-separated distances of spacetime (i.e.~across event horizons) because they naturally possess supersymmetric formulations that provide superpartners for this task. 

Single-qubit stabilizer states, which are are eigenstates of the Paulis, exhibit a quaternion (\(\text{SU}(2)\)) topology under unitary evolution and Pauli measurements. This means that a classical probability theory with an event space (defining a topological space) consisting of the stabilizer states must be doubly larger than the classically natural \(\text{SO}(3)\) group structure. %
Elements in \(\text{Cl}^{[0]}_{3,0}(\mathbb C)\) corresponding to stabilizer states have non-negative real coefficients and so can be interpreted as (classical) probability distributions~\cite{Kocia17_2}. 
A doubly-covered space still allows for classical probability theories, with non-negative real-valued probability distributions and positive state-independent maps, but the double cover may require tracking which ``cover'' is indicated (more precisely, which cover is indicated for one qubit relative to another since (measurement) contextuality requires more than one qubit). For the non-unitary \(\text{SL}(2, \mathbb R)\) group, this is also true but with regards to the double cover of \(\text{SO}^+(2,2)/\text{SO}^+(1,2)\). In terms of \(\text{Cl}_{3,0}(\mathbb C)\), this manifests itself in the requirement to track the probability distribution and dual characteristic function. We make this clearer shortly. %

Overlaps between probability functions and their characteristic function necessarily require introducing the imaginary constant in intermediate arithmetic, even though the final result is real. This is because such dual functions are related by a Fourier transform, which is necessarily a function in complex space. As we showed in Section~\ref{sec:Cliffordhbar1}, the Fourier transform is a \(\mathcal O(\hbar)\) transformation. Together the Weyl symbols and their duals serve as a double-cover of the two-dimensional Clifford group, unlike the odd- or infinite-dimensional Clifford group, which necessitates introducing the imaginary constant.

More precisely, in the three-generator formalism, the presence of the double cover can be found in the Clifford subtheory by noting that since stabilizer states have non-negative real coefficients in their \(\text{Cl}^{[0]}_{3,0}(\mathbb C)\) subalgebra, it follows that their coefficients in \(\text{Cl}^{[1]}_{3,0}(\mathbb C)\) are all imaginary. Hence we can restrict to only looking at the unsigned elements of \(\text{Cl}^{[0]}_{3,0}(\mathbb R)\)
\begin{equation}
  \{\xi_0\} \oplus \{\xi_p \xi_q,\, \xi_q \xi_p,\, \xi_p \xi_r,\, \xi_r \xi_p,\, \xi_q \xi_r,\, \xi_r \xi_q\},
\end{equation}
and when considering their Grassmann Fourier elements we can just look at unsigned elements of \(i \times \text{Cl}^{[1]}_{3,0}(\mathbb R)\)
\begin{equation}
  \{i \xi_p \xi_q \xi_r, i\xi_q \xi_p \xi_r\} \oplus \{ i\xi_p, i\xi_q, i\xi_r \},
\end{equation}
where we organized the elements by their common grade. For single-qubit states (projectors), which consist of a linear combination of the left- and right-hand elements in the equations above, the elements in \(\text{Cl}^{[0]}(\mathbb R)\) only track the sign of the right-hand-side elements through their anti-commuting order and the elements in \(\text{Cl}^{[1]}(\mathbb R)\) only track the sign of the left-hand-side elements. To incorporate a global sign, as necessary to calculate Pauli observable outcomes (but not Clifford unitary evolution), it is necessary to take their mutual overlaps, which as we can see above, requires introducing a parity-four (i.e.~signed imaginary number) phase through the Grassmann Fourier transform. This tracking of both representations is equivalent to tracking which cover of a double cover a state is on.

This is a useful way to characterize the different behavior of the qubit stabilizer Clifford subtheory compared to the odd- and infinite-dimensional case because it makes it clear how the qubit subtheory's non-classicality is due to the mathematical quirkiness of its space being double-covered. This property does not prevent it from being as efficiently simulatable as the classical odd- and infinite-dimensional cases since we have shown that it can be accomplished using the usual efficient classical functions (i.e. classical probability distributions and their characteristic functions) as long as one permits the use of the imaginary number with a polynomial number of \(\Theta(\hbar^1)\) terms. %
In other words, despite being strictly outside classical theory, it is intuitively clear that taking overlaps between classical probability functions and their duals should be properly contained by universal quantum theory as a non-universal subtheory.

\subsection{The \(\Theta(\hbar^0)\) Qutrit Clifford Stabilizer Theory}
\label{sec:hbar0qutritCliffordtheory}

As discussed in Appendix~\ref{sec:Cliffordhbar1}, rotations between \(\{\xi_i\}_{i=1}^3 \in \text{Cl}_{3,0}(\mathbb C)\) Grassmann elements correspond to Clifford operations, which we showed can be described by an \(\mathcal O(\hbar^0)\) refactoring within the \(\text{Cl}_{3,0}(\mathbb C)\). Considering the full octonion algebra, rotations between \(\{\xi_\mu\}_{\mu=4}^7 \in \text{Cl}_{4,0}(\mathbb C)\) Grassmann elements similarly correspond to \(\mathcal O(G^0)\) operations as their Hamiltonians and evolutions \(\lambda\) are equivalent. It is also easy to see that rotations between the sets \(\{\xi_i\}_{i=1}^3\) and \(\{\xi_\mu\}_{\mu=4}^7\) Grassmann elements also correspond to \(\mathcal O(\hbar^0, G^0)\) operations. %

This last set of rotations allows us to form the ``split octonion units'':
\begin{align}
  \label{eq:splitoctonions1}
  u_0 =& \frac{1}{2}(\xi_0 + i \xi_7)\\
  \label{eq:splitoctonions2}
  u_j =& \frac{1}{2}(\xi_j + i \xi_{j+3}),
\end{align}
and their complex conjugates, for \(j = \{1,2,3\}\).

Upon quantization, these satisfy the properties~\cite{Gunaydin73,Gunaydin74}
\begin{align}
  \hat u_0^2 =& \hat u_0,\\
  \hat u_0 \hat u_0^* =& 0,\\
  \label{eq:vacuumentanglement}
  \hat u_0 \hat u_j =& \hat u_j \hat u_0^* = \hat u_j,\\
  \hat u_j \hat u_0 =& \hat u_0^* \hat u_j = 0,\\
  \hat u_i \hat u_j =& \epsilon_{ijk} \hat u_k^*,\\
  \hat u_j \hat u_k^* =& -\delta_{jk} \hat u_0,
\end{align}
and the complex-conjugate equations. This means that these \(\hat u_1\), \(\hat u_2\), \(\hat u_3\) (\(\hat u_1^*\), \(\hat u_2^*\), \(\hat u_3^*\)) operators are Fermi annihilation (creation) operators of \(\text{SU}(3)\) and \(\hat u_0\) is the vacuum state. Their non-associativity means that they have no matrix representation.

It then follows that we can construct qutrit shift \(X\) and boost \(Z\) operators
\begin{align}
  \hat X &= \sum_{i=1}^3 \hat u_{i} \hat u_{((i-1)\oplus 1)+1}^*\\
  \hat Z &= -\sum_{i=1}^3 \omega^{i-1} \hat u_i^* \hat u_i,
\end{align}
where \(\omega \equiv \exp 2\pi i/3\) and \(\oplus\) indicates addition modulus \(3\). These act by ``shifting'' and ``boosting'' the computational basis states:
\begin{align}
  \hat X \hat u_i^* \hat u_0 \ket{0} =& \hat u^*_{i \oplus 1} \hat u_0 \ket{0},\\
  \hat Z \hat u_i^* \hat u_0 \ket{0} =& \omega^{i-1} \hat u^*_i \hat u_0 \ket{0},
\end{align}
where \(\ket{0}\) is the ordinary vacuum acted upon by complex fields. Since they are quadratic operators of \(\hat u_i\), which take \(\hat u_i\) to themselves, it follows that their action can be captured at \(\mathcal O(\hbar^0)\) by the Euler-Lagrange equations of the unquantized Grassmann algebra by factoring out the \(\hbar\) dependence, in the same manner as we did for the qubit Clifford unitary gates. However, unlike for the qubit case, this can also be done for measurements of these generalized Pauli operators as well, by constructing the Weyl-Heisenberg group \(H_3(\mathbb Z/3\mathbb Z)\) from these shift and boost operators~\cite{Kocia17_3,Kocia17}. This means that the full Clifford stabilizer subtheory can be described non-contextually at \(\mathcal O(\hbar^0)\) using these operators.

Note that \(\hat X\) and \(\hat Z\) have matrix representations~\cite{Kocia17_3}; quadratic products of the non-associative \(u_i\) produces associative operators. %
Also note again that we would not have been able to factor out the \(\hbar\) dependence for the qutrit Clifford stabilizer subtheory if we had not expressed the operator product expansion in terms of both the even and odd elements of \(\text{Cl}_{3,0}(\mathbb C)\) so as to make it truncate at \(\mathcal O(\hbar^1)\). 

We deem that we are now justified in associating \(\hbar\) as a dimensionless analog to Planck's constant and we do so henceforth.

In summary, we have shown that the full Clifford stabilizer subtheory for qutrits can be treated at \(\mathcal O(\hbar^0)\). This neatly accords with the non-contextuality of the odd-dimensional Clifford stabilizer subtheory.

\section{Quantum Magic Gates and States}
\label{sec:quantummagicstates}

Supplementing the Clifford stabilizer subtheory --- consisting of Clifford gates acting on stabilizer states with Pauli measurements --- with magic states extends the subtheory to a full quantum implementation~\cite{Bravyi05}. This is convenient as it allows us to consider only a single additional single-qubit gate to sufficiently analyze the full quantum subtheory.

Quantum \(T\) magic gates are unitaries defined by the Hamiltonian
\begin{equation}
  H = -(i /2)\xi_p \xi_r
\end{equation}
which produces the following Hamilton equations:
\begin{equation}
  \der{}{\lambda}(\xi_p, \xi_q \xi_r) = -\hbar (\xi_r, 0, -\xi_p)/4.
\end{equation}
When evolved for \(\lambda = \pi/2\) (by the exponential map) these produce
\begin{align}
  \label{eq:tgatemagicstate1}
  \xi_p &\rightarrow (1-\hbar/4) \xi_p - \hbar \xi_r/4,\\
  \xi_q &\rightarrow \xi_q,\\
  \label{eq:tgatemagicstate3}
  \xi_r &\rightarrow (1-\hbar/4) \xi_r + \hbar \xi_p/4.
\end{align}


The above transformations on \(\xi_p\) and \(\xi_q\) contain two \(\hbar\)-dependent terms. Therefore, the overall evolution is \(\mathcal O(\hbar^1)\) and cannot be reduced through factorization to be \(\mathcal O(\hbar^0)\) like for Clifford unitary evolution. The evolution is still unitary and so described by a quadratic real Hamiltonian. 

In general, given \(BQP\ne BPP\), these number of \(\mathcal O(\hbar^1)\) terms will proliferate exponentially with respect to the number of qubits. (This is true even if approximating T gate Pauli measurement outcomes to additive error).

Due to the two \(\hbar\)-dependent terms, it is also possible to consider this T gate evolution as a linear combination of evolution under two different Clifford Hamiltonians. For instance, in the above \(H = -i \xi_0\) and \(H = -i \xi_q \xi_r\). We will see that this perspective proves useful later when we consider ``gravitational'' magic gates.

We will also see in Appendix~\ref{sec:Airy} that these \(\mathcal O(\hbar^1)\) terms cannot be combined (or ``uniformized''~\cite{Chester57}, as this is called in continuous stationary phase approximation or steepest descent extensions) into fewer higher \(\mathcal O(\hbar^2)\) terms and that the dual of this explains why large thermodynamically stable BTZ black holes in AdS$_3$ spacetime are eternal (i.e. do not evaporate).

\section{Killing Horizons, Wick Rotations, Thermal States and AdS$_3$ Black Holes}
\label{sec:ads3blackholes}

In Section~\ref{sec:SLinoctonions}, we dealt with a \((1+1)\times(1+1)\) Minkowski space on coordinates \(X = (t,x,t',x')\) with metric
\begin{equation}
  ds^2 = -\text d t^2+ \text d x^2 - \text d t'^2 + \text d x'^2.
\end{equation}

The \(\text{SO}^+(1,2) \times \text{SO}^+(1,2)\) symmetry denotes an embedding on a hyperboloid
\begin{equation}
  -t^2 + x^2 - t'^2 + x'^2 = -l^2,
\end{equation}
for some \(l >0\).

Choosing the global coordinates \(R = (\rho, \phi, \tau)\) by setting
\begin{align}
  t =& l \cosh \rho \cos \tau\\
  x =& l \sinh \rho \sin \phi\\
  t' =& l \sinh \rho \cos \phi\\
  x' =& l \cosh \rho \sin \tau,
\end{align}
produces the induced metric
\begin{equation}
  \text d s^2 = l^2 (-\cosh^2(\rho) \text d \tau^2 + \text d \rho^2 + \sinh^2(\rho) \text d \phi^2),
\end{equation}
where we consider the universal covering space with \(\tau \in (-\infty, \infty)\).

AdS$_d$ is a maximally symmetric spacetime and so contain \(n(n+1)/2\) isometries (i.e.~Killing vectors). Therefore, AdS$_3$ has six isometries, \(V^\mu\), that it inherits from its embedding space: \(K^{01}\), \(K^{32}\), \(K^{21}\), \(K^{30}\), \(J^{02}\), and \(J^{13}\), where \(K^{01} = t \partial_x + x \partial_t\), \(K^{23} = t' \partial_{x'} + x' \partial_{t'}\), \(K^{12} = x \partial_{t'} + t' \partial_x\), \(K^{03} = t \partial_{x'} + x' \partial_t\), \(J^{02} = t \partial_{t'} - t' \partial_t\), and \(J^{13} = x \partial_{x'} - x' \partial_x\). (Expressing AdS$_3$ as a symmetric space, \(\frac{\text{SO}^+(1,2) \times \text{SO}^+(1,2)}{\text{SO}^+(1,2)}\), pairs these six isometries into three isometries.) To find these isometries in the global coordinates, we need to calculate
\begin{equation}
  \chi_\mu = P^\nu_\mu V_\nu,
\end{equation}
where \(P_\mu^\nu = \partder{X^\nu}{R^\mu}\) is the projection tensor.

Notably, this produces
\begin{equation}
  J^{02} = -l^2 \cosh^2( \rho) \partial_\tau,
\end{equation}
and
\begin{equation}
  J^{13} = l^2 \sinh^2( \rho) \partial_\phi.
\end{equation}

Also, note that the isometry generated by \(\hat Y = \hat \xi_3 = i (J^{02} + J^{13})\), which contains a global factor of \(i\). The factor of \(i\) is important as it will require us to Wick rotate into imaginary time.

The metric in global coordinates is independent of \(\phi\) and \(\tau\). Therefore, it follows that Killing vectors exist that are coaxial with \(\partial_\phi\) and \(\partial_\tau\), which indeed we identify with \(J^{02}\) and \(J^{13}\), respectively. However, the \(\phi\) coordinate is periodic on \([0,2 \pi)\) while \(\tau\) is defined on the unbounded interval \((-\infty, \infty)\). This means that \(\partial_\tau\) is also associated with a Killing horizon since it changes sign at \(\tau = 0\).

The global metric of AdS reveals that it is a static spacetime. Such spacetimes are understood to contain an event horizon \(\Sigma\) that corresponds to an asymptotically timelike Killing vector field \((\partial_\tau)^\mu\)~\cite{Carroll19}. Therefore, if we perform a Wick rotation into imaginary time \(\tau \rightarrow -i \tau\), then we obtain a Killing horizon from \(\hat \xi_3 = i (J^{02} + J^{13})\).

Such a Wick rotation generates a metric that is related to the Euclidean BTZ black hole metric,
\begin{equation}
  \text d s^2 = (r^2 - r_+^2) \text d T^2 + \frac{l^2}{r^2 + r_+^2} \text d r^2 + r^2 \text d \varphi^2,
\end{equation}
by the coordinate transformation
\begin{align}
  r =& r_+ \cosh \rho,\\
  T =& \frac{l}{r_+} \phi,\\
  \varphi =& \frac{l}{r_+} \tau,
\end{align}
for event horizon at \(r_+\).

A Euclidean BTZ black hole at temperature \(T\) is equivalent to thermal AdS$_3$ at temperature \(T^{-1}\)~\cite{Carlip95,Aharony00,Cadoni10}.

\section{The Information Paradox}
\label{sec:infoparadox}

We take a moment to give a quick overview of what is frequently called the black hole information paradox. This section is meant mainly for non-specialists, for the sake of completeness. Many more in-depth reviews can be found elsewhere~\cite{Page93,Page13}.

Black holes were proposed by Bekenstein~\cite{Bekenstein73} to possess an entropy that scales with the area of its event horizon. Since classically black holes do not radiate, this generally means that the classical entropy of black holes must be a non-decreasing function.

However, this characterization changed after Hawking's discovery that black holes radiate when quantum effects are taken into account~\cite{Hawking76}. Since black holes can evaporate, their area can decrease, and so according to Bekenstein's proposition, black hole entropy should be a decreasing function. However, Hawking showed that the total entropy of a black hole consists of a linear combination of the classical Bekenstein term and the semiclassical Hawking correction from radiation. This latter contribution from external radiation will be an increasing function.

Though the precise nature of Hawking radiation is still undetermined, pairwise entangled modes have been shown to be a characteristic feature of horizons~\cite{Harlow16}. One member of the pair will be infalling and eventually hit the singularity. The other can escape to infinity. An external observer will only measure these external Hawking radiation modes and see a highly mixed final state with large entropy. This presents a contradiction if the initial state of the black hole is a pure state, which has zero entropy. A full quantum mechanical system cannot become mixed without violating unitarity.

\begin{figure}[h]
\includegraphics[]{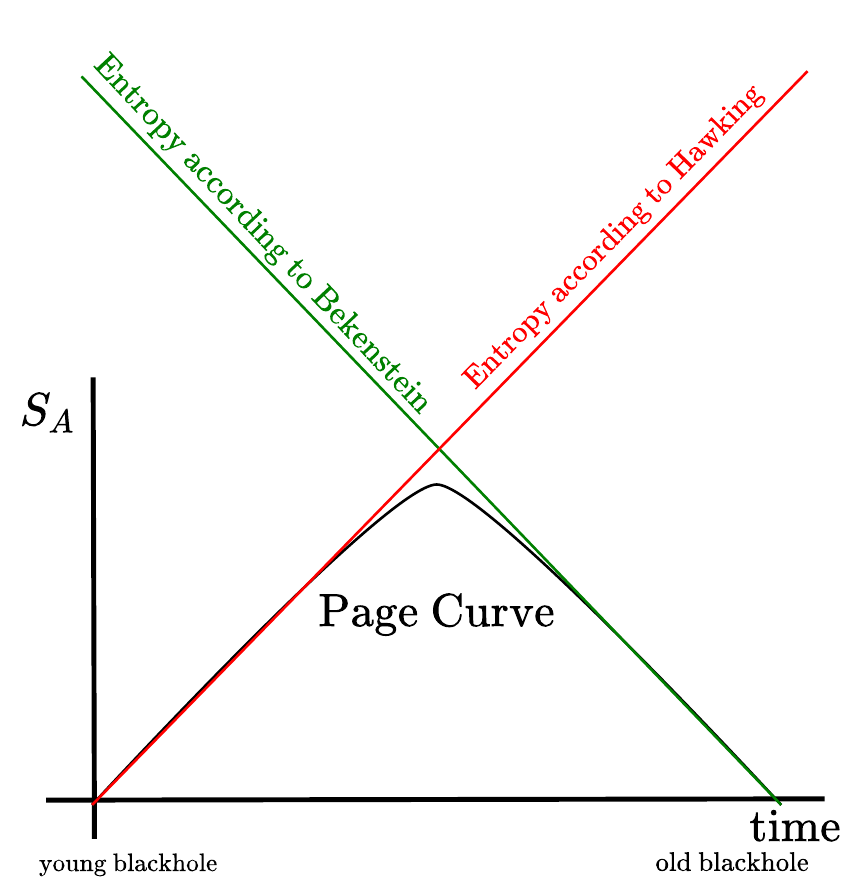}
\caption{The Page curve which follows Hawking's bound initially as the black hole external radiation modes accumulate and then subsequently follows Bekenstein's bound to preserve quantum unitarity. This requires the information entropy of the matter inside the event horizon to somehow escape, thereby posing an information paradox.}
\label{fig:Pagecurve}
\end{figure}

Page argued~\cite{Page93,Page93_2,Page13} that this information lost about the system must eventually escape past the event horizon and thereby restore the unitarity of the dynamics. As a result, though an evaporating black hole that starts out as a pure state will have an entropy that follows Hawking's increasing curve when it is young since the entropy of the external radiation will increase as more and more modes radiate, when the entropy of the radiation is the same as the entropy of the horizon, its entropy will subsequently follow Bekenstein's classical decreasing bound proportional to the decreasing area of the evaporating black hole (see Figure~\ref{fig:Pagecurve}). This means that the early radiation must in some way \emph{purify} the later radiation. Since the event horizon is supposed to present an inescapable barrier, the lack of mechanism for this to be accomplished presents an information paradox.

As mentioned, pairwise entangled modes have been shown to be a characteristic feature of horizons~\cite{Harlow16}. However, their form prevents them from sharing their entanglement across the event horizon and resolving the information paradox~\cite{Preskill92,Maldacena03} without a significant \(\mathcal O(1)\) contribution. It has also been found that small correlations between early and later Hawking radiation is also insufficient. It is deemed unlikely that a remnant from evaporation of Planckian scale could resolve the paradox~\cite{Giddings92}.

\section{Area Subtended by Stationary Phase Points and Uniformization}
\label{sec:Airy}

This section will motivate how higher order \(\hbar\) terms in a stationary phase approximation correspond to the correct solution when the holographic area subtended by stationary phase points becomes smaller, as occurs during black hole evaporation.

We consider the continuous case of a function with symmetric stationary phase points around the origin of the integral~\cite{Berry89,Almeida98,Heller18}:
\begin{align}
  f(\alpha) = \int^\infty_{-\infty} \exp \left[ \frac{i}{\hbar} \phi(x) \right] \text d x,
\end{align}
where
\begin{equation}
  \phi(x) = \left( \frac{x^3}{3} + \alpha x \right).
\end{equation}

We define the second order Taylor series expansion of the exponent around the zeros \(x_\pm = \pm\sqrt{\alpha}\) of its derivative to be
\begin{equation}
  \phi_{\pm\sqrt{\alpha}}(x) = \pm \sqrt{\alpha } \left(x\mp \sqrt{\alpha }\right)^2 \mp \frac{2 \alpha ^{3/2}}{3}.
\end{equation}

Performing a stationary phase approximation around these two stationary phase points produces
\begin{align}
  \label{eq:preuniformizationAiry}
  f(\alpha) \approx& 2 \Re \int^\infty_{0} \exp \left( \frac{i}{\hbar} \phi_{\sqrt{\alpha}}(x) \right) \text d x \nonumber\\
                   & + 2 \Re \int^\infty_{0} \exp \left( \frac{i}{\hbar} \phi_{-\sqrt{\alpha}}(x) \right) \text d x \nonumber\\
                   & + \mathcal O(\hbar)\\
  =& 2 \sqrt{\pi} \hbar^{1/2} \alpha^{-1/4} \cos\left( \frac{2}{3}\alpha^{3/2}\hbar^{-1} - \pi/4 \right) \nonumber\\
                   & + \mathcal O(\hbar).
\end{align}

This solution is only formally correct with the stationary phase solutions are far from the energy shell. Otherwise it diverges.

To handle this case, we can perform a uniformization by doubling the range of integration so that the two stationary phase points interact with each other instead of with the origin~\cite{Almeida98}:
\begin{align}
  \label{eq:postuniformizationAiry}
  f(\alpha) =& \int^{\infty}_{-\infty} \exp \left( \frac{i}{\hbar} \phi(x)\right) \text d x\\
               =& \left(\frac{2}{\hbar^2 \sqrt{\alpha}}\right)^{1/3} \Ai(\alpha)
\end{align}
where \(\Ai\) is the Airy function. %

\begin{figure}[h]
\includegraphics[]{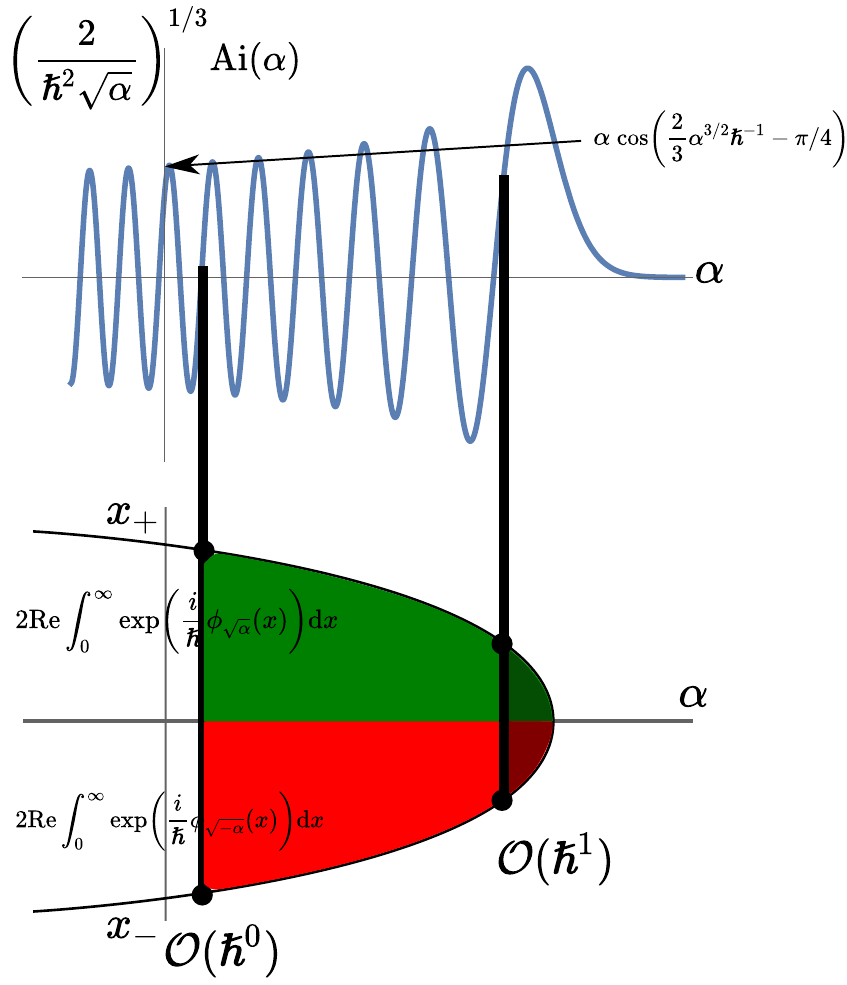}
\caption{uniformization. A very similar figure can be found in~\cite{Heller18} for the wave function in position space.}
\label{fig:uniformization}
\end{figure}

A similar symplectic area association in \(\text{Cl}_{3,0}(\mathbb C)\) can be made by using the integral representation of the OPE (Eq.~\ref{eq:arealaw}) that we introduced in Section \ref{sec:hbarCliffordtheory} and rewrite here comparing it to the OPE (Eq.~\ref{eq:WeylSymbolGroenewoldRuleLeadingTerms_combined}):
  \begin{align}
    &W_1 \star W_2(\bs \xi)  \nonumber\\
    \label{eq:arealaw2}
    =& \int e^{\frac{2}{\hbar}(\bs \xi\cdot \bs \xi_1 + \bs \xi_2 \cdot \bs \xi + \bs \xi_1 \cdot \bs \xi_2)} W_1(\bs \xi_1) W_2(\bs \xi_2) \text d^3 \bs \xi_1 \text d^3 \bs \xi_2,\\
    =& W_1(\bs \xi) W_2(\bs \xi) \nonumber\\
    & - \left(\frac{\hbar i}{2} \sum_i \vec \partial_{{\xi_1}_i} W_1(\bs \xi) \vec \partial_{{\xi_2}_i} W_2(\bs \xi)\right. \\
    & \qquad \left. + \frac{\hbar i}{2} \sum_i \vec \partial_{{\rho_1}_i} \widetilde W_1(\bs \rho) \vec \partial_{{\rho_2}_i} \widetilde W_2(\bs \rho)\right), \nonumber\\
    \equiv& \mathcal O(\hbar^0) + \mathcal O(\hbar^1).
  \end{align}

  It follows an expansion that terminates at \(\mathcal O(\hbar^0)\) term corresponds to a stationary phase evaluation of the integral on line~\ref{eq:arealaw2} at \(\bs \xi_1 = \bs \xi_2 = \bs \xi\),
  \begin{align}
    &W_1(\bs \xi) W_2(\bs \xi) \\
    =& \int e^{\frac{2}{\hbar}(\bs \xi\cdot \bs \xi_1 + \bs \xi_2 \cdot \bs \xi + \bs \xi_1 \cdot \bs \xi_2)} W_1(\bs \xi_1) W_2(\bs \xi_2) \text d^3 \bs \xi_1 \text d^3 \bs \xi_2 \Big|_{\bs\xi_1=\bs\xi_2=\bs\xi}, \nonumber
  \end{align}
  while an expansion that terminates \(\mathcal O(\hbar^1)\) requires the full evaluation of line~\ref{eq:arealaw2}.

\begin{figure}[h]
\includegraphics[]{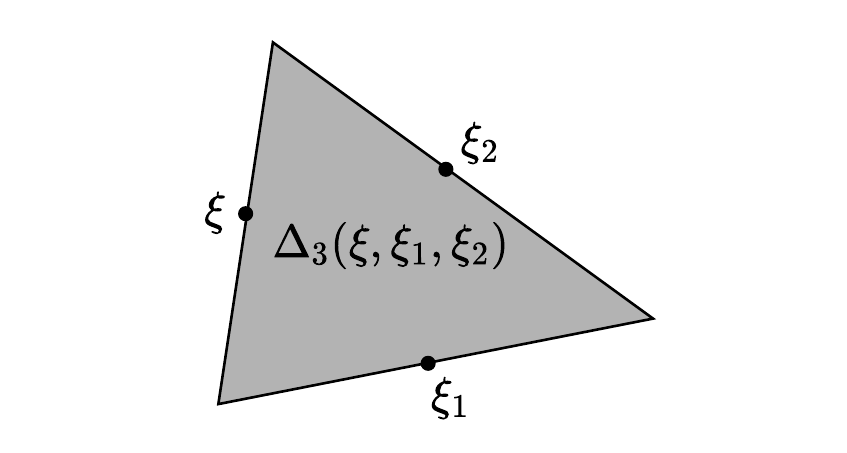}
\caption{The symplectic area associated with \(\Delta_3(\xi, \xi_1, \xi_2)\)}
\label{fig:triangle}
\end{figure}

  The argument to the exponential can be rewritten
  \begin{align}
    \Delta_3(\bs \xi, \bs \xi_1, \bs \xi_2) \equiv &\bs \xi\cdot \bs \xi_1 + \bs \xi_2 \cdot \bs \xi + \bs \xi_1 \cdot \bs \xi_2\\
    =& (\bs\xi_1 - \bs \xi) \cdot (\bs \xi_2 - \bs \xi),
  \end{align}
  which can be interpreted as the ``area of a triangle'' with midpoints \(\bs \xi\), \(\bs \xi_1\) and \(\bs \xi_2\) %
  (see Figure~\ref{fig:triangle}). By putting parentheses around ``area of a triangle'', we mean the Grassmann analog of a signed area associated with a quadratic exterior product. Here this signed area is Grassmann-valued.

This means that the \(\mathcal O(\hbar^0)\) term has null Grassmann symplectic area associated with it, while the \(\mathcal O(\hbar^1)\) term has a non-zero Grassmann symplectic area.

Notice that unlike for the continuous-variable Airy function example, there is no \(\mathcal O(\hbar^2)\) term and the two \(\mathcal O(\hbar^1)\) terms cannot be combined (uniformized) to become such a term. We will see that this captures the property that large thermodynamically stable BTZ black holes in AdS$_3$ spacetime do not evaporate when we consider the dual AdS OPE.

\section{Quantum Extremal Surfaces}
\label{sec:quantumextremalsurfaces}

The Ryu-Takayanagi formula was recently extended to include quantum extremal surfaces \(\chi\)~\cite{Headrick07,Faulkner13,Hubeny13,Engelhardt15,Almheiri20}:
\begin{align}
  S(\rho_A) = \min_\chi \left\{ \text{ext}_\chi \left[ \frac{A(\chi)}{4 G_N} + S_\text{semi}(\Sigma_\text{rad} \cup \Sigma_\text{int}) \right] \right\}.
\end{align}

This is a maximin definition, wherein the minimal maximal surface is selected. When radiating black holes are young, the first term \(A(\chi)/4 G_N\) dominates and is a classical term corresponding to the area subtended by a classical surface \(\chi\). In intermediate ages, the surface \(\chi\) remains classical and a quantum (semiclassical) correction \(\Sigma_\text{rad}\) is added corresponding to the exterior radiation outside the classical area \(\chi\). In late ages, the minimal surface \(\chi\) itself changes due to quantum corrections, though it is still classical. The entropy now has a different classical \(A(\chi)/4 G_N\) term and a new quantum (semiclassical) correction \(\Sigma_\text{rad} \cup \Sigma_\text{int}\), which contains an influence from the interior of the black hole's event horizon.

This narrative is very similar to the changes that occur in the dominance of the \(\mathcal O(\hbar^{0} G^0)\) classical and \(\mathcal O(\hbar^1 G^1)\) quantum (semiclassical) terms in Section~\ref{sec:perturbedAdS3} for the radiating black hole in perturbed AdS$_3$ spacetime. The late age \(\mathcal O(\hbar^0 G^1)\) term can be described as a \(A(\chi)/4 G_N\) term, as it is also zero degree (i.e.~proportional to \(\xi_0\)) and so can be considered to modify the original area \(\mathcal O(\hbar^{0} G^0)\) \(A(\chi)/4 G_N\) term, Since it is a gravitational \(\mathcal O(G^1)\) term, it is a gravitational correction to the classical term inline with a quantum extremal surface. This association as an interior contribution to the entropy of entanglement is strengthened by the reminder that it can be interpreted as joining the two stationary phase points on either side of the event horizon, as described in Appendix~\ref{sec:Airy}.

We now make some of these associations more concrete.

As we saw in Section~\ref{sec:Gsymplecticstructure}, the AdS dual Lagrangian for Clifford Hamiltonians associated with stabilizer states \(\pm \xi_i\) is \(-L = H = i (\bar \xi_\mu \bar \xi_\nu + \bar \xi_\rho \bar \xi_\sigma)\). Similarly, in Section~\ref{sec:RTquantummagicstates} we found that the associated Hamiltonian (and thus Lagrangian) for quantum magic gates is the same as for the Cliffords, just evolved for a different period. So the isometries remain the same.

Since these Lagrangians all have zero \(\mathcal O(\hbar^{0} G^1)\) term, this means that the quantum extremal surfaces associated with their early evolution are null surfaces. On the other hand, their non-zero \(\mathcal O(\hbar^{1} G^1)\) terms identifies isometries that define quantum extremal surfaces for intermediate and late lifetimes. Since there are no more terms in the expansion, it follows that the associated area to their quantum extremal surfaces never decreases.

For the \(\mathcal O(G)\) transformation that produces perturbed AdS$_3$ spacetime in Eq.~\ref{eq:LorentzGboost}, the Lagrangian during early evolution is similarly \(\mathcal L = 0\) and during intermediate evolution it is also similarly a superposition of \(\mathcal L = 0\) and  \(H = -\mathcal L = i(\bar \xi_4 \bar \xi_6 + \bar \xi_5 \bar \xi_7)\). Therefore, it is indistinguishable from evolution associated with a stabilizer state or quantum magic state. However, during late evolution we showed that it was associated with evolution under a superposition of the prior Lagrangians and a \(\delta \tau\) contribution from \(H = -\mathcal L = i \hbar^2 G \xi_0\). This last Lagrangian can be rewritten as the AdS dual element \(H = - \mathcal L = i \hbar^2 G \bar \xi_4 \bar \xi_5 \bar \xi_6 \bar \xi_7\) and therefore corresponds to gravitational actions that are non-linear. %

\begin{figure}[t]
\includegraphics[]{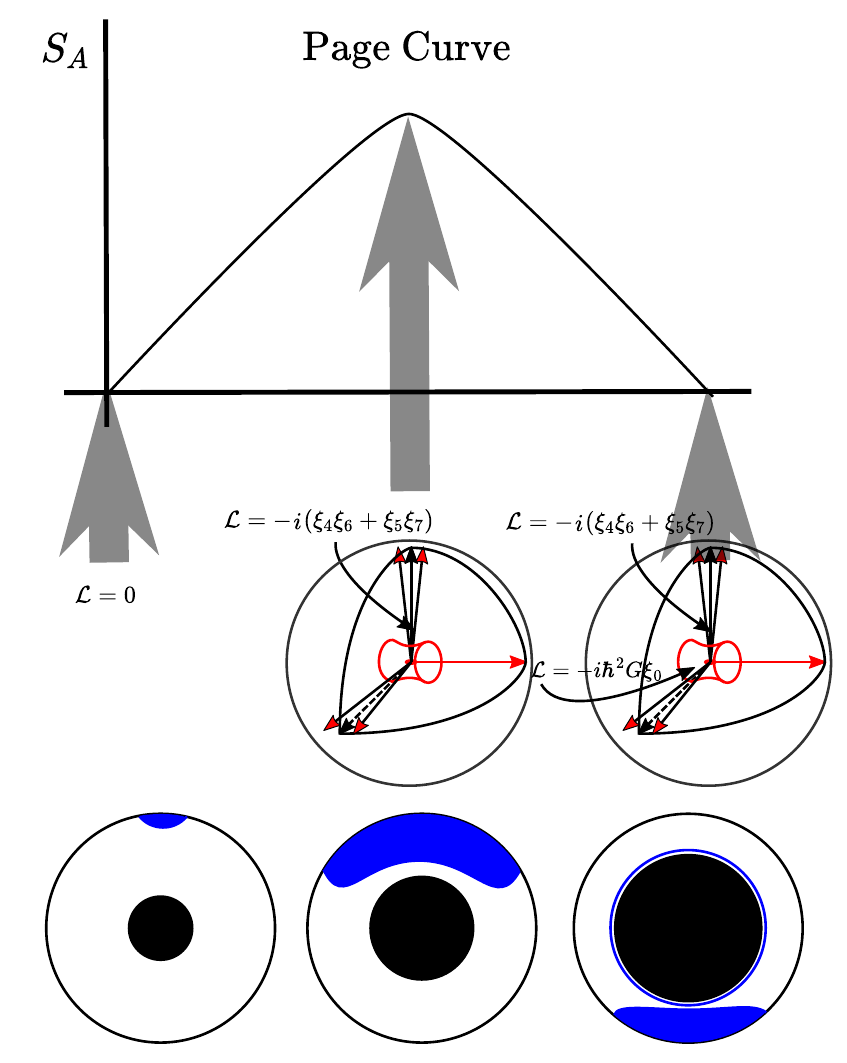}
\caption{Quantum extremal surfaces at distinct times of the Page curve are sketched for AdS$_3$ spacetime  when there is a bath coupled at conformal infinity. In the bottom we have a series of Poincar\'{e} surfaces for perturbed AdS$_3$ spacetime that illustrate the corresponding extremal surfaces \(\chi\).}
\label{fig:quantum_extremal_surfaces}
\end{figure}

These quantum extremal surfaces for the \(\mathcal O(G^1)\) transformation are sketched in Figure~\ref{fig:quantum_extremal_surfaces}. Therein, we speculate that this last term is similar to the final quantum extremal surface that wraps around the singularity~\cite{Hubeny13} associated with an evaporating AdS black hole with CFT-bath coupling~\cite{Almheiri19,Penington20}. If this holds, this explains how the black hole entropy decreases after the Page turning point due to two discontinuous geodesics forming a quantum extremal surface that subtends a smaller overall area.

\section{Emergent Behavior of Gravitons in AdS and Perturbed AdS}
\label{sec:emergent}

It may be worth spending a bit of time discussing the lack of presence of any particle or field directly responsible for gravitation here, i.e.~a graviton. The quaternion generators of the full octonion algebra do define a particle, namely a qubit in \(\text{SU}(2)\), and the full seven octonion generators also define a particle, namely an entangled qutrit particle and anti-particle in \(\text{SU}(3)\). It is only through the internal dualities present within this particle and anti-particle that spacetime is present, namely between the quaternion generators and the non-quaternion generators.

In this sense, for AdS$_3$, the quaternion generators define an ``emergent'' spacetime through their algebraic identity relations with the non-quaternion generators, the properties of which are defined by the quantum state they define. As a result, for AdS$_3$ the non-quaternion generators are not strictly explicitly necessary to define the spacetime isometries as they can simply be written in terms of the quaternion generators using these identity relations. The non-quaternion generators are not adding anything without \(\mathcal O(G)\) transformations. 

On the other hand, the perturbed AdS$_3$ spacetime is formed by a \(\mathcal O(G)\) (i.e.~gravitational) magic gate or (equivalently) state. The associated spacetime isometries explicitly require the non-quaternion generators to express this spacetime's isometries in terms of the entangled particle/anti-particles. Therefore, for perturbed AdS$_3$ the spacetime is still an emergent phenomenon, but defined from the full octonion set of generators.

Just to drive the point home, a graviton needs to be spin \(2\), and no such particle or field exists in this octonion algebra theory. There is no explicit graviton present in this theory.

\end{document}